%% file: sample-authordraft.tex
\documentclass[lettersize,journal]{IEEEtran}

\usepackage{amsmath,amsfonts}
\usepackage{array}
\usepackage{textcomp}
\usepackage{stfloats}
\usepackage{url}
\usepackage{verbatim}
\usepackage{graphicx}

\usepackage{cite}

\usepackage{epsfig,endnotes}
\usepackage{hyperref}
\usepackage{cleveref}
\usepackage{booktabs}
\usepackage{multirow}
\usepackage{todonotes}
\usepackage{tabularx}
\usepackage{colortbl}
\usepackage{comment}
\usepackage[inline]{enumitem}

\graphicspath{{figures/}} 
\usepackage[ruled,norelsize]{algorithm2e}
\SetAlFnt{\small}

\ifCLASSINFOpdf
\else
\fi
\usepackage{amssymb}

\makeatletter
\newcommand{\removelatexerror}{\let\@latex@error\@gobble}
\makeatother

\ifCLASSOPTIONcompsoc
  \usepackage[caption=false,font=normalsize,labelfont=sf,textfont=sf]{subfig}
\else
  \usepackage[caption=false,font=footnotesize]{subfig}
\fi

\usepackage{balance}
\usepackage{xcolor,soul}

\usepackage[most]{tcolorbox}

\newtcolorbox{mytextbox}[1][]{%
  sharp corners,
  enhanced,
  colback=white,
  attach title to upper,
  #1
}

\newcommand{\PaperSubtitle}{An Amplified Covert Channel That Points To Previously Seen Data}

\newcommand{\CCName}{DYST}
\newcommand{\CCFullName}{Did You See That?}
\newcommand{\jkignore}[1]{}

\renewcommand{\hl}[1]{#1}

\begin{document}

%
\title{\CCName{} (\textit{\CCFullName{}}): \PaperSubtitle{}}

%
%
%

\author{Steffen Wendzel~\IEEEmembership{Member,~IEEE}, Tobias~Schmidbauer, Sebastian~Zillien, Jörg~Keller
\IEEEcompsocitemizethanks{
\IEEEcompsocthanksitem S.~Wendzel and S.~Zillien are with the Center for Technology and Transfer (ZTT), Hochschule Worms, Germany\\ E-mail: \{wendzel,szillien\}@hs-worms.de
\IEEEcompsocthanksitem S.~Wendzel and J.~Keller are (also) with the Faculty of Mathematics \& Computer Science at the FernUniversität in Hagen, Germany\\ E-mail: joerg.keller@fernuni-hagen.de
\IEEEcompsocthanksitem T.~Schmidbauer is with the Faculty of Computer Science at the Nuremberg Institute of Technology, Germany\\ E-mail: tobias.schmidbauer@th-nuernberg.de}}


%
%

\markboth{Pre-print (June 2024)}{S.Wendzel, T.~Schmidbauer, S.~Zillien and J.~Keller: \CCName{} (\textit{\CCFullName{}})~\PaperSubtitle{}}
%



\maketitle



\begin{abstract}
Covert channels are stealthy communication channels that enable manifold adversary and legitimate scenarios, ranging from malware communications to the exchange of confidential information by journalists and censorship circumvention.

We introduce a new class of covert channels that we call \emph{history covert channels}. We further present a new paradigm: covert channel \emph{amplification}. 
All covert channels described until now need to craft seemingly legitimate flows or need to modify third-party flows, mimicking unsuspicious behavior.
In contrast, history covert channels can communicate by \emph{pointing} to \emph{unaltered legitimate} traffic created by regular network nodes. Only a negligible fraction of the covert communication process requires the transfer of covert information by the covert channel's \emph{sender}. This information can be sent through different protocols/channels. Our approach allows an \emph{amplification} of the covert channel's message size, i.e., minimizing the fraction of \textit{actually transferred} secret data by a covert channel's sender in relation to the \textit{overall} secret data being exchanged. 
Further, we extend the current taxonomy for covert channels to show how history channels can be categorized. We describe multiple scenarios in which history covert channels can be realized{, analyze the characteristics of these channels, and show how their configuration can be optimized.}
\end{abstract}



%

\section{Introduction}\label{sect:intro}

\begin{mytextbox}[colupper=blue,fontupper=\bfseries\normalsize]
This is a pre-print. The final version of this paper was published by \emph{IEEE Transactions on Dependable and Secure Computing} (TDSC) and is available here (open access):\\
~\\
\url{https://doi.org/10.1109/TDSC.2024.3410679}
\end{mytextbox}

\IEEEPARstart{C}{overt} channels are policy-breaking and stealthy communication channels that are not foreseen in a system's design \cite{Lampson:1973,DoD:1985,petitcolas1999information}. Such channels are regularly used to transfer secret information, e.g., for the purpose of data exfiltration or malware communications \cite{ZanderAB07,OutOfBandSurvey,LCWM:StegoMalware,Luca:StegoMalwareRepo}. However, covert channels can also be applied for censorship circumvention, e.g., by journalists \cite{barradas2020towards,ZanderAB07}. Covert channels have been investigated for different environments, including networks \cite{ZanderAB07,mileva2014covert,CSURpaper}, cyber-physical systems \cite{IoTStego17,Krishnamurthy:2018,hildebrandt2020threat,Lamshoeft22:CPSStego}, local processes/systems \cite{Millen:20Years,XuNWAA19,urbanski2017detecting} and in out-of-band scenarios, such as 
sound, light, vibration, radio-frequency, magnetic fields or temperature \cite{OutOfBandSurvey,BlockNN17,hanspach2014covert,guri2015bitwhisper,guri2017led,guri2018mosquito}.
Similarly to covert channels, some physical-layer security (PLS) methods also aim at providing confidential communication, e.g., through introduction of friendly jamming/noise \cite{gong2021enhancing,mucchi2022security}. {Covert channel signals also aim to hide within noise. The core difference to PLS is that an adversary is explicitly assumed to be aware of the artificial noise, while a covert channel aims to prevent that the \emph{presence} of a secret communication is perceptible to the adversary.}

All known covert channels \emph{embed} secret messages into flows. Since their first appearance {in the 1970s} \cite{Lampson:1973}, authors have performed one of two {embedding} actions: (1) they either relied on the creation of \emph{own} traffic (so-called \emph{active} sending), into which they embed the secret data; or (2) they \emph{modified} legitimate traffic {(or its characteristics)} transmitted by third-party nodes to embed the secret data (called \emph{passive} sending). In both cases, the embedding of secret information renders a channel slightly detectable. This aspect is the central limitation of all previously known covert channels.

To amend this limitation, it would need a covert channel that (at least partially) transfers secret data through \emph{unmodified} legitimate traffic generated by third-party nodes. In other words, it would require the covert channel's sender to craft only a small fraction of the secret traffic (or modify only a small fraction of third-party traffic) while \emph{taking advantage of traffic it neither generates nor modifies}.

In this paper, we present a new covert channel called \CCName{} (\emph{\CCFullName{}}) that fulfills this criterion for the first time since Lampson founded this research area in 1973. In particular, our contributions are as follows:

\begin{enumerate}
  \item \textbf{Novel Class of Covert Channels and Concept of Covert Channel Amplification:} We introduce \emph{history covert channels}. These covert channels advance over state-of-the-art {as they do not modify existing traffic and generate only minimal covert channel traffic. This traffic is solely used for informing (\emph{signaling}) a receiver that a secret message \emph{appears}, while \emph{no} own data traffic (represents the \emph{actual} secret message) is transferred.} This is used to \emph{amplify} the size of the secret message{, i.e., the signal is smaller than the actual secret message and thus more challenging to detect}.

  \item \textbf{Scenario Provision:} We provide multiple scenarios for local network and remote communications to show where the application of history covert channels can be beneficial.
  {In particular, we} show that history covert channels can be applied in highly constrained environments, where a covert sender is incapable of creating any new/modifying any existing traffic and is thus only allowed to send packets with pre-defined content.
  These features render history channels challenging to detect.
  
  \item \textbf{Taxonomy Extension:} We extend the existing taxonomy for network covert channels with a new category called \emph{fully-passive sending} to reflect all components of this new class of covert channels.

  \item \textbf{Functionality Description for Multiple Methods:} As history covert channels comprise a whole family of variants, we provide a description of their functionality using different variants called \CCName{}-Basic (for local networks), \CCName{}-Ext (also for local networks), \CCName{}-Remote-Smarthome (for local-to-remote connections) and \CCName{}-Remote-RTCP (for local-to-remote connections).
  
  \item \textbf{Theoretical Analysis:} {We
  conduct a theoretical analysis of the performance and optimization of
  \CCName{}.}
  
  \item \textbf{Implementation:} {
  We describe} several ways 
  to implement a history covert channel in a network environment.
  {As an example}, we provide the first implementations of such a channel for \CCName{}-Basic and \CCName{}-Ext. 
  Our implementations contain a \emph{data} channel that requires no own or modified traffic and a \emph{signal} channel that consists solely of rarely sent 
  packets with legitimate content (representing only 1 covert bit).
  
  \item \textbf{Evaluation:} Using both local network implementations, we evaluate \CCName{}'s robustness, detectability, and optimization under different settings and show that \CCName{}-Basic and \CCName{}-Ext allow the transfer of variable secret data bits through only $1$ covert signaling bit, which state-of-the-art covert channels do not achieve.
  We further simulate a remote version of \CCName{}-Basic called \CCName{}-Remote-Smarthome to show the feasibility of history channels outside of local networks {as well as a throughput-enhanced multi-pointer variant}.
\end{enumerate}

The remainder of this paper is structured as follows. Sect.~\ref{sect:relwork} provides background information and discusses related work while Sect.~\ref{sect:method} presents the functioning as well as the theoretical description and optimization of \CCName{}. We describe our experimental testbed and the evaluation of \CCName{} in local networks in Sect.~\ref{sect:eval:LAN}. Next, we conduct a feasibility-analysis of a remote scenario for \CCName{} in Sect.~\ref{sect:eval:remote} and discuss a performance optimization in Sect.~\ref{sect:DYSTMultiHash}. A discussion is provided through Sect.~\ref{sect:disc}. Finally, Sect.~\ref{sect:concl} concludes. 
{The electronic supplement covers additional optimization aspects of \CCName{}.}

\section{Background \& Related Work}\label{sect:relwork}

\paragraph*{Covert Channels}
A \emph{Covert Channel} exchanges information in a stealthy manner between a \emph{covert sender} (CS) and one or more \emph{covert receiver(s)} (CR). A covert channel is one that is not foreseen in a system's design \cite{Lampson:1973} and relies on the concept of policy-breaking communication \cite{DoD:1985}. If the covert information can be received by more than one CR, the communication can be considered as a multicast or even broadcast covert channel. In computer networks, the covert channel nests into a network protocol, e.g., by manipulating bits of a packet header or by adjusting the delays between successive network packets. 
{Similar to covert channels, some physical-layer security methods utilize nearby network nodes' emitted (artificial) noise} \cite{gong2021enhancing}, {and steganography tools typically aim to hide within noise, which also applies to \CCName{}'s signaling channel (but not its data channel).}

In \cite{Cabuk06,CabukTimingChanDet_2009,Cabuk:2004:ICT:1030083.1030108}, Cabuk et al.\ propose the idea of an advanced version of a covert channel based on delays between packets: the covert channel transmits the hidden data by modulating the delays between consecutive network packets.
The advanced version of their covert channel mixes covert transmissions with sections of real, legitimate network traffic. This helps to skew the statistics and makes the detection of the covert channel harder.
The difference to our approach is that Cabuk et al.\ use sections of legitimate traffic solely to introduce noise into the actual covert channel signal. The legitimate traffic carries no hidden information at all.

Several additional methods work similarly to the one of Cabuk et al. For instance, \emph{JitterBug} by {Shah and Molina} \cite{JitterBug} adds random delays to legitimate Telnet traffic. Walls et al.\ proposed \emph{Liquid} \cite{Walls2011Liquid}, an extension of JitterBug, in which they split the channel into ``transmitting'' and ``shaping'' delays (shaping delays carry no information but manipulate the statistics of traffic). Similarly, Gianvecchio et al.\ \cite{ModelbasedCTC} tailor traffic automatically based on the statistical characteristics of legitimate traffic.
{In all these cases}, artificial modifications are performed to transfer secret information, even if based on legitimate traffic, which is the key difference to \CCName{}.

There are also approaches that work on a more abstract level. Yarochkin et al.~\cite{DBLP:conf/prdc/YarochkinDLHK08} proposed the so-called \emph{network environment learning} phase. This approach was used solely to determine which protocols occur regularly in a network to succeedingly exploit only these protocols for covert communication.
No work is known that exploits legitimate traffic for a covert channel. Moreover, the covert channel of Yarochkin et al.\ did not split the covert channel's control channel from its data channel as we do.

Image steganography uses a variant called \emph{cover selection}, where a database of images is used, a hash function is applied to each image, and if the secret message matches the hash value, the image is sent by the covert sender \cite{Fridich2009}. Several methods have been proposed to conduct or optimize the cover selection process, e.g., \cite{CoverSelection2,CoverSelectionImageSimilarity,CoverSelection3,CoverSelection4,CoverSelection5,CoverSelection6}.
Similarly, \emph{coverless image steganography} utilizes a database of image patches to (partially) reconstruct an original secret message using image patches, see, e.g., \cite{CoverlessImageStego} and \cite{CoverlessImageStego2}.
In contrast to cover selection and coverless image steganography, our method operates in a network and it relies on network traffic \emph{that is transmitted anyway}, i.e., independent of the steganographers, and only uses signals{, e.g., ARP or RTCP} requests, instead of generating extra traffic such as images, even if those images are innocent.

{The Address Resolution Protocol (ARP) is used to associate a link layer address, e.g., MAC (Medium Access Control) address to an IPv4 address, and to this end uses request and reply packets on the link layer. 
ARP had been utilized in} \cite{DBLP:conf/wcnis/JiFM10} {to implement a local covert channel by encoding covert information into the target IP field of an ARP request, sending the request directly to the covert receiver. Another approach utilizing ARP for covert communication had been described in} \cite{Schmidbauer:DeadDrops:ARP}{, wherein ARP requests are exploited to store covert information within ARP tables of uninvolved intermediary third-party nodes. The covert information have to be fetched by the covert receiver utilizing SNMP. The Real-Time Protocol (RTP) and the Real-Time Control Protocol (RTCP) are used for media streaming. We use these protocols for our PoC remote scenario. These protocols had been exploited in} \cite{DBLP:conf/otm/MazurczykS08,DBLP:conf/iih-msp/BaiHHX08,lizhiRTP} { for covert communication. In} \cite{DBLP:conf/otm/MazurczykS08} {, the authors utilized various ways to directly send covert information to a covert receiver by crafting and manipulating RTP traffic. The techniques include beside others the manipulation of timing behavior, and embedding information in the padding, the extension, and the sequence number field. In} \cite{DBLP:conf/iih-msp/BaiHHX08}{, the jitter field is manipulated for directly sending information to the covert receiver, and in} \cite{lizhiRTP}{, the authors significantly improve RTP and RTCP based timing channels. While all these covert channels had been implemented with ARP or RTP/RTCP, every former implementation had to \textit{craft or manipulate} packets of these protocols to transport covert information to covert receivers. The approach described in \textit{our} paper significant differs from  these former implementations as there is no need to craft a packet carrying the covert data to transmit to the covert receiver, as legitimate packets carry the covert, but still legitimate information that are solely signalled to be part of a secret message.}

There is one method that actually splits the control channel from the data channel: as shown by Wendzel and Keller in \cite{wendzel2014hidden}, several covert channels propose to utilize internal control protocols. Therefore, a covert channel is nested into the utilizable bit areas of a network packet. Some of the utilizable bits are used for the control protocol, while others are used for the data channel. However, that approach has a major limitation compared to ours as both, the control and the data channel, reside in the same packet and modify legitimate packets or craft new packets instead of exploiting solely unaltered legitimate traffic for the data channel.

Finally, there is one proposal by Caviglione et al.\ from botnet research that works by waiting for a pre-defined network packet sequence \cite{BotnetBookCh2019}. If the sequence occurs, the botnet nodes would perform a certain action. The idea was solely described on a conceptual level and was not implemented by the authors. Further, their concept did not involve the option to influence which secret message is transferred, which makes it fundamentally different from our history channels.

{HICCUPS }\cite{HICCUPS}{ points to data \emph{within} a network packet. Therefore, a sender corrupts a frame checksum, which indicates the presence of a secret message in the frame's payload. HICCUPS is tied to LAN environments and the frame-level. It further does not point to historic data but solely \emph{places} the pointer \emph{and} the secret data inside the packet (instead of finding matching third-party data). Thus, HICCUPS is not a \emph{history} covert channel and provides no amplification.}

\paragraph*{Network Covert Channel Detection}
Several detection methods for covert channels have been proposed throughout the years. Popular ones are, e.g., compressibility score \cite{CabukTimingChanDet_2009}, $\epsilon$-similarity \cite{Cabuk:2004:ICT:1030083.1030108}, regularity metric \cite{Cabuk:2004:ICT:1030083.1030108}, a method from Berk et al.~\cite{BerkEtAl}, as well as classical methods, such as  Kullback-Leibler divergence test \cite{zhang2018covert}, Kolmogorov-Smirnov \cite{zhang2018covert} or entropy-based analyses \cite{CSURpaper}. All of these methods require at least a few hundred covert channel packets to provide somehow reliable detection of covert channel flows. In contrast, history channels send few \emph{signaling} packets per time, resulting in only a minimal influence on a flow while the data flow of history channels is entirely legitimate and thus indistinguishable. We evaluate two common detection methods on \CCName{} in Sect.~\ref{sect:eval:LAN}.

\section{The History Covert Channel Method}\label{sect:method}

In this section, we first discuss the requirements of our history covert channel, followed by a description of the detailed functionality of \CCName{}. Further, we explain the chosen parameters for \CCName{}, including the optimization, and finally extend the existing taxonomy of active and passive covert channels.

\textbf{Definition 1.} A \textbf{history covert channel} is one that \emph{points} to already existing (live or stored) data that matches a secret message instead of \emph{sending} a secret message itself. The only covertly transferred information is the pointer. $\blacksquare$

\begin{figure}[ht]
  \centering
  \includegraphics[width=0.99\linewidth]{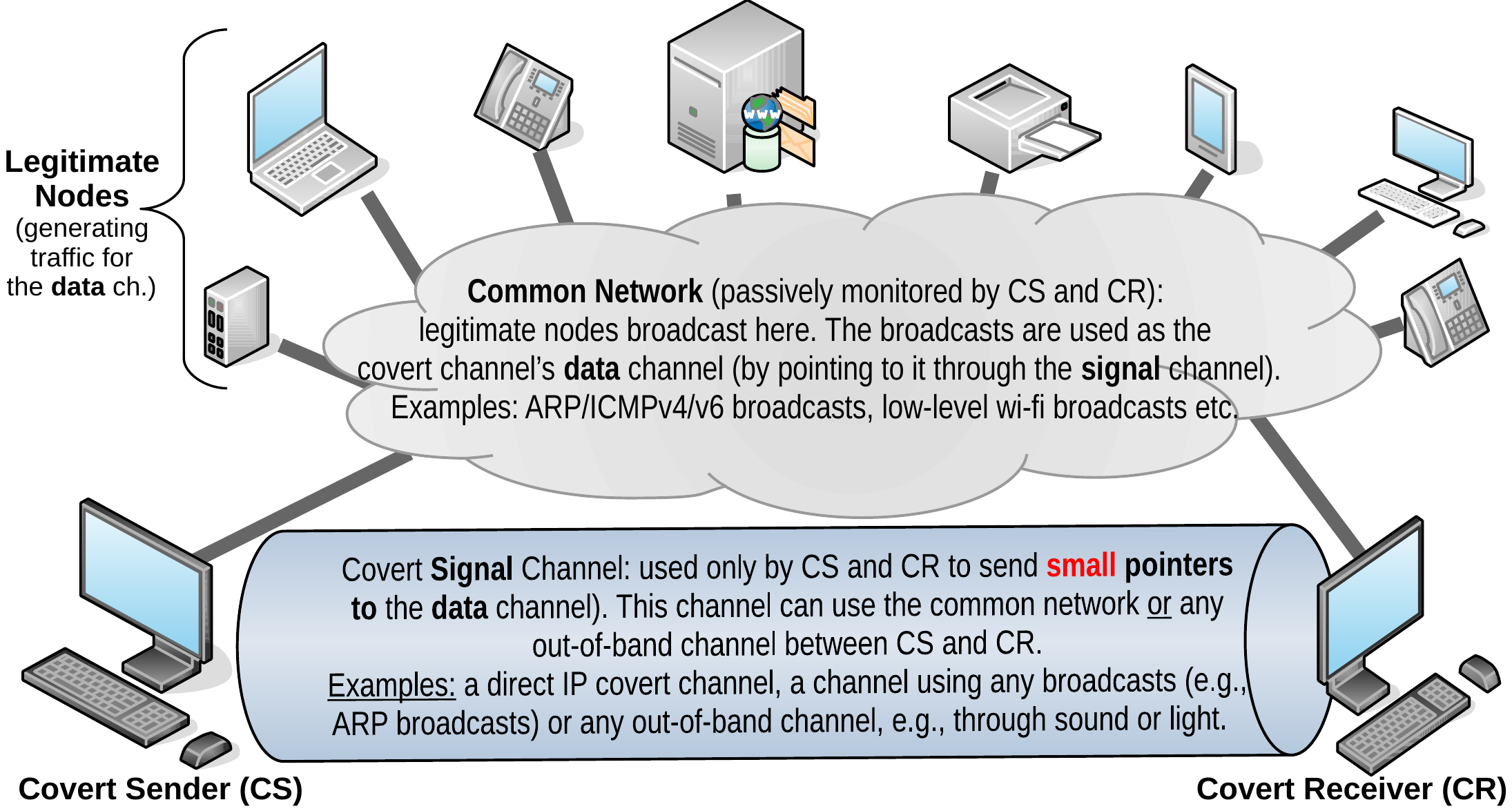}
  \caption{General functioning of a history covert channel {that uses broadcasting}. {The amplification is achieved by sending small pointers that refer to larger data pieces (e.g., packets or their hash values) of the data channel}.}
  \label{fig:histcc:general}
\end{figure}

{Data, such as packets, that do \emph{not} match a secret message, are not processed further by a covert sender, i.e., no pointer will refer to these. Note that}
pointers may be considered as a modification of third-party or a crafting of own traffic. However, the content {that is pointed to is} legitimate. Sending small-bit pointers to secret data with \textit{more} bits thus minimizes the fraction of covert traffic to be sent/modified in comparison to covert data transferred overall, which we believe is a key novelty in this domain. This means that history covert channels perform a form of covert message size \emph{amplification}. {We call the size-increasing factor for the covert message the \emph{covert amplification factor} (CAF).} 
{The functioning of a history covert channels that uses legitimate broadcasts as a data channel} is visualized in Fig.~\ref{fig:histcc:general}. {Note that non-broadcast data channels can be used too, which we will detail for remote scenarios in Sect.}~\ref{sect:threatmodel}{.} 

\textbf{Note.} We chose the specific \textit{proof-of-concept} (PoC) implementation discussed in the remainder to simplify experiments and explanations. We note, however, that history covert channels are not restricted to local networks and more variants are possible, cf.~Sect.~\ref{sect:disc}.

\begin{table}[ht]
    \caption{Table of used notations in this paper.}
    \label{tab:symbols}
\centering
\begin{scriptsize}
    \begin{tabular}{cp{6.4cm}}\toprule
     \textbf{Symbol}  & \textbf{Definition}\\
     \midrule
          $|.|$        & length of a string $.$ or cardinality of a set $.$\\
          $bw_{basic}$ & bandwidth of \CCName{}-Basic\\
          $bw_{ext}$   & bandwidth of \CCName{}-Ext\\
          $C(.)$       & checksum function on input $.$\\
          $c$          & number of bits used for error correction \\
          $D$          & sender-side robustness timeframe between two \textit{packets of interest} (PoI)\\
          $d(x,y)$     & hamming distance between bitvectors $x$ and $y$\\
          $dist_{basic}()$ & signaling distance of \CCName{}-Basic\\
          $dist_{ext}()$ & signaling distance of \CCName{}-Ext\\
          $extchksm(x)$& function to extract the last $c$ bits from a bitvector\\
          $extmsg(x)$  & function to extract the first $h-c$ bits from a bitvector\\
          $F(a)$       & cumulative distribution function of a random variable $X$, i.e., $P(X\leq a)$\\
          $H(.)$       & cryptographic hash function using input $.$ \\
          $h$          & length of hash value \\
          $h_i$        & hash value $i$ \\
          $\Im$        & compressing function \\
          $\kappa$     & compressibility score \\
          $k$          & number of message $M$'s fragments \\
          $len(M)$     & length of message $M$ \\
          $M$          & secret message \\
          $\tilde{M}$  & encoded message $M$ with concatenated checksum \\
          $P(X)$       & probability function for $X$\\
          $p_i$        & network packet $i$ \\
          $R$          & receiver-side robustness timeframe between two PoI\\
          $S$          & string representation of concatenated inter-packet delays\\
          $S_{h_i}$    & set of bit vectors with distance $\leq t$ to $h_i$\\
          $T_{h,t}$    & number of possible modifications of a hash value to test\\
          $t$          & number of non-matching bits of a message \\
          $U_{h,t,c}$  & chance that message $M$ has the first fit among $T_{h,t}$ possibilities\\
          $X$          & random variable \\
          $x$          & bit vector of length $h$\\
         \bottomrule
    \end{tabular}
    \end{scriptsize}
\end{table}

\subsection{Threat Model and Requirements}\label{sect:threatmodel}

We assume that it is beneficial for a CS to send as few secret information bits to a CR as feasible. In particular, we consider a scenario where only legitimate-\textit{content} messages (e.g., regular ARP broadcasts with their unaltered content) are allowed to be sent from CS to CR, and that only the \textit{timing} of these legitimate-content messages is used as a \emph{referrer} to some third-party packets containing the actual secret content. Thus, we assume a hybrid covert channel, split into a signaling channel (from CS to CR) and a data channel (traffic of third party nodes) that both contain solely legitimate content (see Fig.~\ref{fig:histcc:general}). Our channel does not match any of the recent patterns of indirect covert channels found in \cite{SchmidbauerWendzel:IndirectCCSurvey}.

\noindent Several scenarios are imaginable where using such a covert channel would be beneficial:

\begin{figure}[ht]
  \centering
  \includegraphics[width=0.49\textwidth]{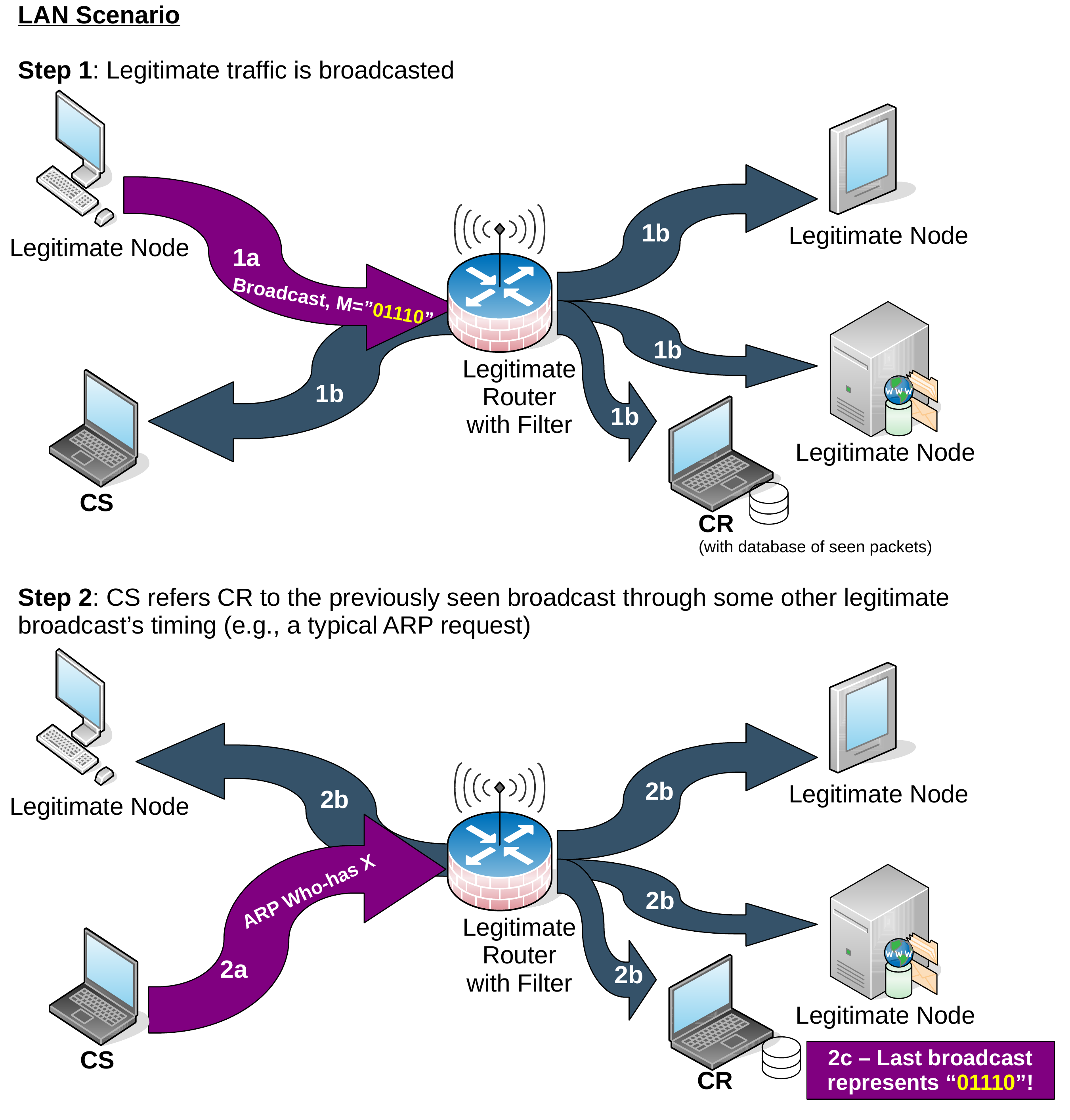}
  \caption{Confined communication in LAN: CS is not capable of placing the secret message content by itself due to restrictions.}
  \label{fig:DYST:scenario:LAN}
\end{figure}

\begin{figure}[ht]
  \centering
  \includegraphics[width=0.49\textwidth]{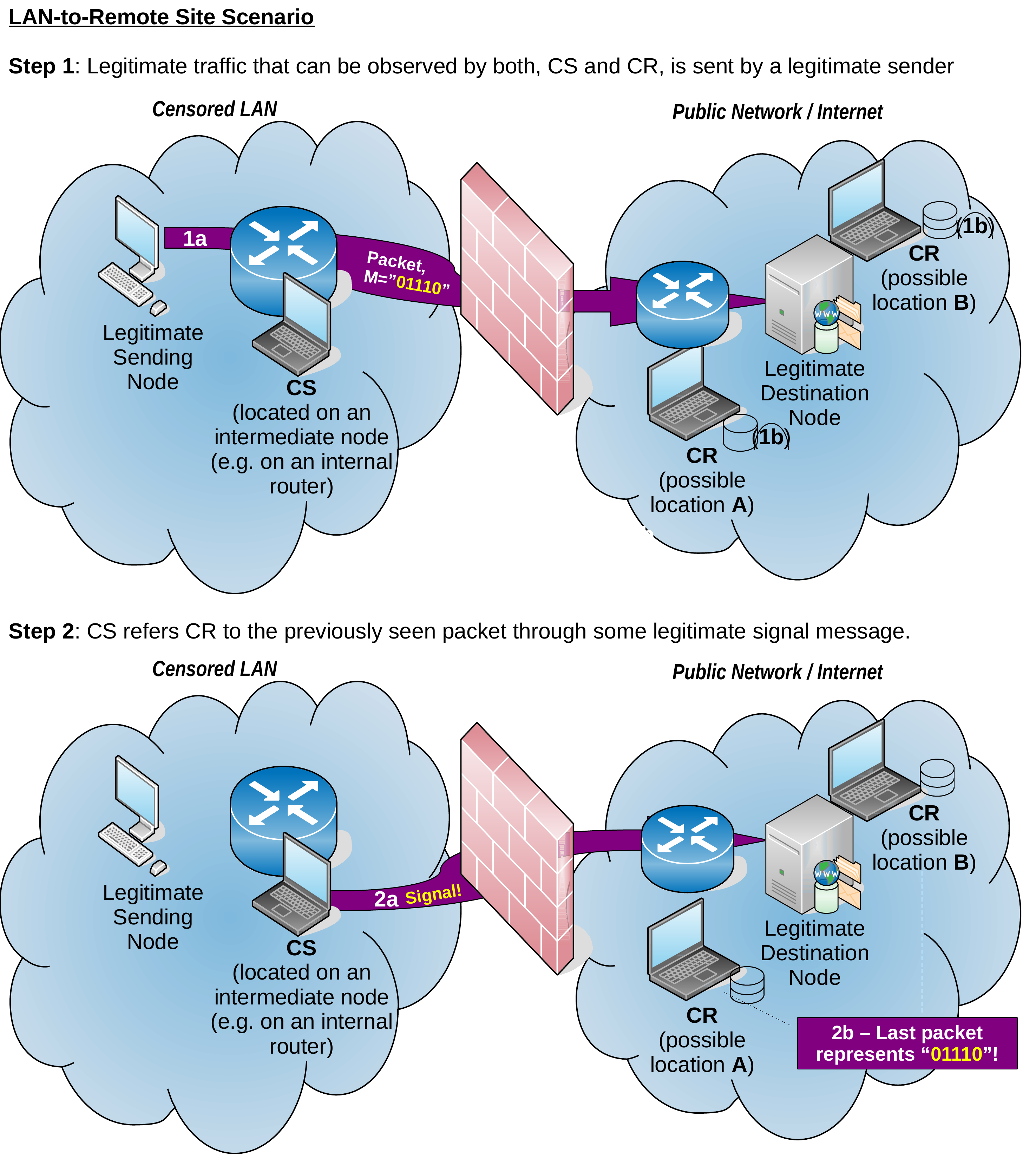}
  \caption{Confined communication from a LAN to a remote site: CS resides inside the local network and CR resides in an uncensored public network.}
  \label{fig:DYST:scenario:remote}
\end{figure}

\textbf{\textit{Local scenario 1/1 (\CCName{}-Basic and \CCName{}-Ext):} Highly constraint LAN environments}, where a CS cannot freely craft the content of network packets at will due to a filter and can barely manipulate the timing of some legitimate messages. This scenario is shown in Fig.~\ref{fig:DYST:scenario:LAN}. CS could try to establish a covert timing channel with CR in such a scenario. However, typical covert timing channel detection methods might detect a classical timing channel. For this reason, \CCName{}'s timing channel does not contain the actual secret information. Instead, it is bound to the occurrence of legitimate third-party messages called \textit{packets of interest} (PoI) exchanged within the network environment and visible to CS \textit{and} CR. In our LAN scenario such PoI messages are broadcasts. Legitimate nodes send such broadcasts (e.g., ARP or ICMPv6 packets) on a regular basis (step 1a in Fig.~\ref{fig:DYST:scenario:LAN}). PoIs are then received by the nodes of the local network, including CS and CR (step 1b in Fig.~\ref{fig:DYST:scenario:LAN}). The CR keeps track of all recently seen broadcasts in a local database (packets older than a few seconds can be discarded).
Once a received third-party broadcast \emph{matches} the secret message, CS sends an own legitimate broadcast message (step 2a). This message can be any kind of broadcast message that CS would send on a regular basis (e.g., ARP or ICMPv4/v6). Note that CS only sends such a timing signal to CR if a match was found. For this reason, the timing channel messages (signal messages) of CS only occur after such legitimate messages appear, and thus do not reflect a typical timing channel pattern. 
In step 2c, CR loads the most recent legitimate broadcast (that was \textit{not} sent by CS) from its database and extracts the secret information from that packet.
Note that due to the secret message being represented by some third-party traffic, CS does not need the capability to craft such messages by itself, since it \textit{points} to these messages instead using the signal.

\noindent\textbf{Remote Scenarios}
Second, there are non-local scenarios imaginable, where \CCName{}-Basic would be applicable but utilizes different protocols.

\noindent\textbf{\textit{Remote scenario 1/2:} Highly-constrained remote access/censorship circumvention (\CCName{}-Remote-Smarthome).} Similarly to the local scenario, we assume that CS operates in a restricted environment, where only pre-defined packets can be sent by CS and where CS can influence the time at which these are sent. For instance, some smart devices regularly communicate from internal networks to public services, such as cloud providers. This is a \textit{Man-in-the-Middle} (MitM) setup, where an on-path CS would be located in the LAN (alternatively on a router or on the path to the destination) and where CR would be located down-path outside the LAN, e.g., in the public Internet, close to the destination or residing directly at the destination. In other words, CR must have direct access to the traffic sent by the legitimate sender. This scenario is shown in Fig.~\ref{fig:DYST:scenario:remote}. Note that \emph{no broadcast traffic} could be used in this scenario as it would not reach the CR.
Therefore, CS needs to wait for regular unidirectional (or multicast) messages (step 1a) from the legitimate sender to a destination where the traffic passes CR or where CR \textit{is} the destination. As in the \emph{local} scenario, CR records all recent \textit{packets of interest} (PoI) in a local database (step 1b). Afterward, CS needs to send some unprohibited \emph{signaling} packet observable by CR (step 2a), such as a whitelisted (i.e., not filtered) DNS or HTTP request.
For instance, if CR is a web service that has the current news or the current canteen menu on its webpages, then CS could time the specific allowed request to CR.\footnote{We assume that HTTP requests face deep packet inspection or are normalized so that CS has indeed no option to modify pre-defined requests of a set of allowed network services CS can use.} Alternatively, if CS is not allowed to send packets to CR directly, CS might send data to a third-party node observed by CR. For instance, CS and CR could use a network entity, e.g., a feed where CS is allowed to ``like'' postings (but not posting any content) so that the timestamp of a like is observable by CR to establish the signal channel. Finally, CR interprets the last observed packet's secret content (step 2b) to extract the secret message.

\noindent\textbf{\textit{Remote scenario 2/2:} Censorship circumvention using RTCP (\CCName{}-Remote-RTCP).} 
This scenario is a special form of \textit{remote scenario 1} and added solely to underpin the existence of this option. This scenario is called \CCName{}-Remote-RTCP and can be imagined in settings where sender and receiver can establish a communication which includes at least a partial time synchronization, e.g., a media communication via 
RTP in conjunction with 
RTCP \cite{RFC3550}. The PoI comprise the media traffic itself which is transported via RTP. A time synchronization is achieved via Sender Reports (SR) in RTCP which are regular messages that include a Network-Time Protocol (NTP) timestamp and counters for how much traffic has been sent in the last period. Pointers are signalled within such SR.
CS and CR listen for traffic on both protocols, and CR records from RTP (which hosts the majority of data traffic) all PoI of the time interval since the last SR.
CS only needs to signal within such an SR if a suitable piece of information has been sent within a PoI since the last SR.
The pointer to this information may consist of an offset, or may be a part of the information itself that is sufficient to uniquely identify the information within all PoI since the last SR.
Thus, the history channel at least reduces the fraction of secret bits that the covert sender needs to transfer explicitly through such a covert channel in contrast to a covert channel that uses RTCP alone (cf., e.g., \cite{interprotocolstego}), and still has no need to modify traffic in RTP protocol itself.

\textbf{Environment Characteristics for History Covert Channel:} \CCName{} relies on some characteristics of the environment in which CS and CR operate:

\begin{enumerate}
    \item CS and CR must be able to observe some messages which they both receive (almost) at the same time. There are different options to achieve that: \textit{i)} CS and CR could read broadcast messages in a local WiFi network that both of them receive; \textit{ii)} CS and CR could only evaluate messages that pass through the routing path of CS \emph{and} CR (e.g., when both act as routers); \textit{iii}) CS \emph{and} CR are members of a multicast-group, receiving frequent updates (for instance IGMP, like exploited for an active indirect covert channel in \cite{OvadiaOMGO19}). In any case, CR must be able to store information about the recently received packets in a local database (older packets can be discarded from the database);

    \item To assign the same secret data to observed packets, CS and CR need to utilize the same procedure. This means that they need to read the same input parameters (i.e., which packets and which header \emph{and} body fields of these packets, timestamps, and so forth are used) of the network traffic and also calculate the same function over these data (e.g., a hash can then represent secret data);

    \item Moreover, CS must be able to send the signal so that it is observable by CR in a timely manner, e.g., CS might be able to send a legitimate ARP {(or, e.g., DHCP, ICMPv6)} broadcast to the local WiFi network or a remote message. While it is not within our scope, CS might alternatively use some an out-of-band covert channel \cite{OutOfBandSurvey} to signal CR.
\end{enumerate}

Optionally, CS and CR may even reside on two existing legitimate nodes as there is no need for CS and CR to utilize their own dedicated nodes. The above-mentioned conditions can also be fulfilled outside the scope of networks.

\textbf{Adversary Capabilities:} We assume that the adversary (e.g., a government-level Internet censor) has highly restrictive capabilities:
\begin{enumerate}
    \item The adversary is able to \emph{observe all} traffic exchange in the local network, where CS resides.
    \item The adversary can further \emph{block any} traffic sent by CS if its content (that is a combination of packet header and payload) does not match an element of a set of pre-defined legitimate messages. For instance, in a local network scenario, CS might be able to only send one specific ARP broadcast that calls for the address of the local network's router.
    \item The adversary is also able to observe \emph{all} traffic that is exchanged on the path between CS and CR, even if CR resides in a remote network. If the adversary detects a suspicious communication, it can decide to block the communication from CS to CR for an arbitrary time.
    \item {The adversary is not capable of locally monitoring actions or resources on CS or CR, i.e., there is no adversary malware on CS or CR.}
\end{enumerate}

\subsection{Functioning of \CCName{}}\label{subsec:functioning}
In this section, we primarily describe two essential variants of \CCName{}: first, we explain \CCName{}-Basic, a simple variant where all bits of an observed hash need to match the chunk of the covert message.
Second, we will describe \CCName{}-Ext, which is an extended variant capable of transferring correct information, though the hashes are not matching perfectly.
This approach creates more variants to choose from in the tradeoff between steganographic bandwidth and detectability. Further, we describe derivatives of \CCName{}-Basic: \CCName{}-Remote-Smarthome and \CCName{}-Remote-RTCP for the remote scenarios 1 and 2 (matching Fig.~\ref{fig:DYST:scenario:remote}). Note that an overview of the notations used in this paper is given by Tab.~\ref{tab:symbols}. {Further, Tab.} \ref{tab:DYSTanalysis:overview} {summarizes how each \CCName{} variant is handled in this paper.}

\begin{table*}[ht]
    \caption{Overview of \CCName{} variants in this paper.}
    \label{tab:DYSTanalysis:overview}
\centering
\begin{scriptsize}
    \begin{tabular}{p{1.1cm}|p{4.5cm}|p{2.5cm}|p{2cm}|p{5.5cm}}\toprule
     \textbf{\CCName{} variant}  & \textbf{Matching Scenario (Sect.~\ref{sect:threatmodel})} & \textbf{Functionality Description (Sect.~\ref{subsec:functioning})} & \textbf{Optimization (Sect.~\ref{sect:parameterChoice})} & \textbf{Evaluation Method for Covert Channel's Characteristics (Sect.~\ref{sect:eval:LAN} / Sect.~\ref{sect:eval:remote})}  \\
     \midrule
          \textbf{Basic}  & Constrained LAN & ~~~~~~~~~~\checkmark & ~~~~~~~~~~\checkmark & Implementation, with partial simulation \\
          \textbf{Ext}    & Constrained LAN & ~~~~~~~~~~\checkmark & ~~~~~~~~~~\checkmark & Implementation, with partial simulation \\
          \textbf{Remote}    & Constrained Network with Remote Link / Censorship Circumvention &  \multicolumn{2}{c|}{~~~\checkmark (essentially the same as \CCName{}-Basic)} & Due to overlap with \CCName{}-Basic: only simulation with testbed traffic from a smart home \\
         \bottomrule
    \end{tabular}
    \end{scriptsize}
\end{table*}


\subsubsection{\CCName{}-Basic}
\begin{figure*}[ht]
  \centering
  \includegraphics[width=0.9\textwidth]{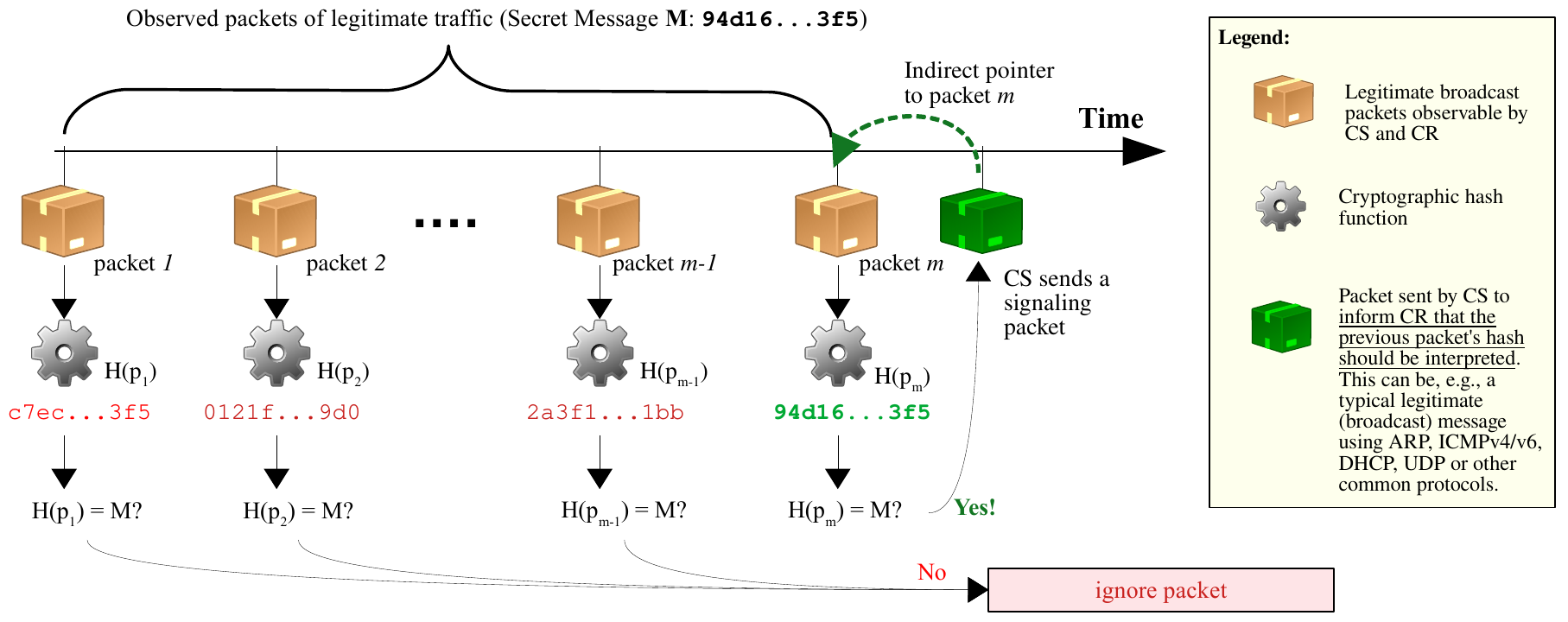}
  \caption{The \CCName{}-Basic Sending Process. {\emph{Hashes} representing a secret message can be calculated over whole packets or just parts of packets (e.g., selected header fields). Note that packets with non-matching hash values are not processed further. \emph{Signaling} packets can take the form of any typical (legitimate) broadcast, such as ARP or DHCP requests.}}
  \label{fig:CC_schema_Wifi:Backwards}
\end{figure*}

For \CCName{}-Basic, the following steps must be performed continuously until a whole message is transferred (see Fig.~\ref{fig:CC_schema_Wifi:Backwards}):
\begin{enumerate}
   \item CS and CR both record legitimate traffic with their network interfaces connected to a shared network.
   \item For each 
   packet{ of interest} $p_i$ that both CS and CR can observe (e.g., broadcast messages), they apply a hash function $H()$ to the input values
   they have agreed on,
   to generate a hash value $h_i = H(p_i)$ of length $h$.
   
\end{enumerate}

CS and CR can exchange a secret message in a bidirectional manner. In the remainder, we describe the sending process from CS to CR. However, CR can simultaneously operate as a CS and CS as a CR to send/receive data.

The sending process can be described as follows (see Fig.~\ref{fig:CC_schema_Wifi:Backwards}) and requires that each secret message $M$ to be sent has the length $h$. To transfer a secret message of length $k\cdot h, k \in \mathbb{N}, k\geq 2$, the message must be split into $k$ fragments and the sending process must be performed successively for each of the $k$ fragments.
\begin{enumerate}
    \item To signal a secret message of length $h$, CS waits for a 
    packet{ of interest} $p_i$ with a hash value $h_i=H(p_i)$ which equals $M$.
    \item After the CS observed such a
    packet,
    it sends a signal to CS. In our \emph{example}, we use legitimate ARP requests with which CS asks for the address of some legitimate node (e.g., a router) in the network. 
    \CCName{}'s {example} ARP request can be replaced by any other seemingly legitimate unicast, broadcast, multicast, or anycast message, observable by the CR.
    Other possibilities for signals are discussed in Subsections \ref{subsubsec:dyst-remote-Smarthome} and \ref{subsubsec:dyst-RTCP}.
\end{enumerate}

Finally, the receiving process is conducted as follows:
\begin{enumerate}
    \item CR interprets the occurrence of 
    {the signal message}
    from CS as a 
    {prompt}
    that the expected message can be observed in the data channel. 
    \item CR interprets the 
    hash value
    of the previous packet of interest
    that 
    represents the covert message.
\end{enumerate}

Obviously, the channel is noisy and requires error detecting (and correcting) bits or mechanisms to ensure a robust transmission of the correct information.

Moreover, CS reaches multiple CR simultaneously, if desired, rendering the channel a multicast or broadcast covert channel.

The major advantage of such a history covert channel's sending procedure is that CS needs to send only one bit of covert information (represented by {the existence of the signal message, e.g., an ARP request or RTCP packet}) to transfer $h$ bits of secret content, i.e., the CAF for the covert message's size is $h$ (or a fraction of $h$ if more than one signaling bit must be sent).
\CCName{} {modifies the inter-packet gaps by introducing signal messages, while existing timing channels modify timing of existing packets to transmit a bit of the secret message.
We will treat this in more detail when analyzing detectability, cf. Sect.}~\ref{sect:detection}.

Assuming that each of the $2^h$ possible hash values is equally likely, then on average an exact match between secret message $M$ and hash value $h_i$ will be achieved after $2^h$ packets. If an exact match is not required but up to $t$ bits can be wrong, i.e.\ the hamming distance between secret message $M$ and hash value $h_i$ can be at most $t$, then the chance of such a partial match would increase by a factor $\sum_{i=0}^t \binom{h}{i}$. To enable the CR to still decode the message correctly, the message would have to be encoded with an error-correction code that allows correction of up to $t$ bit errors. This means that only $h-c$ bits are available for the secret message itself and the remaining $c$ bits are used for the error-correction bits \cite{ClarkECC1981}. For a binary block code, we have $c>2t$ as $c=2t$ is a bound achieved by \emph{maximum distance separable} (MDS) codes, which however only exist in trivial form for binary\footnote{Using non-binary codes decreases match probabilities further and thus is no option, either.} codes \cite[Prop.~9.2]{Vermani1996}. Furthermore, besides a strong restriction on the number of message bits (the larger $c$, the smaller $h-c$), only some combinations of $h$, $c$, and $t$ 
are available for applicable code families such as binary \emph{Bose-Chaudhuri-Hocquenghem}
(BCH) code, see e.g.\ \cite[App.~A]{ClarkECC1981}.
Our initial investigations revealed that error-correction codes only in some cases match the performance (in terms of bandwidth and average signal distance) of \CCName{}-Basic, and mostly perform worse.
Yet we carried over the idea of using partial matches to using checksums instead of error-correction codes, creating an extended version of \CCName{}.

\subsubsection{\CCName{}-Ext}
The functionality of \CCName{}-Ext works similarly to \CCName{}-Basic, but the secret message $M$ now only comprises $h-c$ bits (this reduces the CAF by $c$ bits in comparison to \CCName{}-Basic), and is concatenated by CS with a $c$-bit checksum to an encoded message $\tilde{M}$ of length $h$.
When comparing $\tilde{M}$ with $h_i$, CS allows up to $t$ non-matching bits.
Thus, the advantage of \CCName{}-Ext comes from the fact that it can utilize a larger fraction of observed messages, leading to a shorter waiting time for fitting packets.

CR, upon receiving a signal (such as an ARP request) indicating a secret message transfer, again picks up the latest hash value $h_i$ from the hash database. It then tries out all possibilities to flip up to $t$ bits in $h_i$, until the checksum of the first $h-c$ bits in the modified $h_i$ matches the last $c$ bits in modified $h_i$ (called a \textit{hit}) for the first time. CR accepts the first $h-c$ bits as secret message $M$.

CS knows the order in which CR will apply modifications to the hash value $h_i$. Thus, CS can check if the first hit really will produce the message $M$. CS will only send an ARP request as a signal for CR 
if at most $t$ bits of $h_i$ and $\tilde{M}$ do not match \textbf{and} the first hit found by CR will produce $M$. So, not all $t$ bits matching packets can be utilized for this approach.

\begin{figure}[ht]
  \centering
  \includegraphics[width=0.99\linewidth]{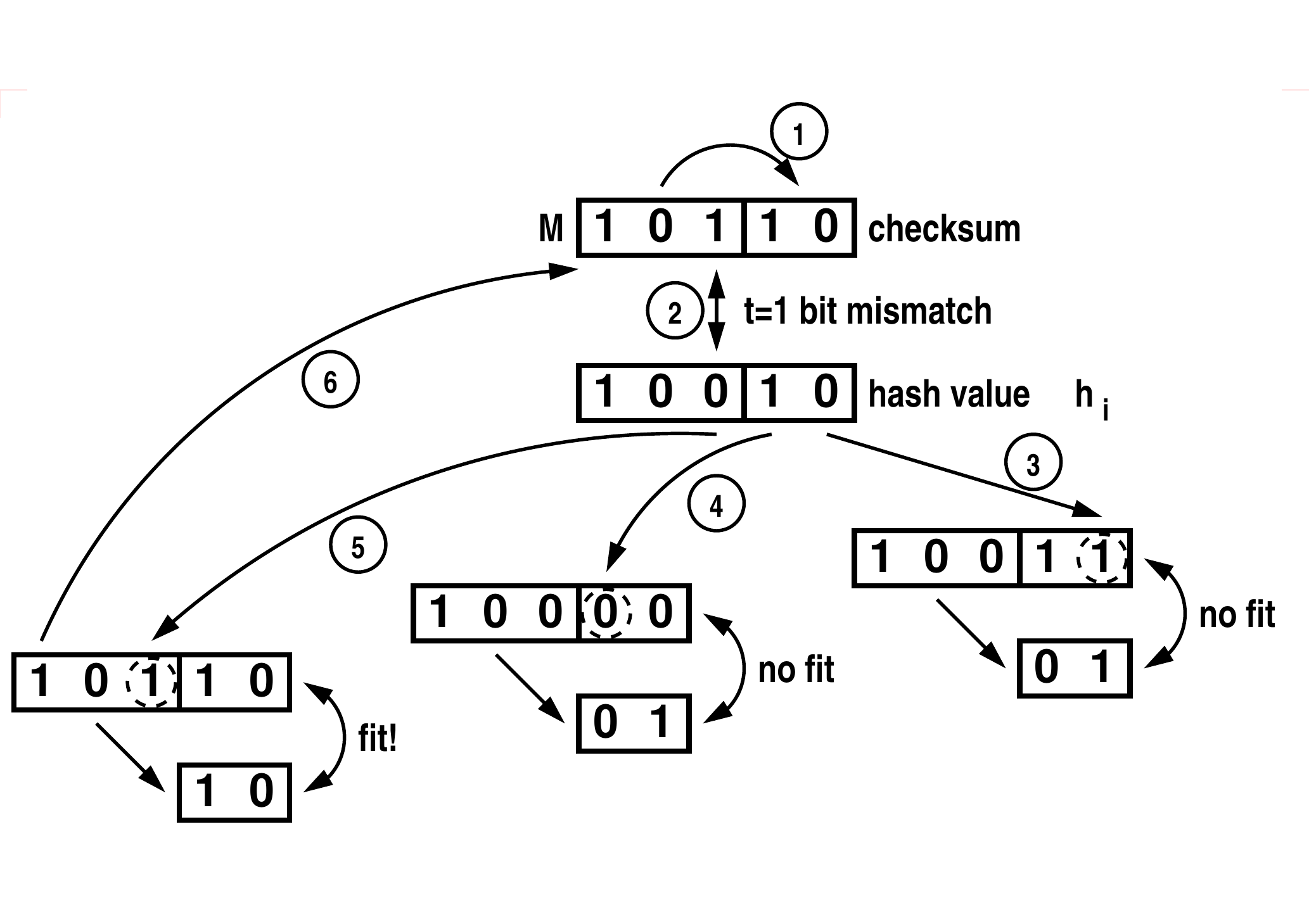}
  \caption{Functioning of \CCName{}-Ext. For illustration, we use a message chunk $M$ of $h-c=3$ bit length and a checksum of $c=2$ bit, which simply represents the number of ones in the message chunk as a binary number (step 1). In the hash value $h_i$ with $h=5$ bits, only $t=1$ bit do not match (step 2). CS thus checks all modifications of the hash value until a hit between the reconstructed message and reconstructed checksum occurs (steps 3 to 5, flipped bits marked by dashed circles, order of bit flips: right to left). As the reconstructed message equals message chunk $M$ (step 6), CS will send an ARP request to CR. CR will perform the same computation and reconstructs $M$.
  For message chunk $M=110$ (same checksum), the same fit would apply but not reconstruct $M$, and no ARP is sent.}
  \label{fig:exampledecode}
\end{figure}

Fig.~\ref{fig:exampledecode} illustrates the working of \CCName{}-Ext.

Formally, if $C:\{0,1\}^*\to\{0,1\}^c$ is the checksum function, and $d(x,y)$ is the hamming distance between two bitvectors $x$ and $y$ of equal length, CS first checks if %
\begin{equation}\label{eq:testeq}
    d(M||C(M),h_i) \leq t\; .
\end{equation}
If so, CS enumerates the set
\begin{equation}\label{eq:seteq}
S_{h_i}=\{x\in \{0,1\}^h\;|\; d(x,h_i)\leq t\}
\end{equation}
in a pre-defined order, i.e., it generates a sequence of distinct bitvectors $x^{(1)}, x^{(2)},\ldots$ that together form $S_{h_i}$.
Let $extmsg(x)$ be a function to extract the first $h-c$ bits from a bitvector of length $h$, while $extchksm(x)$ extracts the last $c$ bits.
For $j=1,2,\ldots,|S_{h_i}|$, CS checks if
\begin{equation}\label{eq:checkeq}
    C\left(extmsg\left(x^{(j)}\right)\right) = extchksm\left(x^{(j)}\right)
\end{equation}
and stops with the first hit at index $j^*$.
If
\begin{equation}\label{eq:exteq}
  M=extmsg\left(x^{(j^*)}\right)\; ,  
\end{equation}
then CS sends an ARP request.

Upon receiving such an ARP request, CR looks up $h_i$, and also enumerates $S_{h_i}$ according to Eq.~(\ref{eq:seteq}), does the computations from Eq.~(\ref{eq:checkeq}) and stores secret message $M$ according to Eq.~(\ref{eq:exteq}).

\jkignore{
As an alternative to using a checksum, we also investigate the use of an error-correction code (ECC) which adds $c$ bits to a $(h-c)$-bit message, and can correct up to $t$ bit errors. This restricts the choice of parameters, as $t$ will depend on $h$ and $c$, and not all values for $h$ and/or $c$ might be possible depending on the ECC, but using an ECC will guarantee that CR can reconstruct $M$ from a hash value $h_i$ which CS has signaled. Thus, CS can send an ARP request whenever Eq.~(\ref{eq:testeq}) is fulfilled. We will call this variant \CCName{}-ECC.
}

Please note that both \CCName{}-Basic and \CCName{}-Ext
are \textit{families} of variants, because they are parameterized in $H$ and $h,c,t$, respectively.

\subsubsection{\CCName{}-Remote-Smarthome}\label{subsubsec:dyst-remote-Smarthome}
In case of \CCName{}-Remote-Smarthome, the same considerations apply as for \CCName{}-Basic with the exception that no local broadcasts are considered as PoIs and no ARP broadcasts are sent as signal packets. Instead, PoIs are solely such messages that can be observed by both, CS and CR, i.e., both must be on the path of a third-party packet. Further, signal packets can be any legitimate packets generated by CS that pass CR directly (e.g., if CR is a router) or that can be inferred by CR (e.g., because CS pulls a \textit{Git} code repository while CR can monitor such pulls).

\subsubsection{{\CCName{}-Remote-RTCP}}\label{subsubsec:dyst-RTCP}

Finally, we like to highlight the imagined scenario for a censorship circumventing channel that uses RTP/RTCP. This is another approach to transform DYST-Basic into a remote scenario. It can be realized via a media communication, where CS only signals if the first PoI in the time interval since the last SR is the one that serves as the actual secret message. Alternatively, CS might include a short index which of the PoIs in the time interval is the one to be used as the actual secret message.
Assume for example that the complete communication on RTP is split into pieces of length $2h$ bits each, and the PoIs are comprised of the $h$-bit hash values of those pieces. For a data rate of $r$ bits per second and an SR interval of $t$ seconds (usually 5 seconds \cite{RFC3550}), there are  
$T=rt/(2h)$ PoIs of $h$ bits each. To uniquely describe the intended PoI, $\log_2 T$ bits are necessary. Thus, to really save bandwidth compared to a direct covert channel of $h$ bits, we require $\log_2 T\ll h$. Given $r$ and $t$, a suitable value for $h$ can be determined.

\subsection{Parameter Choice}\label{sect:parameterChoice}

This section explains the optimization of \CCName{}-Basic and -Ext. Since \CCName{}-Remote-Smarthome and \CCName{}-Remote-RTCP are solely deviations of \CCName{}-Basic, their optimization is not detailed separately.

\subsubsection{\CCName{}-Basic}
We first consider the situation of \CCName{}-Basic, where CS signals to CR if the hash value of a network packet matches the secret message exactly.
If the hash function $H$ has optimal properties, each bit of the hash value 
has
a value of 0 or 1 with probability 50\%, respectively.
The probability that $h_i$ equals a given secret message $M$ then is $2^{-h}$, as all bits of the hash value can be considered independent.
As 
the hash values of the diffe\-rent packets can be considered uniformly distributed and independent, the number of packets until a match between hash value and secret message occurs follows a geometric distribution with success probability $2^{-h}$, i.e., with expectation $2^h$.

\subsubsection{Pareto-optimal variants}\label{sssec:pareto-optimal}
A covert channel such as \CCName{}-Basic can be characterized by two properties: 
the \textit{distance}, i.e., the average number of observed packets between two 
signal messages,
which will influence detectability, and the 
steganographic \textit{bandwidth}, i.e., the number of secret message bits transmittted via one signal message in relation to the distance.
%
Thus, \CCName{}-Basic with parameter $h$ is a family of covert channels with distance $dist_{basic}(h) = 2^h$ and bandwidth $bw_{basic}(h) = h/2^h$. 
{%
The unit of distance is number of packets of interest, while the unit of bandwidth is number of bits of the secret message per packet of interest.
By multiplying the bandwidth with the number of packets of interest per hour in a particular scenario, cf.~Sect.}~\ref{sect:scenariotestbed}, {an absolute bandwidth is obtained in number of bits of the secret message transported per hour.
Similarly, distance must be divided by the frequency of PoI to obtain an absolute distance between signal messages measured in hours.
}

We would like to maximize both: \textit{signaling distances} because there is a threat of detectability when {signal messages such as} ARP requests occur too often, and \textit{bandwidth} to increase applicability.\footnote{%
The third parameter of the covert channel magic triangle, robustness, is considered a fixture in this optimization, as it will be considered before going into this tradeoff, cf.\ Sect.~\ref{ssec:robustness}, and will reduce the number of observable packets per time so that it affects all variants in a uniform way.}
Yet, increasing distance will reduce bandwidth and vice versa.

Hence, to achieve an optimal compromise between the two parameters, we search for a Pareto front, i.e., a set of non-dominated variants\footnote{A covert channel variant A is dominated by variant B if distance$(A)\leq$~distance$(B)$ and bandwidth$(A)\leq$~bandwidth$(B)$.}. Alternatively, 
we impose a constraint on one parameter and search the optimal value of the other parameter, i.e., we cut the Pareto front with a vertical or horizontal line into two halves, and search the point on the front closest to the border (in the ``allowable'' half).
It is obvious that variants of
\CCName{}-Basic with different values for $h$ do not dominate each other, as improving one parameter makes the other worse. 

\subsubsection{\CCName{}-Ext}
In \CCName{}-Ext, only $h-t$ or more bits of the hash value must match the encoded message, comprised of message chunk and checksum. 
%
The number of matching bits is a random variable $X$ that is binomial distributed with $h$ trials and success probability 0.5, and thus the probability that at most $t$ of the $h$ bits do not match is
\begin{equation}\label{eq:binom}
P_h(X\geq h-t) = \sum_{j=0}^t \binom{h}{j} / 2^{h}\; .
\end{equation}
CS and CR must try out
\begin{equation}\label{eq:probeq}
    T_{h,t}=\sum_{j=0}^t \binom{h}{j} = P_h(X\geq h-t)\cdot 2^h
\end{equation}
%
possible modifications of the hash value. For each, the chance that the checksum of the message part of such a modified hash value equals the checksum part of the modified hash value, i.e., the chance that Eq.~(\ref{eq:checkeq}) is fulfilled, is $2^{-c}$. Thus, the number of trials until a fit will occur is geometrically distributed with parameter $2^{-c}$, 
yet with a limited range of $T_{h,t}$ trials.
The chance that the true message has the first fit thus is
\begin{equation}
 U_{h,t,c} = \frac{1}{T_{h,t}}\cdot \sum_{k=0}^{T_{h,t}-1} (1-2^{-c})^k
 = \frac{1-(1-2^{-c})^{T_{h,t}}}{T_{h,t}\cdot 2^{-c}}\; .
\end{equation}
As both events (at most $t$ non-matching bits, secret message chunk is re-constructed in first fit) 
can be considered independent, the chance that a message transfer can be signaled by 
{a signal}
message is their product, $P_h(X\geq h-t)\cdot U_{h,t,c}$.
Taking Eq.~(\ref{eq:probeq}) into account, we see that this product is approaching $2^{h-c}$, the probability of a signal in \CCName{}-Basic with a message of length $h-c$, yet the additional solutions can still be non-dominated.
As the product probability is independent of the particular hash value, the number of packets until a signal occurs is again geometrically distributed, with the expectation
\begin{equation}\label{eq:distext}
   dist_{ext}(h,t,c) =  \frac{1}{P_h(X\geq h-t)\cdot U_{h,t,c}}\; ,
\end{equation}
and the covert channel has a bandwidth
\begin{equation}\label{eq:bwext}
   bw_{ext}(h,t,c) =  (h-c)\cdot P_h(X\geq h-c)\cdot U_{h,t,c}\; ,
\end{equation}
as $h-c$ bits of the secret message
can be decoded by CR
with each signal.

\subsubsection{Pareto Front}

We have computed distance and bandwidth for \CCName{}-Basic with $h=14,\ldots,20$ and for \CCName{}-Ext with $h-c=14,\ldots,18$, $c=6,\ldots,10$ and $t=1,\ldots,5$, both analytically, and supported by simulations of $5\cdot 10^7$ hash values, where we counted how often CS signals in each variant considered.
The hash values in simulations were generated by successively encrypting a 128-bit value with AES and a fixed 128-bit key, starting with value 0 and using the first $h$ bit of the value as hash value. We used three different checksum functions in simulations (we only use first $c$ bits of longer results): SHA-3, CRC8, and a handcrafted function that cuts the encoded message into pieces of length $c$, adds those pieces as binary numbers, adds 9, and takes the $c$ lowermost bits of the result.
All simulations were repeated with a second seed and results were manually compared to exclude the possibility of artifacts, which however did not show.
Raw data, Pareto front data, and the simulation code are available via a repository: \url{https://github.com/NIoSaT/DYST}

The values for $h,t,c$ were chosen to allow comparison between different variants in a restricted range, and to illustrate development of distance and bandwidth over the range for a particular parameter.

\begin{figure*}[t]
  \centering
  \includegraphics[width=0.32\textwidth]{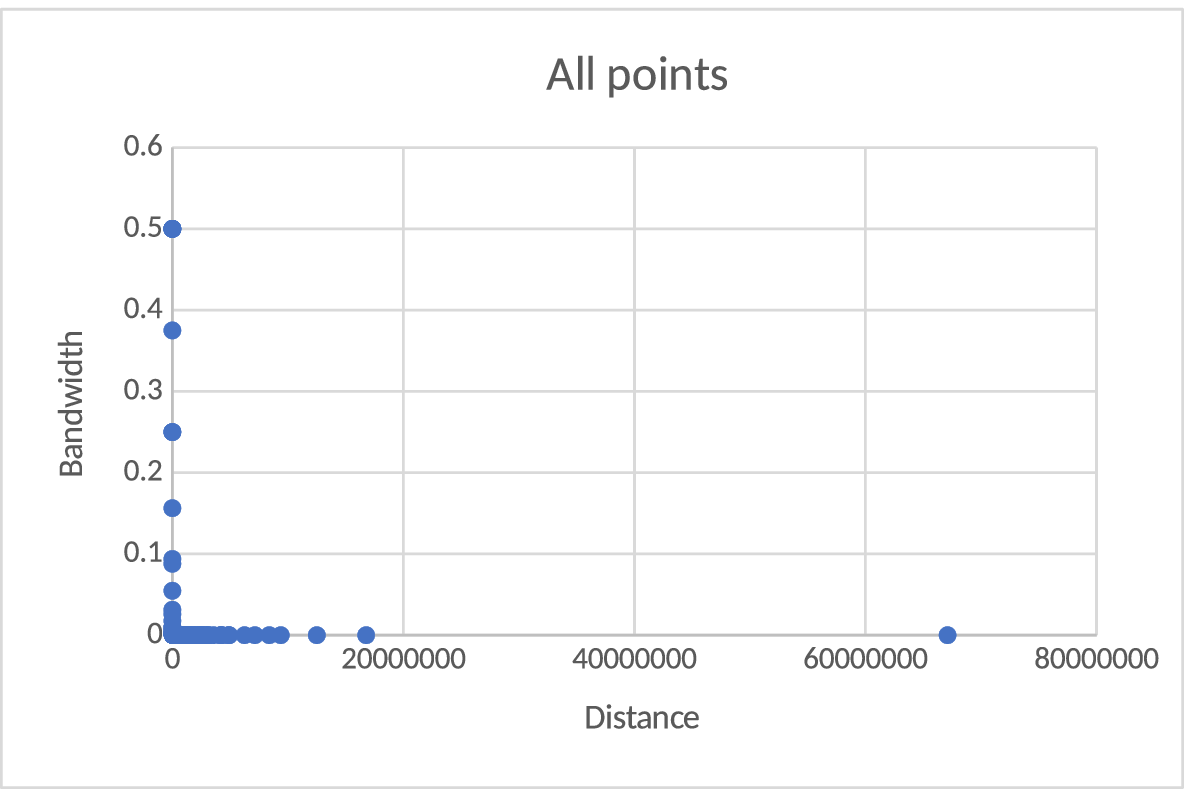}\quad%
  \includegraphics[width=0.32\textwidth]{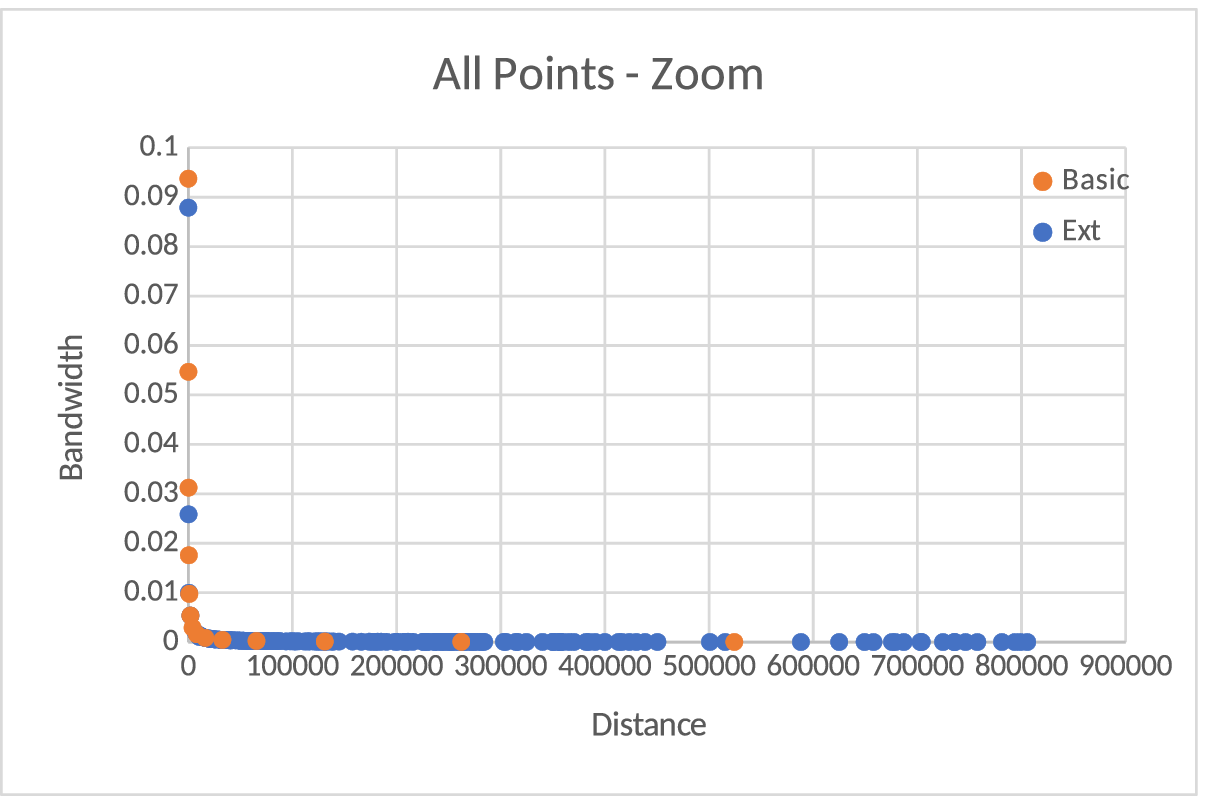}\quad%
  \includegraphics[width=0.32\textwidth]{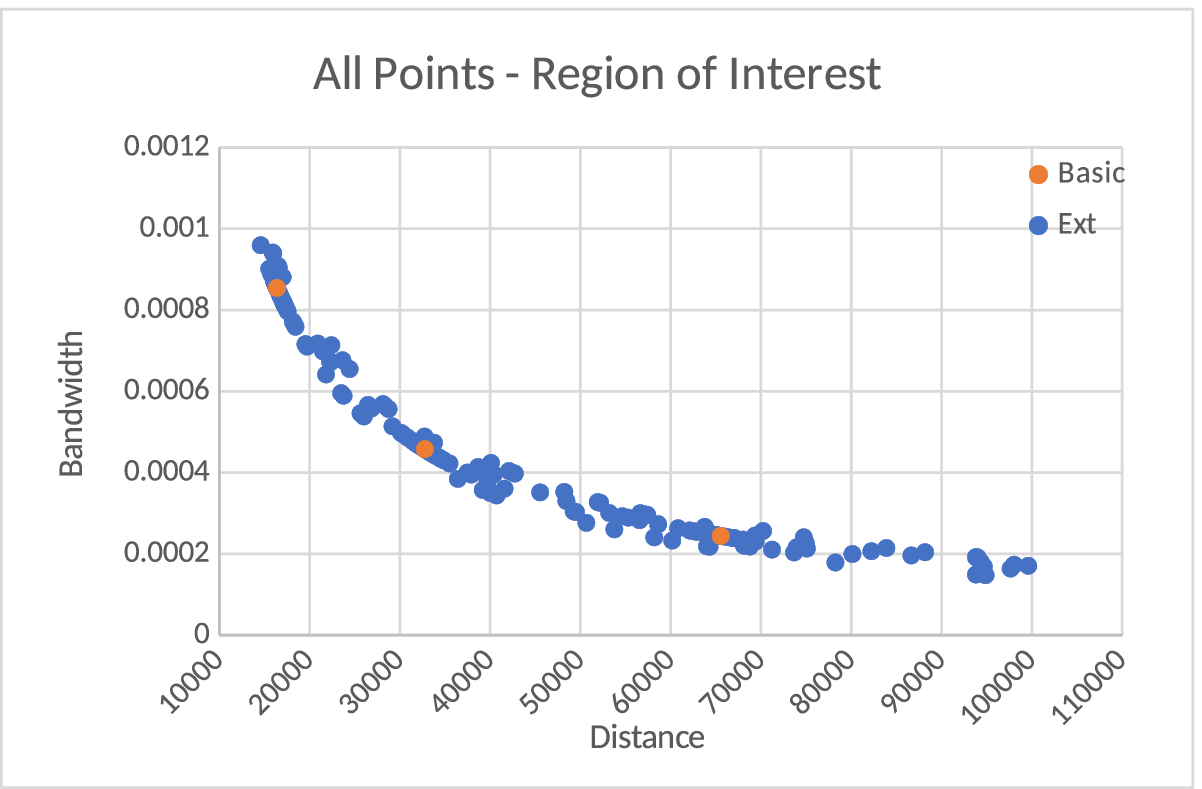}\quad%
  \caption{Simulation results for \CCName{}-Ext (left), and zooms for \CCName{}-Basic and \CCName{}-Ext, respectively.
  Axes give distance, i.e., the average number of observed packets between signalled transmission of secret message parts, and steganographic bandwidth, i.e., the number of secret message bits transmitted with one signal in relation to the distance.
  }
  \label{fig:sim1}
\end{figure*}

Fig.~\ref{fig:sim1}(left) provides all of the above variants as points in the plane with distance on the x-axis and steganographic bandwidth on the y-axis, as defined in Section~\ref{sssec:pareto-optimal}. All points are quite close together, so we do not have a point ``cloud'' but still 
a bit ``thicker'' line.
Thus, \CCName{}-Ext extends \CCName{}-Basic in the sense that the user has more choices in the tradeoff between distance (stealthiness) and steganographic bandwidth than with \CCName{}-Basic alone.
This is illustrated in Fig.~\ref{fig:sim1}(mid) and (right) that depict zooms into $(10^6,0.1)$ and $(10^5,0.001)$,
respectively.
The points that represent variants of \CCName{}-Ext (blue) fill the 
gaps between the points representing variants of \CCName{}-Basic (red).
Fig.~\ref{fig:sim1}(right) depicts the region of interest, i.e., the region where the actual tradeoff between distance and bandwidth can be seen.
Put otherwise, this is the region with the bend in the bottom-left quadrant of Fig.~\ref{fig:sim1}(mid).

The Pareto front contains only about one-third of the variants.
As its shape is similar to Fig.~\ref{fig:sim1}(left), we refrain from showing another figure.
Among the \CCName{}-Ext variants in the Pareto front, both SHA-3 and ad hoc checksums show quite often. CRC8 shows only seldomly. CRC8 and BCH are often doubles, i.e., they have the same distance and bandwidth as \CCName{}-Basic or -Ext with SHA-3. 
%
Quite some variants from the theoretical analysis are not on the Pareto front, indicating that sometimes the simulations gain a little in practice.

\subsection{Throughput-optimized \CCName{} Using Multiple Pointers}\label{multihash}
{For \CCName{}-Basic and \CCName{}-Ext, only one hash is calculated per PoI and eventually pointed to. \CCName{}'s throughput can be increased by using multiple pointers, signalling that a secret message might be found after calculating one of multiple hashes using a counter $i$, so that $H_1 = H(M||i=0), \dots, H_n = H(M||i=n-1)$, i.e., the signal (if sent) tells the receiver what counter $i$ needs to be used to obtain the secret message. This requires $n$ pointers in the form of $n$ distinguishable inter-packet times between PoI and signaling packet (or a classical covert storage channel with $\log_2 n$ pointer bits).}

\subsection{Taxonomy}

\begin{figure}[ht]
  \centering
  \includegraphics[width=0.9\linewidth]{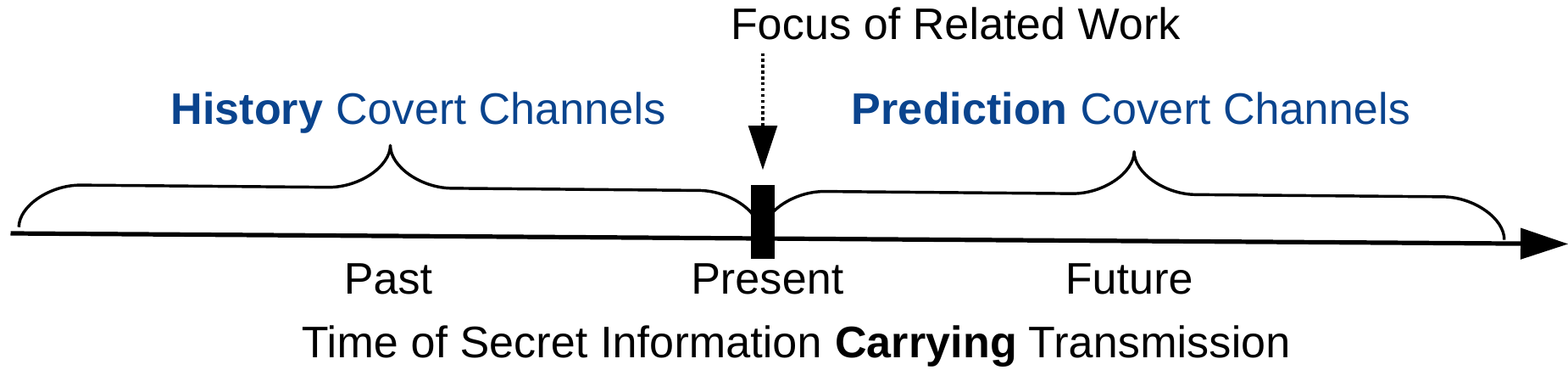}
  \caption{History and prediction covert channels, differentiated by the secret data carrying transmission they \emph{point} to.}
  \label{fig:histpredictcateg}
\end{figure}

Existing publications on network covert channels exclusively focus on network traffic that is live traffic to be modified or generated. Some covert channels also replay traffic recordings enhanced with secret data. \emph{History} covert channels \emph{point} to secret data in a carrier that was transmitted in the past (Fig.~\ref{fig:histpredictcateg}), i.e., the carrier traffic of the data signal is not altered. While not a core aspect of this paper, we still like to point out that it would be imaginable to create \emph{prediction} covert channels, which point to anticipated future data. For instance, ARP requests in LANs and sensor value readings in CPS occur on a regular basis and are thus possible to predict and utilize. The only difference between prediction and history covert channels is whether they point to old or upcoming data. However, predictions of future traffic are less reliable than pointing to already-seen traffic. For this reason, we solely focus on history channels.

\begin{figure}[ht]
  \centering
  \includegraphics[width=0.99\linewidth]{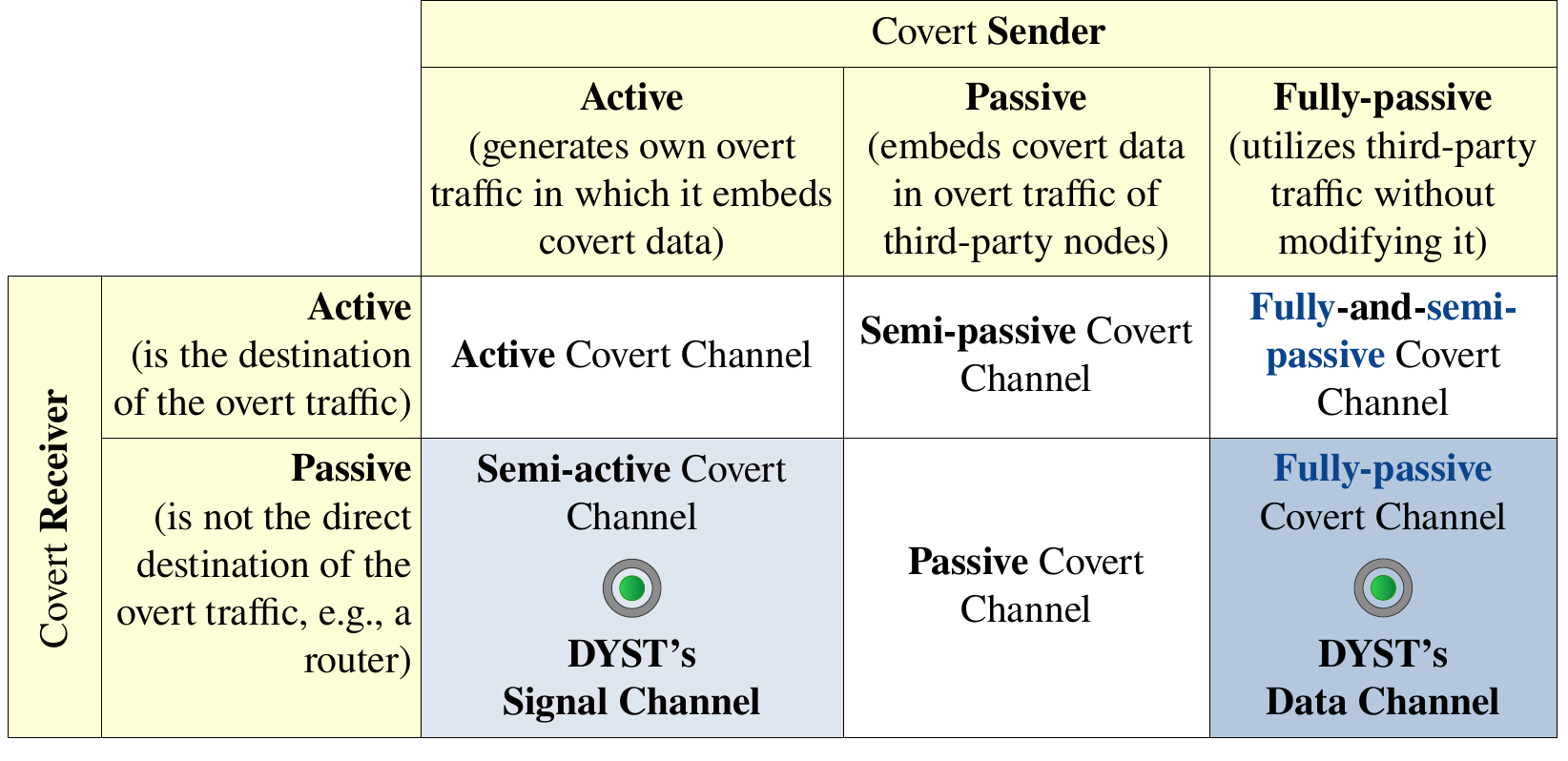}
  \caption{Categorization of \CCName{}'s data and signal channels.}
  \label{fig:taxonomy}
\end{figure}

The literature differentiates covert channels into active and passive ones (Fig.~\ref{fig:taxonomy}). An active sender generates the traffic in which the secret data is embedded while a passive sender modifies third-party traffic for this purpose. Usually, a passive sender is an intermediate network node, such as a router. The receiving process can also be performed in an active or passive manner. Here, the terminology considers
a receiver as active if it is also the destination of the overt traffic. If it passively observes the traffic (which is directed to another hop), the receiver is considered passive.

Lamshöft and Dittmann recently added a further differentiation in \cite{SemiPassiveSemiActive}, which is also shown in Fig.~\ref{fig:taxonomy}. They consider covert channels as semi-active if the covert sender is active but the covert receiver is passive. In contrast, a channel is called semi-passive if the covert sender is passive but the covert receiver active.

As the current differentiation between active and passive covert channels does only represent the \emph{signaling} channel of \CCName{}, we add a new category of covert communication, which we call \emph{fully-passive} because of its truly passive handling of third-party traffic (which is not modified). Because of the broadcast nature of the utilized messages, the receiver of \CCName{}'s \emph{data} channel is a passive one.

As our channel's sending and receiving processes are decoupled in a way that the sender does not directly address the receiver, our data channel can further be considered as an \emph{indirect} one, while the signaling channel can be considered as a \emph{direct} covert channel.

Finally, a covert channel could also be a \emph{fully-and-semi-passive} one, which is at least theoretically feasible and reflects a channel where the covert receiver waits for  pre-defined packets directed to it by some third-party node (fully-passive covert sender). Such a channel could be configured by having \CCName{} to operate with directed messages instead of broadcast messages and would be less functional.

\jkignore{The data channel shares characteristics of a so-called Redirector, introduced in \cite{...}, however, no intermediary node is exploited to forward messages. Henceforth, we propose to extend the taxonomy introduced in \cite{...} as follows
}

\jkignore{
\paragraph*{\textbf{Takeaway}} \emph{History channels transfer covert data by signaling which previously seen legitimate traffic matches a given secret using unobtrusive messages. We theoretically analyzed different configurations for our implementation \CCName{}, of which \CCName{}-Basic and \CCName{}-Ext turned out to be useful. We further defined a new category of covert channels that we call fully-passive. Our history channel is a first variant of such a fully-passive covert channel.}
}

\section{Evaluation Using Local Network Setup}\label{sect:eval:LAN}
This section presents the evaluation of the 
\CCName{} implementations Basic and Ext for local area networks using two different scenarios: a private smart home network and a university network.
After presenting our implementation and the experimental setups for different scenarios in which we evaluated \CCName{}, 
we analyze the robustness and detectability of our covert channel. 

\subsection{Implementation}\label{sect:PoCImplementation}

The PoC for \CCName{} using local area networks (\CCName{}-Basic/-Ext) was implemented with Python 3 and utilizes the \emph{scapy} library for eavesdropping legitimate traffic and crafting signal packets. Our implementation utilizes the following packets as they can be received by CS and CR when residing in the same network:
    \begin{itemize}
        \item IPv6 anycast packets with IPv6 destination \texttt{ff0*::} or to IPv6 link layer address \texttt{33:33:*}
        \item IPv4 broadcasts to the subnet broadcast address
        \item ARP requests to broadcast addr.\  \texttt{ff:ff:ff:ff:ff:ff}
    \end{itemize}

For hashing, we utilized the SHA3 hash algorithm with a bit length of 512, provided by the Python 3 library \emph{hashlib}. The input values contained the source IPv4 and IPv6 address, depending on the packet type. As the same input of a hash function results in the same hash, we additionally utilized the CS and CR packet receiving timestamp in seconds. The utilization of the timestamp results in a new hash for the same source addresses each second. As packets are not received at the same time by all devices in a network, we filtered packets that were received 
closer than 0.05 seconds to a full second, i.e.\ having a fractional time value of less than 0.05 or higher than 0.95 seconds, to minimize the possibility of CS and CR using a different timestamp when hashing.

For signaling, we utilized an ARP broadcast request, sent by the CS requesting the MAC address containing the target IPv4 address of an uninvolved third-party system. The CR interpreted this request as the signal to extract the latest hashed value. 
The process of the CS implementation is shown in Alg.~\ref{alg:DYST}. An additional check for PoI-collisions is performed with and without activated robustness mode (cf.~Sect.~\ref{ssec:robustness}).

\begin{algorithm}
\caption{DYST CS-side implementation}\label{alg:DYST}
    \textbf{Input:} Secret message $M$\\
    Split secret message $M$ into chunks $chunk_0, \ldots, chunk_n$\\
    $desired\_hash$ = $chunk_0$\\
    \For{Observed $packet$ in $traffic$}{
        \If{$packet$ is $PoI$}{
            $hash = H(packet || timestamp)$\\
            \If{$hash$ == $desired\_hash$}{
                \If{NOT Robustness Mode}{
                    Send $signal$\\
                    }
                \Else{
                    \If{No collision within timespan $D$}{
                        Send $signal$ \\
                        $desired\_hash = chunk_{next}$\\
                        }
                    \Else{
                        Ignore $packet$\\
                        }
                    }
                }
            \Else{
                Skip $packet$ \\
                }
        }
        \Else{
            Ignore $packet$\\
            }

    }
\end{algorithm}

\subsection{Scenarios and Testbed}\label{sect:scenariotestbed}

To evaluate \CCName{} under different circumstances, we came up with several scenarios which provide different traffic characteristics. These scenarios are described in detail in the following paragraphs. 

\paragraph{Scenario 1: University Network}
Traffic for this scenario was recorded in a university 
network from regular office workstations. We did not use any port mirroring or a prominent location in the network to 
see
how a regular device would see traffic. The environment itself is composed of 75 to 100 devices, around 50 of which are used on a daily basis. The network mostly consists of office laptops, printers, and some smart devices. All 
major operating systems (Linux, macOS, Windows) are present.
For our intents and purposes, the 
university network
resembles that of a company, thus this scenario applies to use cases in both settings.

\textit{Example use case:}
In an APT, 
an attacker might infect multiple clients in the network and use \CCName{} as a means of internal communication between 
the
compromised clients.
If one 
infected machine got access to an account with higher privileges, it could share the credentials with all other instances in the 
network 
for a faster spread of malware. Similarly, \CCName{} could be used as a command and control channel between multiple compromised clients.

\jkignore{
\paragraph{Scenario 3: High-speed Train}
We recorded traffic in three high-speed trains' guest WiFi networks through 
100, 95 and 65 minutes, respectively. 
For privacy, 
only ARP broadcasts 
were recorded.

\textit{Example use case:} Several dissidents want to leak information to the world press. To this end, they all take the same train, while they place themselves into the same or in succeeding coaches. While performing legitimate actions (browsing the web etc.), their devices monitor broadcast traffic and send out typical ARP broadcasts, e.g., to poll the address of a WiFi router.
}

\paragraph{Scenario 2: Home Network}
This scenario represents a typical home network with mixed devices, permanently connected to a WiFi router. The utilized router was a Speedport Smart 3 with current firmware, extended by two mesh repeaters to cover a larger area. In total, up to 30 devices were connected simultaneously, 
consisting
of classical IT devices (
three laptops, two raspberries, a network printer, smartphones), IoT devices (SmartTV, vacuum cleaning robot, coffee machine) as well as home automation (various Google Nest Mini). All devices were commonly used and, except for laptops and smartphones, connected permanently to the home network. All devices were connected within one /24 IPv4 subnet.

\textit{Example use case:}
Several compromised smart home devices exchange information under the radar, e.g., to collaboratively collect surveillance data and profile inhabitants. It is not probable that such a network is monitored for covert channel detection, so a less sophisticated approach will also be applicable. 
Anyhow, we decided to analyze this scenario because it shows the flexibility of \CCName{}, even if there are few changing devices. In this scenario, especially the throughput of \CCName{} can be optimized as there are no wardens.

Note that further scenarios are imaginable, e.g., journalists exchanging secret information through a WiFi hotspot in a bullet train or at a public airport.

\subsection{Match Distribution}\label{sect:matchdistribution}
\jkignore{\begin{figure*}
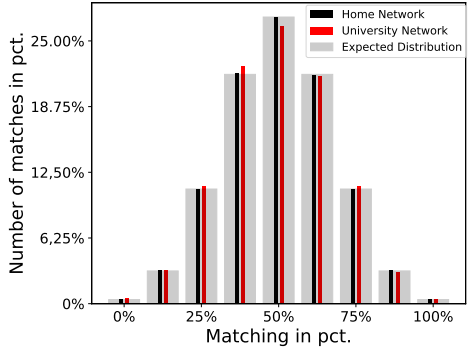

    \centering
    \subfloat[Office\label{fig:dirstibution:office}]{\includegraphics[width=0.24\textwidth]{figures/old/hs_worms_office_00006_20220420225826.pcapng.result_plot.pdf}} 
    \subfloat[BACnet\label{fig:distribution:BAC}]{\includegraphics[width=0.24\textwidth]{figures/old/wireshark-testlog-new_00003_20160522170052.pcapng.result_plot.pdf}}
    \subfloat[Train\label{fig:distribution:ICE}]{\includegraphics[width=0.24\textwidth]{figures/old/ICE_Ulm-Str-Mannh-2022-03-29_Duration_ca_95min_including_a_Videochat.pcapng.result_plot.pdf}}
    \subfloat[Home Network\label{fig:distribution:homelan}]{\includegraphics[width=0.24\textwidth]{figures/old/2022-04-16_tobias_homelan_1byte_fixed_plot.pdf}}
    \caption{Match Distribution}
    \label{fig:distribution}%
\end{figure*}
}

The utilized input parameters of the hash function generate different hashes $h_i$ for each modified bit, i.e., new packet, if the hash function is collision-resistant. These generated hashes $h_i$ are compared by the CS to a specific pattern $M$ (the data it wants to signal), which is constant until a suitable match is found. As explained in the derivation of Eq.~(\ref{eq:binom}), the number of matching bits follows a binomial distribution with $h$ trials and a success probability 0.5. 

We compare the actual frequency for the number of matching bits in relation to the total number of hashes generated to the expected results in Fig.~\ref{fig:distribution} for both scenarios. Therefore, we searched for one specific pattern equal to 8 bits and expected the distribution to follow the binomial distribution with $N=8$ and $p=0.5$, which is represented by the gray histogram. Further, we assume that the more hashes are observed, the more the actual distribution will follow the expected distribution. The black and red bars represent the actual distribution of matching bits in the home and university network scenarios, respectively. Both actual distributions follow the expected distribution, however, the university network scenario differs slightly. This can be explained by the number of observed hashes (89,959 observed hashes for the home network; 2,075 for the university network), drawing both assumptions correct. This points out our hash-generation methodology is correct.

\begin{figure}[bht]
  \centering
  \includegraphics[width=0.7\linewidth]{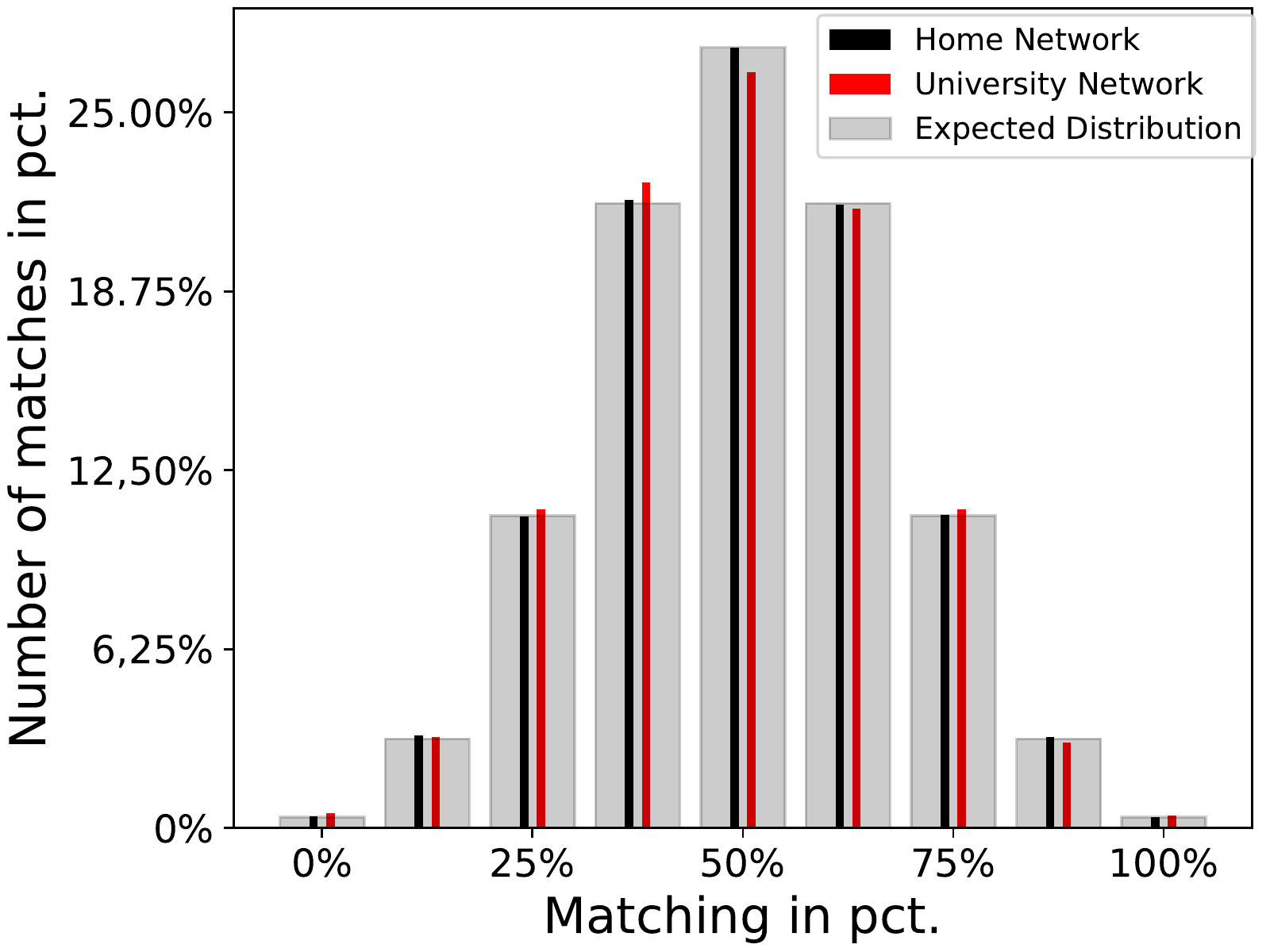}
  \caption{Match Distribution}
   \label{fig:distribution}
\end{figure}

Besides the evaluation in our testbeds, also the number of matches for different bit lengths is evaluated. As the CS is searching for a 100 percent match in \CCName{}-Basic, we calculated $P_h(X\geq h)$.
As for \CCName{}-Ext also partial matches can be utilized, we add 
$P_h(X\geq 0.8\cdot h)$ for at least 80\% matching bits, both calculated by Eq.~(\ref{eq:binom}).

The results of the example experiments for the home network scenario are presented in Tab.~\ref{tab:dist}. The 8-bit and 16-bit experiments were performed in a live scenario, while the 12-bit and 21-bit match distributions were simulated. The results indicate that for $h=8$ bits, slightly more hashes than expected show $h$ matching bits, while for at least 80\% matching bits, slightly less were observed. According to the deviation, 281 perfect matches should be detected, while actually 343 were detected. For the \CCName{}-Ext mode, 3,232 potential matching hashes should be detected, while actually 3,162 were detected. For our $h=12$ bit experiment, the hit rate for both 100 and (at least) 80 percent matches slightly performed worse than expected, resulting in 150 matches instead of 164 and 1,068 instead of 1,086, respectively.
For $h=16$ bit and a 100 percent match, the observed rate neared the expected value, while the number of actual 80 percent matches was slightly higher than expected. There should have been 5 matches, while actually 5 had been found for \CCName{}-Basic and for \CCName{}-Ext potential 2,379 hashes with at least 80 percent match should be observed, while 2,413 were actually detected.
For our $h=21$ bit experiment, no matches were found. As 0.43 packets are expected for the number of observed packets, the expectation is met. Further, 152 packets had at least 80\% matching bits, while there should have been 170 packets.

\begin{table}[ht]
    \caption{Expected and actual match results (Home Network)}
    \label{tab:dist}
\centering
\begin{scriptsize}
    \begin{tabular}{l|rrrr}\toprule
      $h$   & \textbf{8 bits}&\textbf{12 bits} & \textbf{16 bits} &\textbf{21 bits} \\
         \midrule
\# observed packets & 89,959 & 56,105 & 226,488 & 43,325 \\
$P_h(X\geq h)$ & 0.3125\% & 0.2930\% & 0.0024\% & 0.0001\% \\
freq. pkts w. 100\% match & 0.3813\% & 0.2673\% & 0.0022\% & 0.0000\% \\ 
$P_h(X\geq 0.8 \cdot h)$  & 3.5937\% & 1.9360\% & 1.0504\% & 0.3917\% \\ 
freq. pkts w. $\geq 80$\% match & 3.5160\% & 1.9035\% & 1.0654\% & 0.3508\% \\ 
         \bottomrule
    \end{tabular}
    \end{scriptsize}
\end{table}


\subsection{Robustness}\label{ssec:robustness}
The main concern for the robustness of \CCName{} lies in which messages are seen and interpreted as signaling.
As \CCName{} only uses legitimate network packets, it relies on the general robustness of network transmissions (e.g., Ethernet frame checksums) and timestamps (ensured by time synchronization mechanisms like NTP for example). Such effects are caused by active and inactive jitter-influencing factors causing disorder, retransmissions, dropouts, and delays amongst others. Neither CS nor CR can control jitter factors of their transmission as they are not \textit{directly} communicating with each other. To mitigate such factors caused by the physical layer we need to introduce the so-called \textit{robust mode}.

\begin{figure}[bht]
    \centering
    \includegraphics[width=0.6\linewidth]{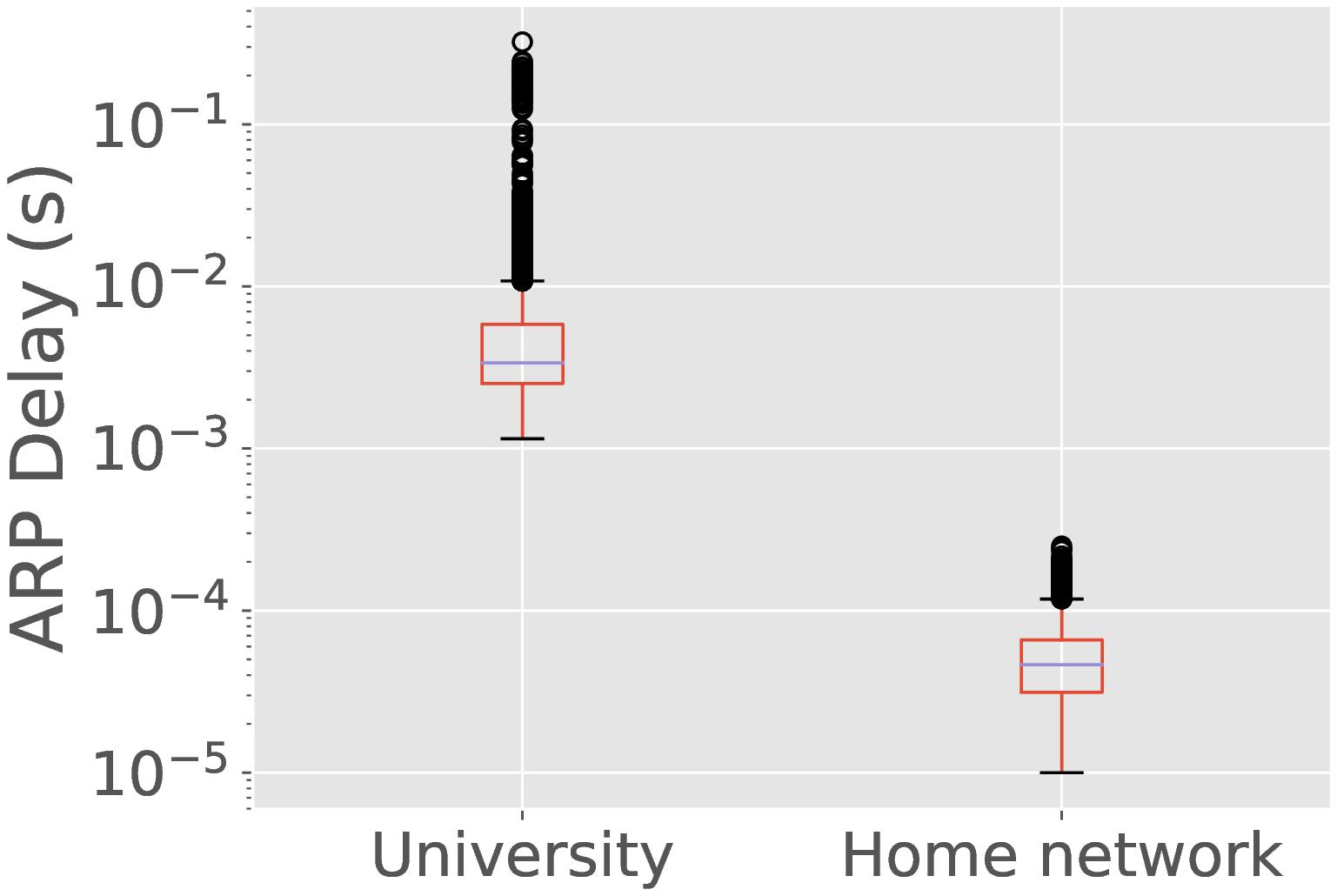}
    \caption{ARP delays for University and Home network setups. }
    \label{fig:jitter_box}
\end{figure}

We evaluated the jitter in our two local network environments for \CCName{}-Basic by investigating the delays between ARP requests and the corresponding replies. Fig.~\ref{fig:jitter_box} shows the results for the university 
and home network.

We can see that the mean values and standard deviations for the ARP delays, which are by nature two-way-delays, are orders of magnitude lower than the waiting periods that were used for the robustness measure (see further below). The standard deviation for the university 
is at $9.1\cdot 10^{-3}$s and for the home network at $2.1\cdot 10^{-5}$s. While the university 
network has generally higher values, we still only have 35 outliers above $0.1$s out of over 18,500 measurement points. Therefore, we can conclude that the jitter of our test networks did not have a relevant impact on the robustness of \CCName{}. Further, it is also viable to adjust the waiting periods for sending signal packets according to the characteristics of the network in which one chooses to deploy \CCName{}.

If the same packet arrives at CS and CR, we can assume that the content will be the same, resulting in the same input parameters for the hashing function.
We do have other concerns about the robustness of \CCName{} that need to be addressed:
\begin{itemize}
    \item[a)] CS and CR must receive the same 
    PoI. 
    \item[b)] CS and CR must receive the PoIs in the same order.
    \item[c)] Consecutive PoIs must have a sufficient delay to allow reliable signaling.
\end{itemize}

\subsubsection*{a}
To address this issue, we have to carefully select which packets are used by \CCName{} to ensure that both CS and CR receive the same packets.
Depending on the deployment scenario, we might choose different sets of packets to achieve this goal:
In a local scenario like a university network, a smart home network or an open Wi-Fi like a café, we can focus on local broadcast packets.
If \CCName{} is used between two routers, we can use our knowledge about the routing topology to filter for packets that will pass through both CS and CR.
Similarly, if both CS and CR are part of the same multicast group, one could filter for packets from that group to ensure synchronization.
Moreover, there is a tolerance regarding the received PoI. As only a fraction of the packets are actually used for signaling, a packet received by CR but not received by CS would simply not be checked for a match and thus also not used as a carrier. However, if CS receives a packet and both of the following conditions are met: CR does not receive the packet and CS actually points to this packet for signaling, then CR might receive an incorrect message as it interprets the wrong packet (see point b)).
All in all, it is possible to choose a robust set of packets to be used with \CCName{}, with only a little prior knowledge about the deployment scenario.

\subsubsection*{b}\label{subsub:order}
\begin{figure}
    \centering
    \subfloat[Covert Sender Side\label{fig:sending_delay_CS}]{\includegraphics[width=0.43\linewidth]{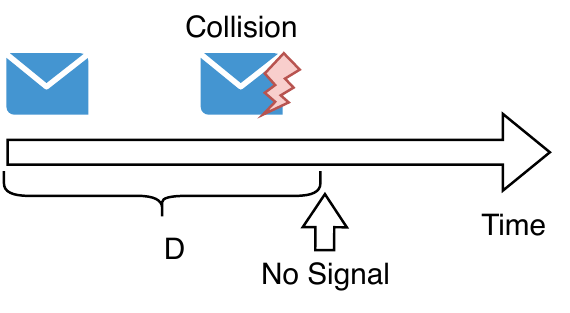}} 
    \quad
    \subfloat[Covert Receiver Side\label{fig:sending_delay_CR}]{\includegraphics[width=0.43\linewidth]{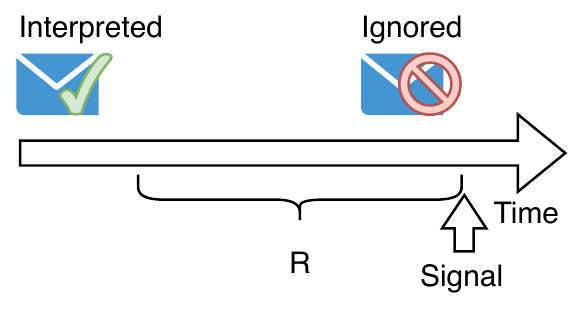}}
    \caption{Robustness Measures}
    \label{fig:sending_delay}
\end{figure}

This variable is outside the control of \CCName{} as we cannot influence the routing or buffering behavior of other parts of the network.
We performed an evaluation in a home network by running \CCName{} between two different clients on the same network.
During our evaluation, we observed significant problems when testing our scripts. In our first test, 3 of 65 characters of the message were transmitted correctly.
This was due to the fact that a significant portion of the PoIs did not arrive in the same order for the CS and the CR.
These PoI packets were in the wrong order because of their tight succeeding timing and different networking delays for CS and CR.
Therefore, 
two additional configuration parameters for \CCName{} 
control the mandatory delay between received packets to reduce this issue: 
%
If the CS receives two or more PoIs in less than $D$ milliseconds, the CS will ignore all PoIs received in that timeframe. So only isolated PoIs will be considered for \CCName{}.
If only one PoI is received in $D$ milliseconds, the CS will send out the signal. The CR will ignore all PoIs that arrived less than $R$ milliseconds before the signal and only interpret earlier PoIs. This gives the CS enough time to calculate and send the signal without risking a race condition (see part c) below). 
Fig.~\ref{fig:sending_delay} illustrates this process.
Fig.~\ref{fig:sending_delay_CS}, shows the side of the covert sender. This example uses two packets that arrive close to each other, which leads to them being ignored by the CS and not considered for \CCName.
Fig.~\ref{fig:sending_delay_CR} shows the receiver-side. Here we can see that the receiver ignored a packet that arrived too close to the signal and instead interpreted the older one.
This will decrease the potential throughput, as we ignore more packets that could potentially be useful. 
But we significantly decrease potential errors in cases where packets arrive in a different order at the CS as at the CR.
Varying scenarios will require differently tuned values for $D$ and $R$, see Electronic Supplement. 
Additionally, we can counter possible errors with an error-correcting code that covers multiple transmissions. Different approaches are possible, e.g., simply transmitting the same message multiple times and taking a majority vote on gained hash values, can increase robustness.

\subsubsection*{c}\label{subsub:race}
Similarly to b), \CCName{} might encounter errors if many PoIs arrive in a short amount of time.
The CS might receive a PoI that creates a hit. While the CS is evaluating and preparing to send out the signal, another PoI arrives at the CR even before the signal from the CS reached the CR. In such a case, the CR would interpret the latest PoI and not the correct one.
This error source is also countered by the option $R$ of the multistage delay introduced in b). 
Since \CCName{} only considers PoIs that are isolated and the CS has $R$ milliseconds to perform the signaling, it is far less likely for this race condition to appear for any PoIs that can be seen by CS and CR.
Similar to b), a larger delay will result in a lower bandwidth but higher robustness.
Depending on the scenario, drastically different configurations for $D$ and $R$ are possible. If CS and CR are both routers with a fixed route between them, it is easy to see that the CS will have better knowledge about the order and delays in which the CR will receive the PoIs and can therefore choose a lower value for $D$ and $R$.

\paragraph*{Impact of Robustness}\label{subsub:robust_impact}
We evaluated the timing between PoIs in both scenarios of our local network testbeds to get an overview of network behavior and the impact of our robustness measure. Fig.~\ref{fig:IAT_PoI} shows the plots for the different scenarios.
\jkignore{
\begin{figure*}
    \centering
    \subfloat[Office\label{fig:IAT_box_office}]{\includegraphics[width=0.24\textwidth]{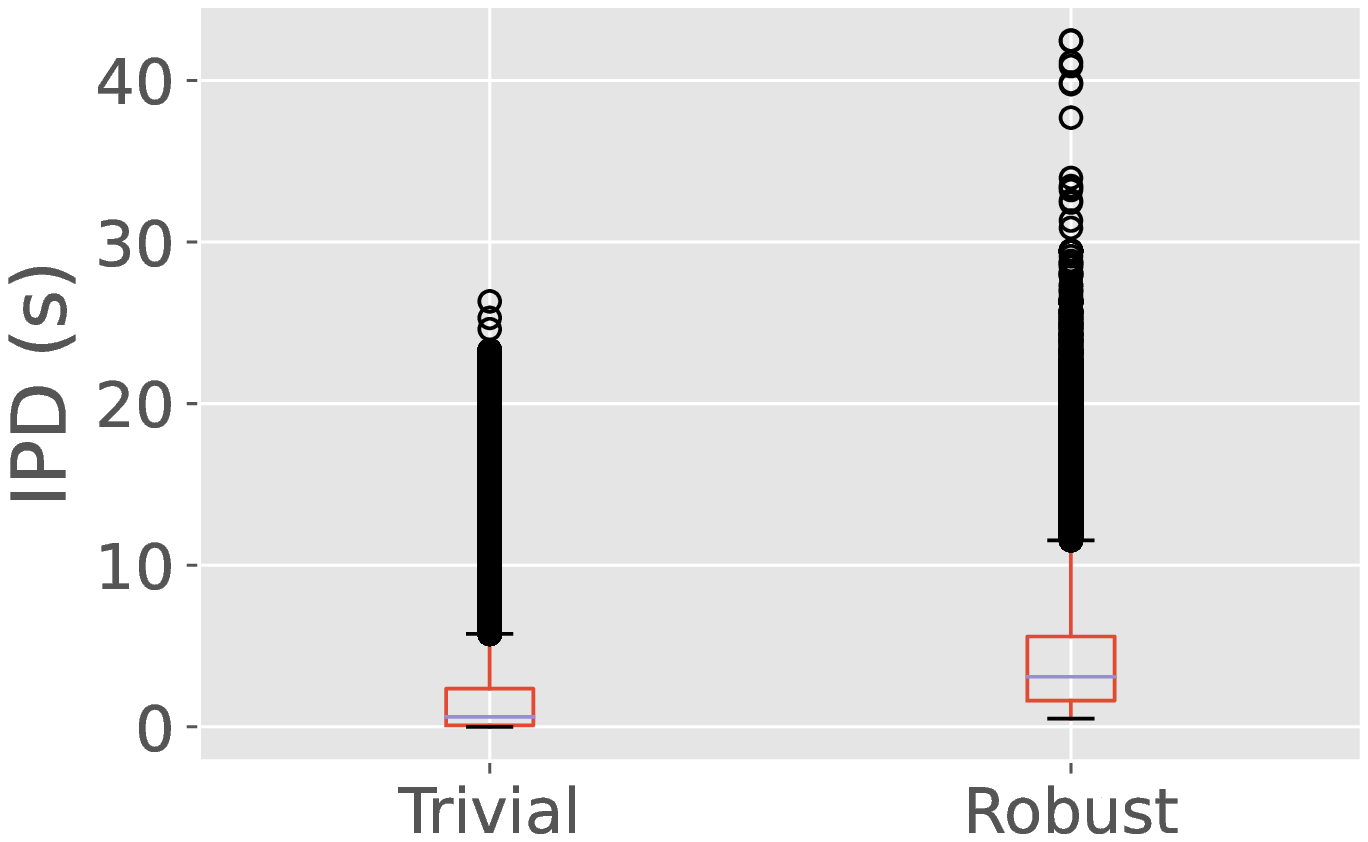}}
    \subfloat[BACnet\label{fig:IAT_box_bacnet}]{\includegraphics[width=0.24\textwidth]{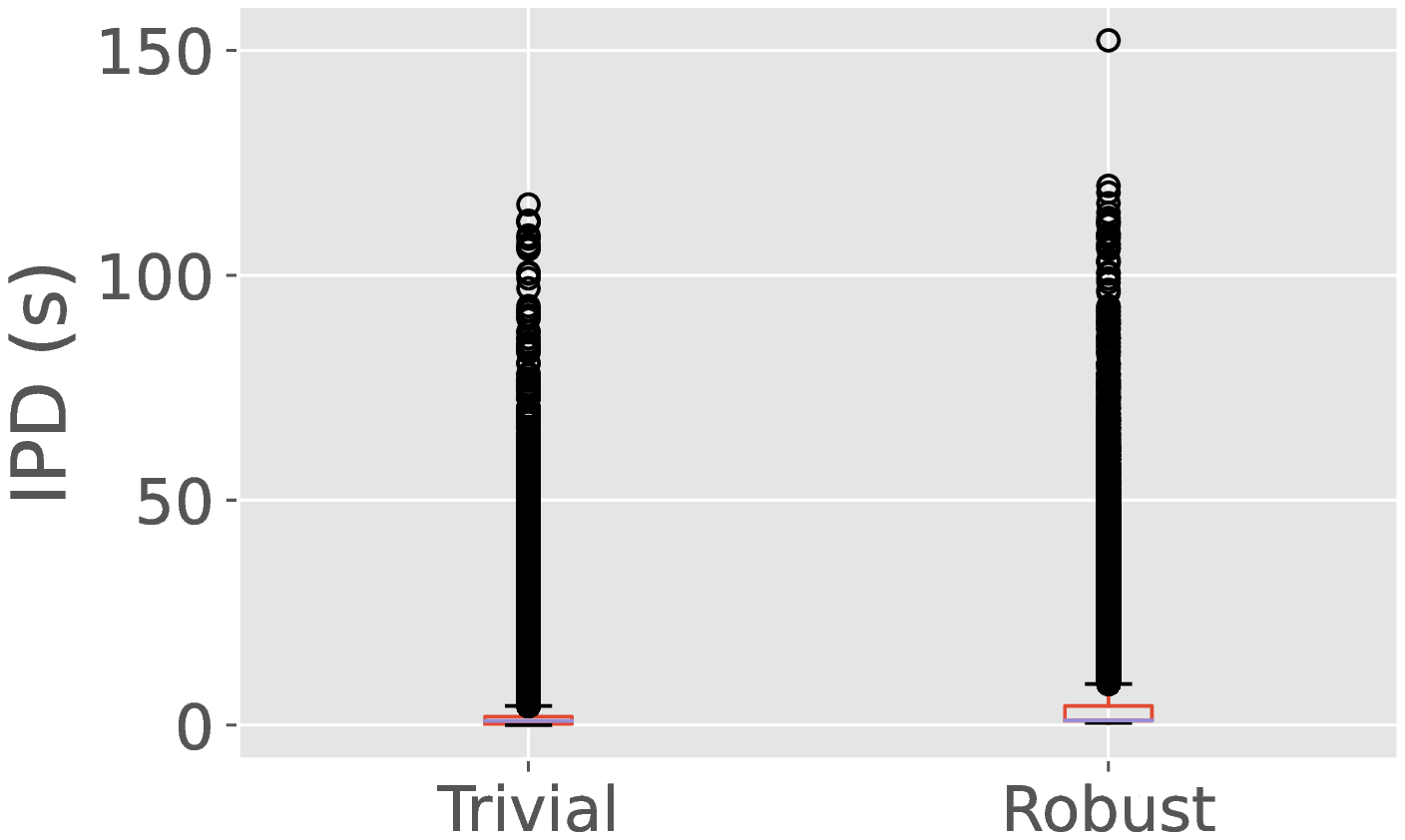}}
    \subfloat[Train\label{fig:IAT_box_ICE}]{\includegraphics[width=0.24\textwidth]{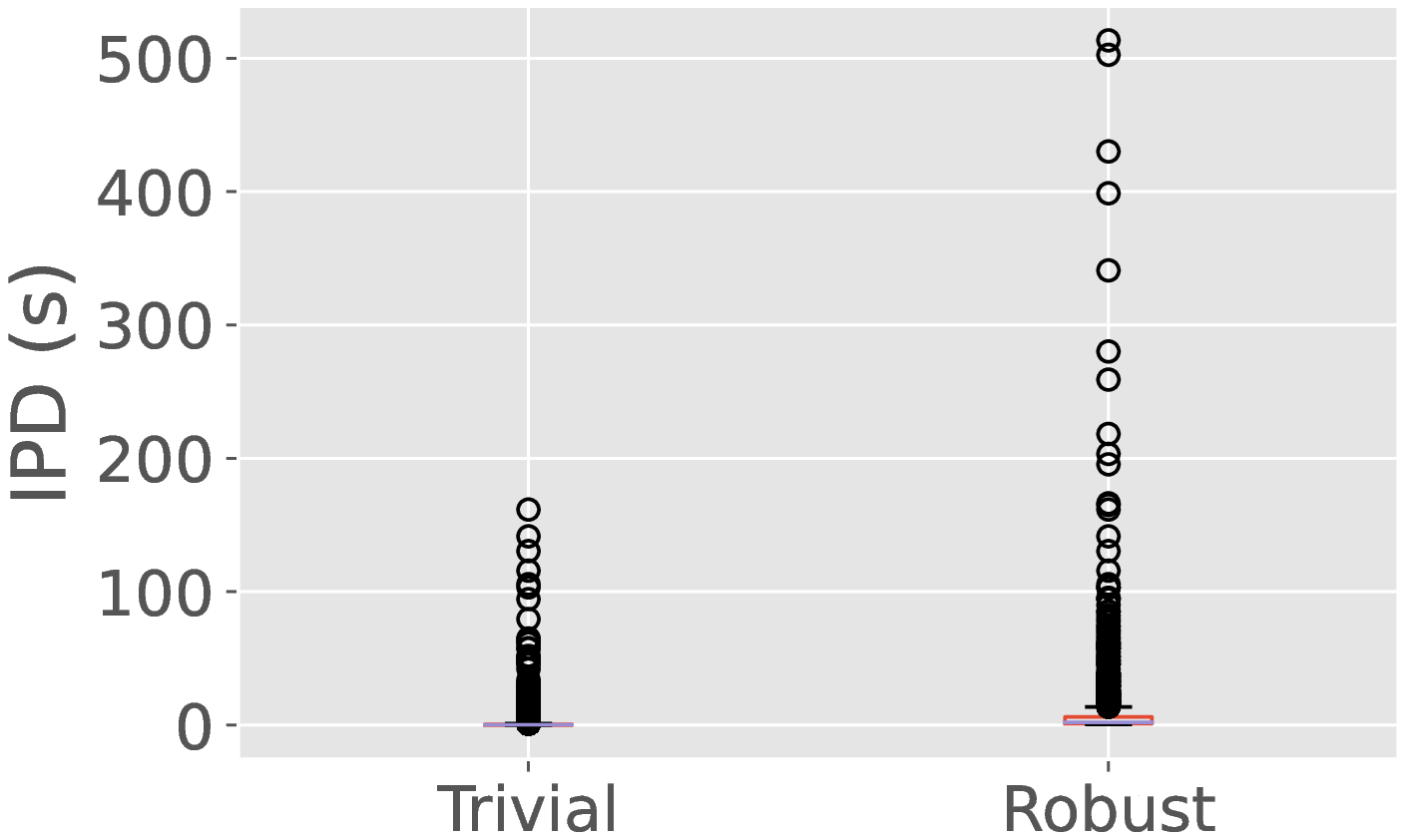}}
    \subfloat[Home Network\label{fig:IAT_box_home}]{\includegraphics[width=0.24\textwidth]{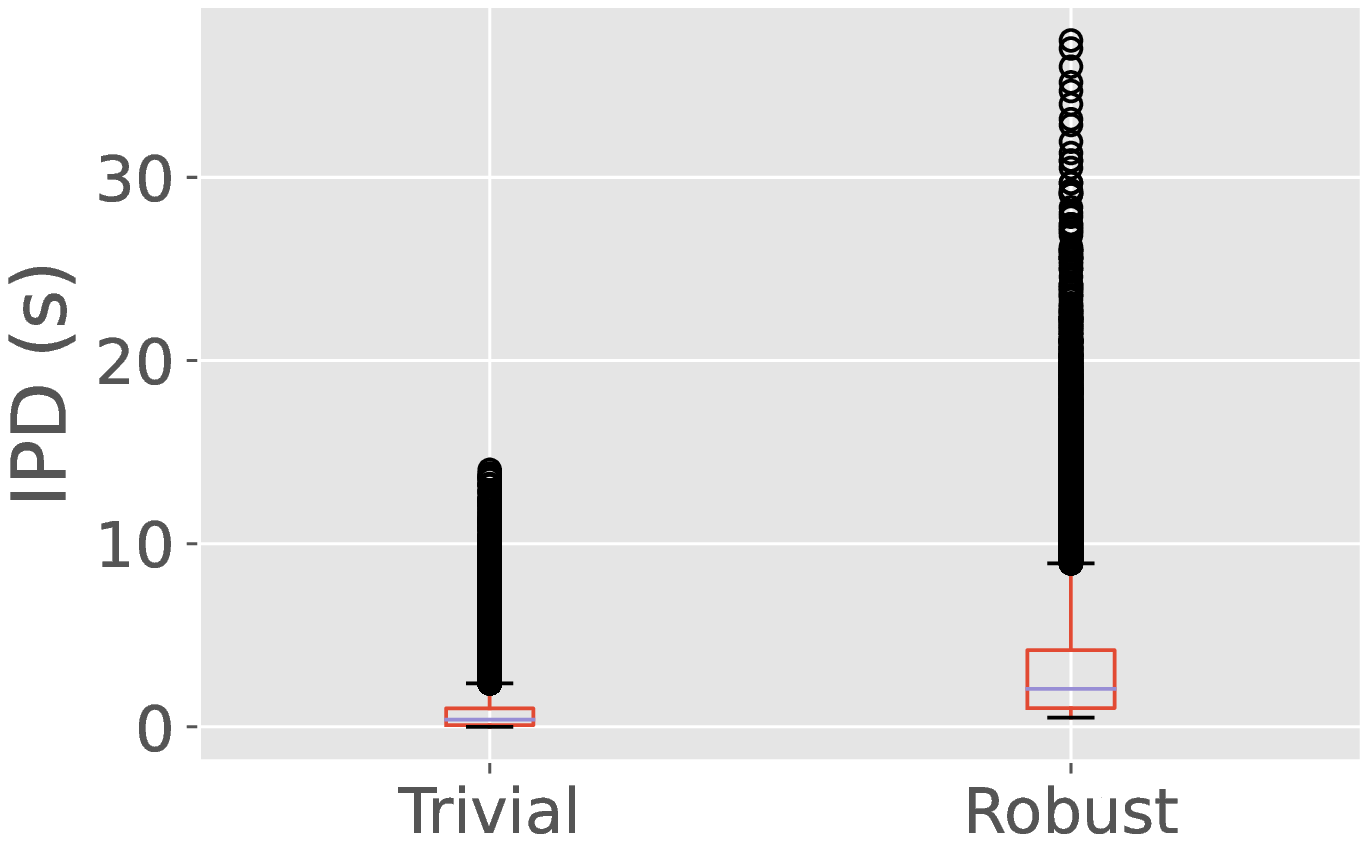}}
    \caption{Boxplot comparison of inter-packet delays (IPDs) between PoIs (SIMULATED)}
    \label{fig:IAT_PoI}%
\end{figure*}
}

\begin{figure}
    \centering
    \subfloat[University Network\label{fig:IAT_box_office}]{\includegraphics[width=0.48\linewidth]{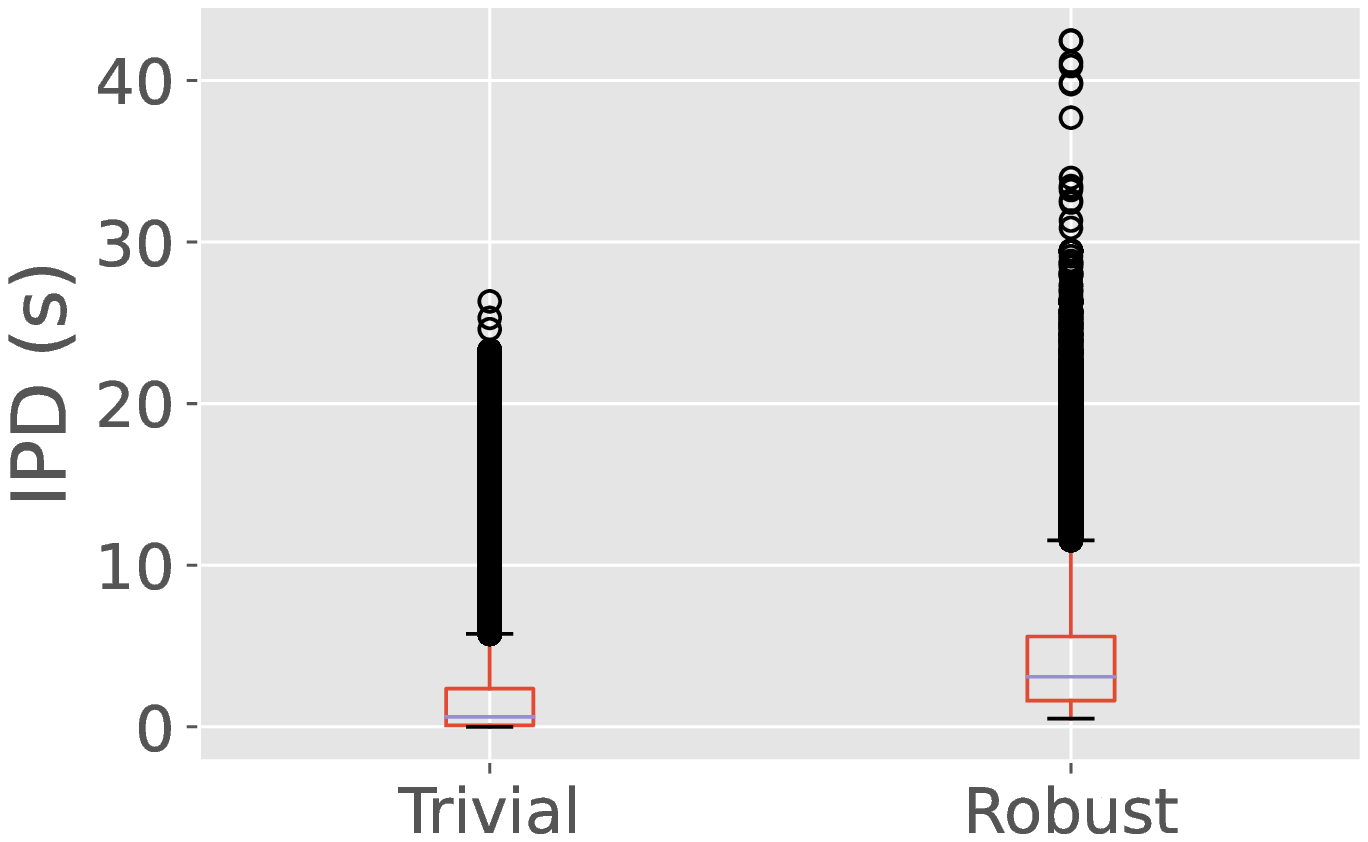}}
    \quad
    \subfloat[Home Network\label{fig:IAT_box_home}]{\includegraphics[width=0.48\linewidth]{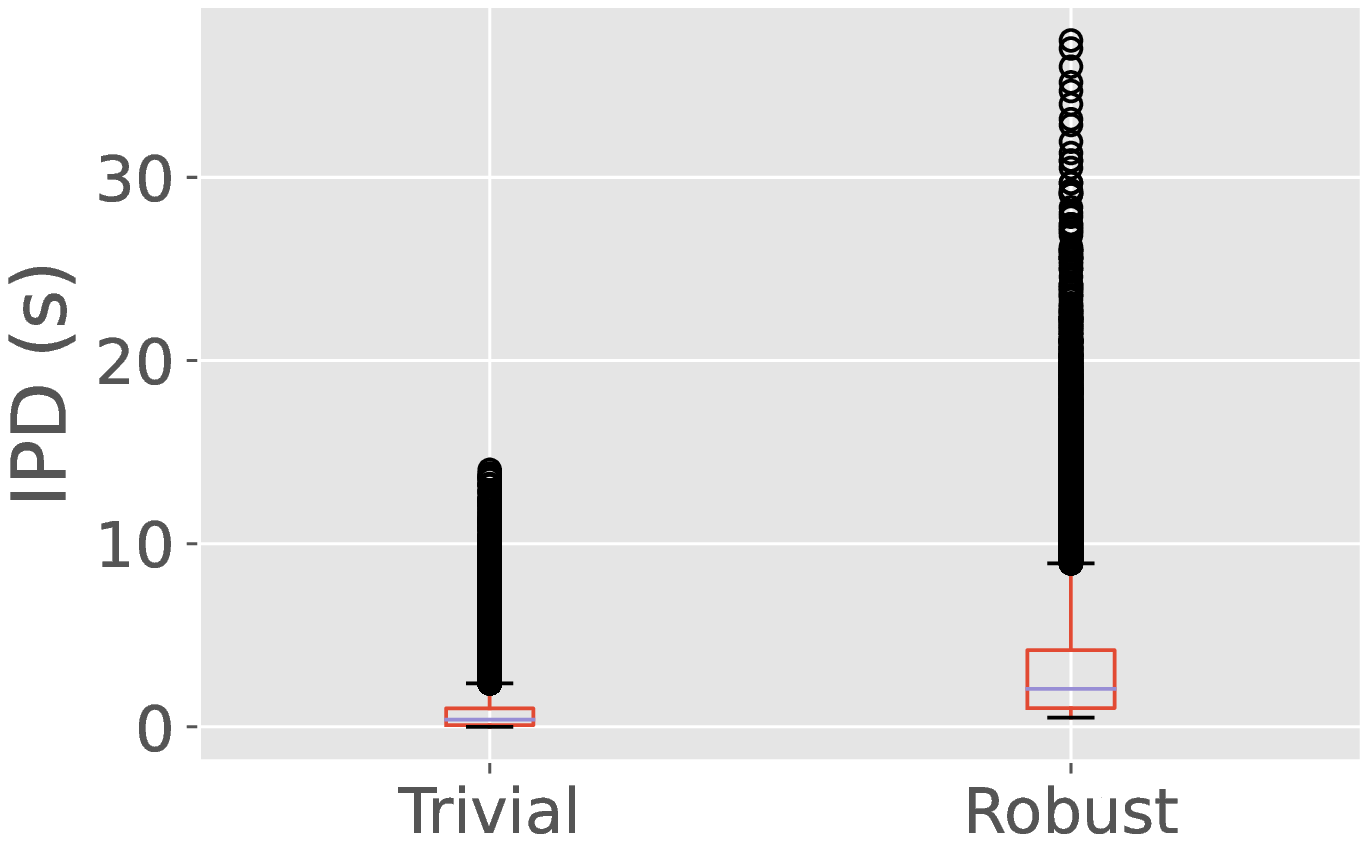}}
    \caption{Comparison of inter-packet delays (IPDs) between PoIs (SIMULATED)}
    \label{fig:IAT_PoI}%
\end{figure}

We can see that both scenarios show a generally similar picture:
the mean value for the inter-packet delays (IPDs) of the PoIs for \CCName{}-Basic is close to 0 with a significant number of outliers. The outliers are beneficial for the robustness of \CCName{}, as they separate PoIs.
If we look at the delays with the robustness filter ($D=0.5s$, the simulation only happened for the CS side), we again see a similar image for all scenarios. We generally see higher delays between PoIs, as \CCName{} ignores some PoIs and therefore the delays between evaluated PoIs are higher.
This means that \CCName{} will have fewer packets at its disposal to transmit the message, but we gain in reliability (see Electronic Supplement for details).

We evaluated the effectiveness of our robustness approach in 
the home network and university network scenarios.
Tab.~\ref{tab:robust} shows the results for two scenarios (home and university network). For each scenario, we conducted a test run with and without robustness measures and recorded the percentage of characters that were correctly transmitted.
\begin{table}
    \caption{Robustness Evaluation - Matching Transmitted Characters }
    \label{tab:robust}
    \centering
\begin{tabular}{lcc}
\toprule
        & Non-robust Matches & Robust Matches \\ \midrule
University Network  &     100\%        &    100\%  \\
Home Network &     4\%       &    56\%  \\
 \bottomrule
\end{tabular}
\end{table}

We can see that the home network had significant problems without robustness measures, while the university network setup had no problems during our tests.
The university network setup showed very little activity{, i.e., very low frequency of PoI,} during our robustness tests, which resulted in a reliable transmission but also in a low {absolute} bandwidth (8 characters in 8 hours) {despite usual frequency of matching PoI}.
The home network setup had drastically more activity and higher bandwidth (48 characters in 7 hours), which in turn resulted in a significantly less reliable transmission.
Initially, we tested with $D=0.3\;s$ and $R=0.015\;s$ which resulted in ca.~31\% correctly transmitted characters. We further tested the parameters $D=0.5\;s$ and $R=0.3\;s$, which resulted in 56\% correctly signaled characters.
We can therefore see that the home network setup benefits from more aggressive configurations.
However, the robustness would increase using a higher $D$ or additional robustness measures, such as the redundant transfer of secret messages. 

\begin{table}
    \caption{Robustness Evaluation - Total PoIs (SIMULATED) }
    \label{tab:robust_poi_count}
        \begin{tabular}{rrrr}
        \toprule
                & Non-robust & Robust ($D=0.5s$) & Fraction (\%) \\ \midrule
        University Network  &     8,130       &  4,595    & 56.5\% \\ 
        Home Network &   79,319       & 12,689    & 15.9\% \\ 
        \bottomrule
        \end{tabular}
\end{table}

In Tab.~\ref{tab:robust_poi_count}, we show how many PoIs were observed in a simulated offline run. We used real recordings and ran the pcaps through an offline version of \CCName{} (only CS side is simulated).
We can see that in all scenarios, the number of usable PoIs is significantly reduced.
This means there will be fewer PoIs available for \CCName{}, which in turn can reduce the potential bandwidth of the covert channel. 
It is noticeable that the home network setup suffered more than the university network setup, this can be explained by the higher activity in the home network compared to the university network. This again explains the worse performance in the home network without robustness measures.

On the other hand, a reduction of possible bandwidth aids the undetectability, by spreading signals even further apart.

\subsection{Detectability}\label{sect:detection}

{%
Detectability of a timing channel is analyzed by comparing the distributions of inter-packet gaps of packet streams with and without timing channel. This can be done with statistical and information-theoretic means. We have used Kolmogorov-Smirnov-test for the former and Cabuk's compressibility score for the latter.
The results of these tests are not absolute: the timing channel can only be detected if the difference between results with and without timing channel is \textit{larger} than the variation of test results for different packet streams without timing channel.
}

To evaluate the detectability of \CCName{}, we gathered legitimate reference and covert channel recordings from two different scenarios and with six different configurations for the covert channel: 1 and 2 byte basic mode ($h$ has 8 or 16 bits, respectively), 1 and 2 byte robust mode (also 8 and 16 bits), and 1 and 2 byte extended mode (where $h$ is also 8 or 16 bits, respectively. The checksum of length $c$ is added on top).
We performed recordings in two different networks, a home network and a university network (see Tab.~\ref{Tab:Rec_Data} for an overview).

\begin{table}[ht]
    \centering
    \scriptsize
    \caption{Overall Recorded Traffic}
    \label{Tab:Rec_Data}
    \begin{tabular}{@{}llccc@{}}
    \toprule
                                &          & \# Pkts & \# ARP Requests & Rec.\ Time \\ \midrule
    \multirow{2}{*}{Legitimate} & Home Network     & 17,608,213     & 8,619,952            & 40 Days          \\
                                & University Network   & 2,096,387     & 218,707             & 15 Days           \\ \midrule
    \multirow{2}{*}{Covert}     & Home Network   & 5,604,205     & 3,023,422             & 11 Days           \\
                                & University Network & 5,380,286     & 522,451             & 56 Days           \\\bottomrule
    \end{tabular}
\end{table}

As the data channel itself cannot be detected, a defender relies on detecting the signal channel.
Since our \CCName{} implementation uses ARP requests for signaling, our detection focuses on the IPDs of ARP requests.

\subsubsection{KS-test}
To gauge the potential detectability, we chose a two-sample KS-test \cite{KS_Test}. The KS-test is a general measure of similarity between two samples, with high test results indicating difference of the samples \cite{NIHbook}. Given empirical cumulative distribution functions $F_1$ and $F_2$ for random variables $X_1$ and $X_2$, i.e., $F_i(a)=P(X_i\leq a)$, the similarity of the two functions is computed by $\sup_a |F_1(a)-F_2(a)|$, with $\sup$ being the supremum of the set of distances over all real values $a$ \cite{NIHbook}.

We use the KS-test to measure the similarity between the ARP IPD distribution of two recordings. In addition to our original recordings, we also filtered the covert channel recordings to remove any signals produced by the covert channel to produce a second (but synthetic) source of legitimate traffic. We then performed a cross validation between all possible recordings of a scenario (home and university networks).
This provides us with 3 different classes of combinations, which we considered important for our analysis:
    (1) covert vs.\ filtered recordings,
    (2) pairs of legitimate recordings,
    and (3) covert vs.\ legitimate recordings.
To get a better understanding of the detectability, we focused on several combinations, which we describe separately.

First, we compared different \emph{legitimate} recordings against each other, while excluding exact matches. For this, we used several legitimate recordings from the home network. This gives us a mean p-value of $3.84\cdot 10^{-11}$ with a standard deviation of $4.01\cdot 10^{-10}$ and a mean D-value of 0.19 with a standard deviation of 0.13. The results for the legitimate university network scenario are almost exactly the same. This points towards a significant difference between all the legitimate recordings.
With that, we can already see that legitimate traffic drastically varies depending on the time of day and the activity of participating nodes that do not follow repetitive behavior. This already points to a low possibility of detection.

Next, we compared the \emph{covert channel} recordings with the corresponding \emph{filtered covert channel} recordings. This gives us a mean p-value of 0.99 with a standard deviation of 0.04 and a mean D-value of 0.0008 with a standard deviation of 0.001. This on the other hand shows that our covert channel barely alters the characteristics compared to legitimate traffic.

If we compare \emph{covert channel} recordings with \emph{legitimate} recordings, we obtain a mean p-value of $1.53\cdot 10^{-95}\sim0$ with a standard deviation of $1.12\cdot 10^{-94}\sim0$ and a mean D-value of $1.87\cdot 10^{-1}$ with a standard deviation of $1.31\cdot 10^{-1}$. These results again point towards significant differences between the two scenarios. However, these differences are comparable to those of two legitimate recordings.

\begin{figure*}[tb]
    \centering
    \subfloat[p-value results from KS-test\\(All Scenarios)\label{fig:ks_p-values}]{\includegraphics[width=0.24\textwidth]{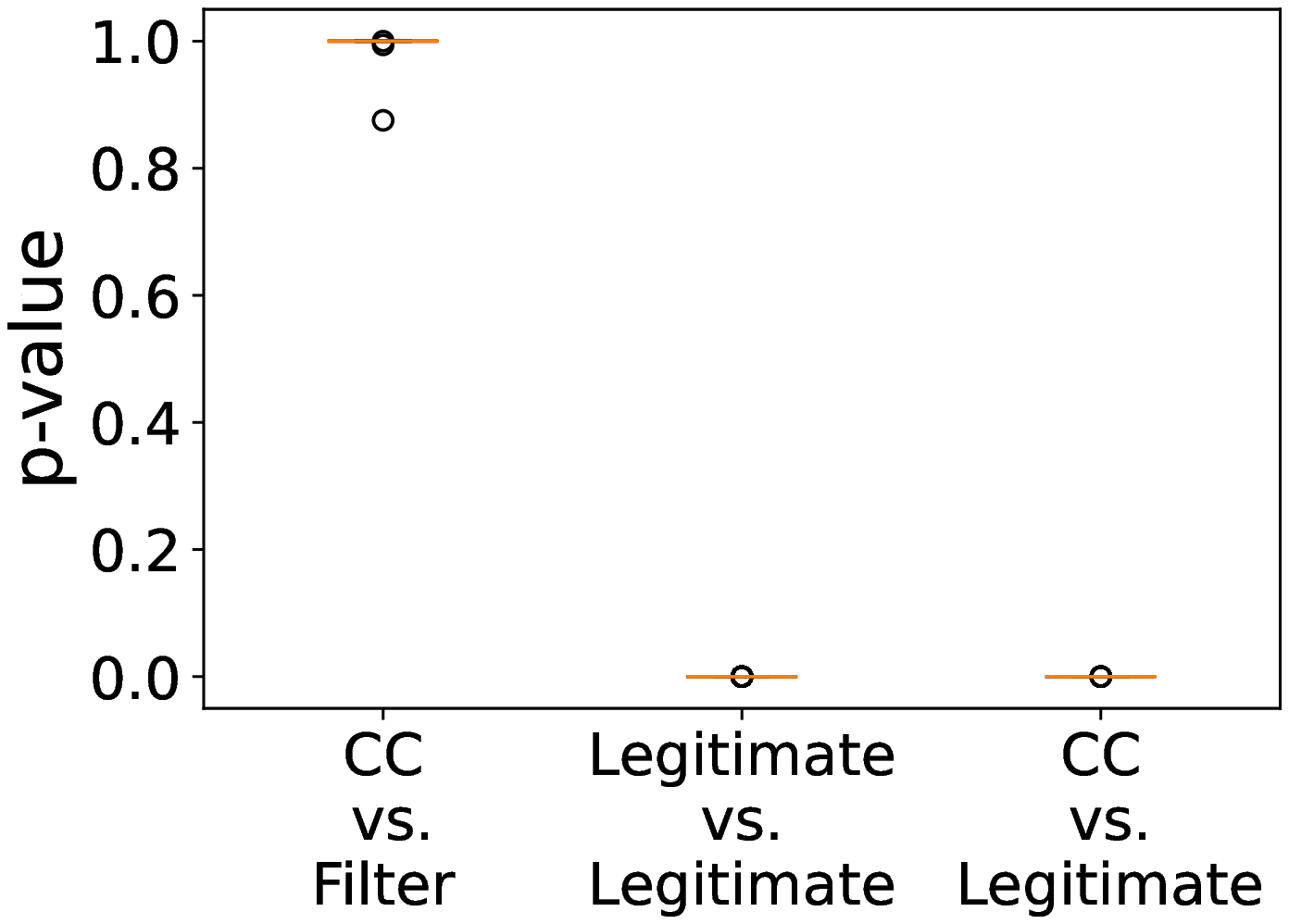}}
    \subfloat[Covert vs.\ Filtered Covert\\(Home Network)\label{fig:cc_vs_filter}]{\includegraphics[width=0.24\textwidth]{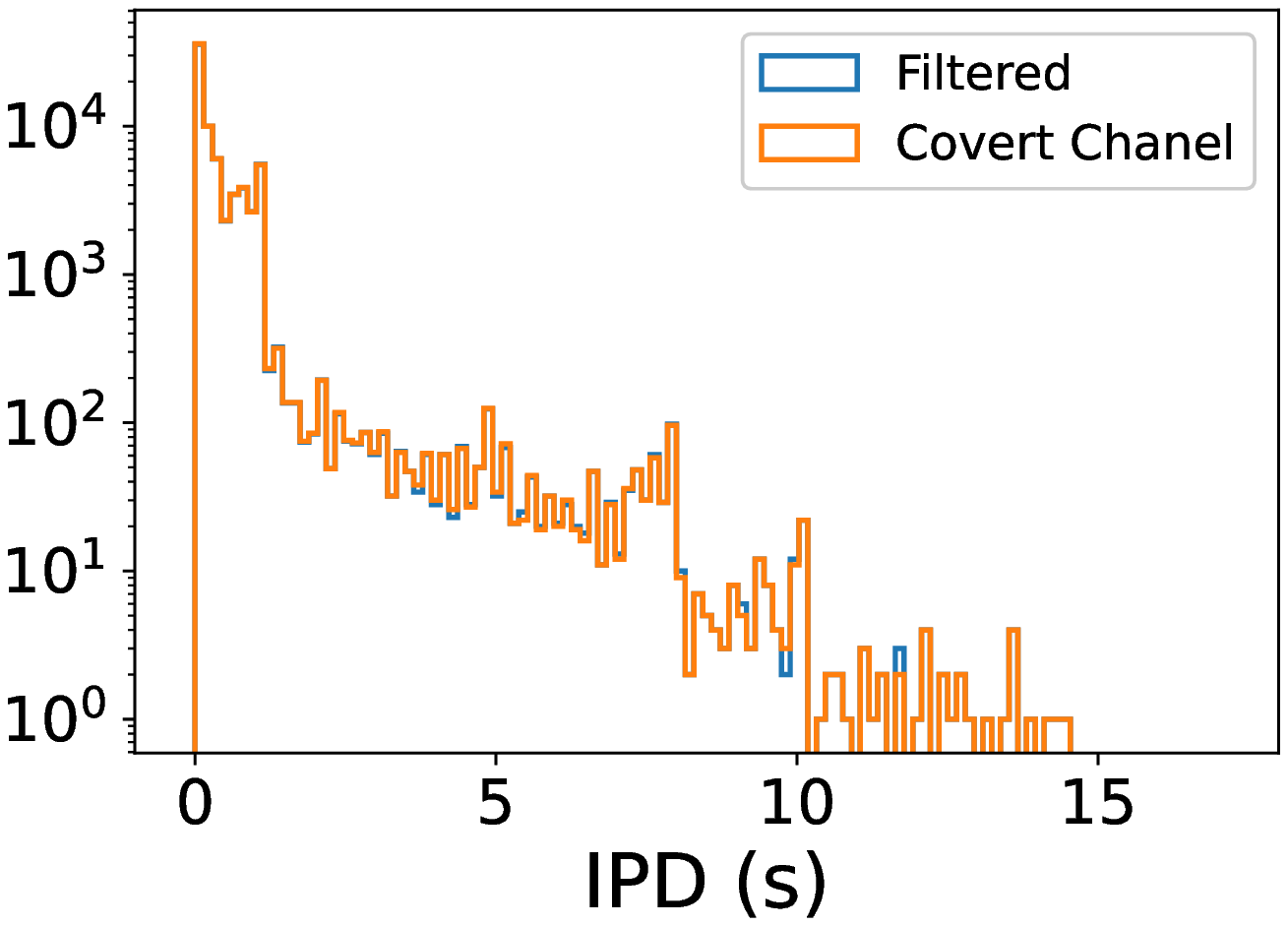}}
    \subfloat[Legtimate vs.\ Legitimate\\(Home Network)\label{fig:leg_vs_leg}]{\includegraphics[width=0.24\textwidth]{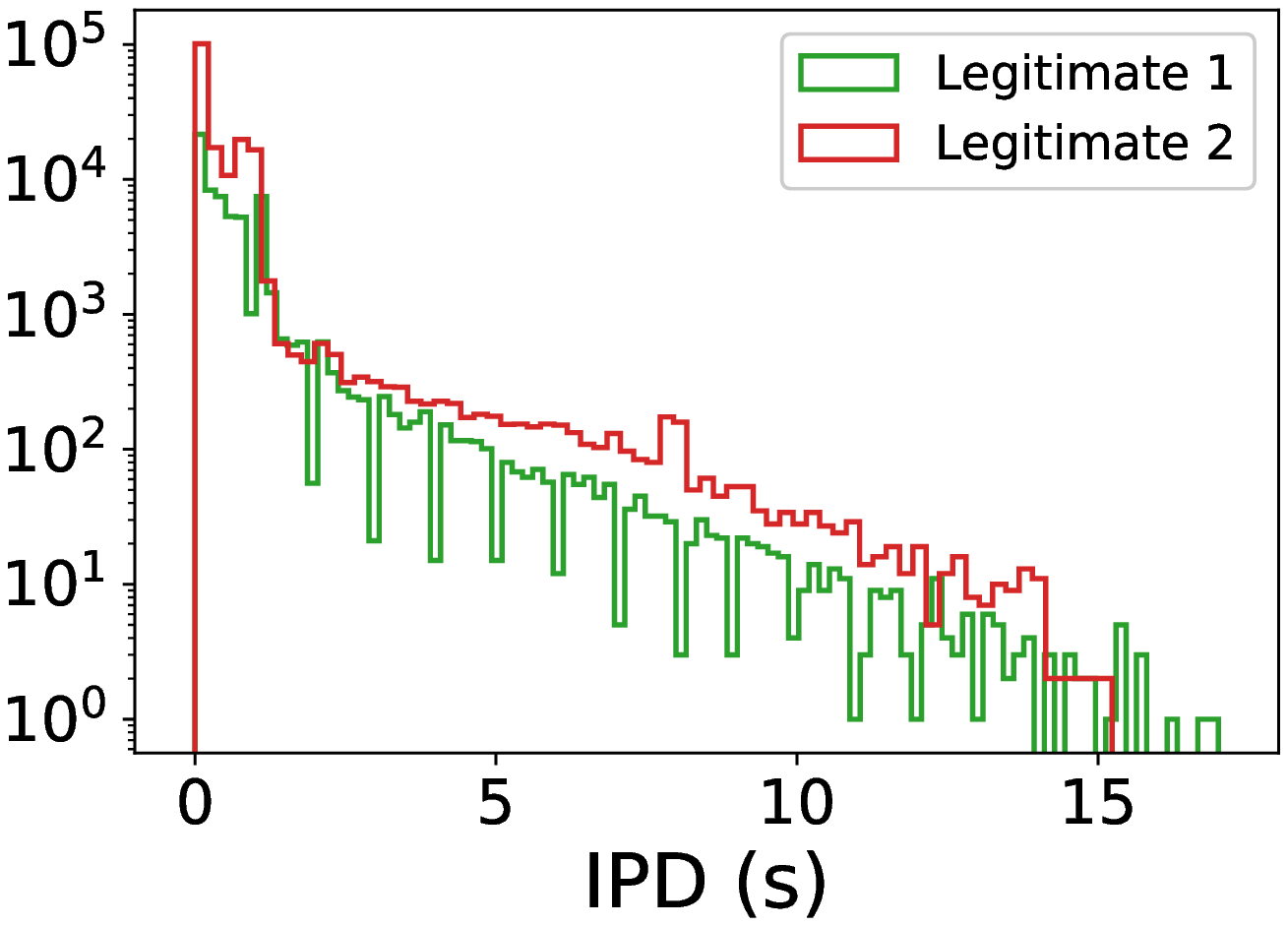}}
    \subfloat[Covert vs.\ Legitimate\\(Home Network)\label{fig:cc_vs_leg}]{\includegraphics[width=0.24\textwidth]{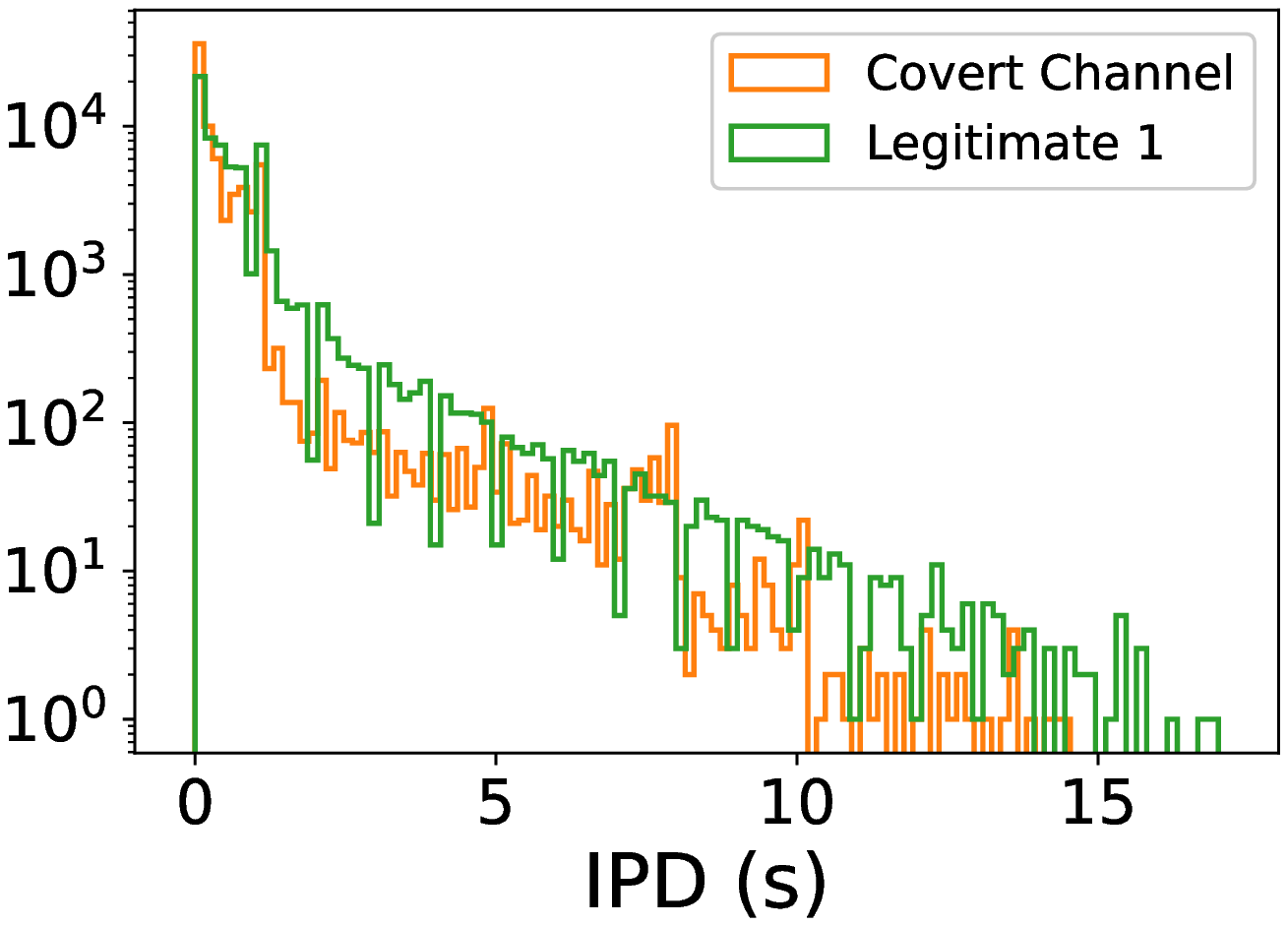}}
    \caption{Overview of legitimate and \CCName{} traffic's characteristics, with detailed examples for IPD values of the home network.}
    \label{fig:IAT_comp}%
\end{figure*}

Fig.~\ref{fig:ks_p-values} shows the p-values of the KS-tests and we can see a strong similarity of the CC vs.\ filter scenario and a strong dissimilarity for the other two scenarios.
Fig.~\ref{fig:cc_vs_filter}, \ref{fig:leg_vs_leg} and \ref{fig:cc_vs_leg} show exemplary histogram plots for the ARP request IPDs for each of the considered scenarios from the home network setup. Again, we can determine a high level of similarity between the covert channel recording and the filtered recording and slight differences between two legitimate recordings as well as between covert and legitimate recordings. 

These three results 
in combination
point towards an almost impossible detection of \CCName{}, as the differences between different legitimate recordings are larger than the difference between covert channel and filtered recordings.

\subsubsection{Compressibility Score}
In addition to the KS-test, we also used a widely known detection method from covert channel research: the compressibility score as proposed by Cabuk et al.~\cite{Cabuk06}.
Again, we used the IPDs of ARP requests for the detection and focused on the same three classes of combinations with the same recordings.
To calculate the compressibility score, we divided the recordings in windows of fixed length (1,000 IPDs). Each window was then transformed into a string representation $S$ of concatenated IPDs, which was compressed using a compressor $\Im$, in our case \texttt{gzip}. The final compressibility score $\kappa=|S|/|\Im(S)|$ for a window is the compression ratio between the original string and the compressed string. Since covert timing channel flows use similar IPD values in a re-occurring manner, their compressibility score is typically higher than the score of legitimate flows.

\begin{figure*}[tb]
    \centering
    \subfloat[$\kappa$-values for all recordings\\(All Scenarios) \label{fig:comp_box}]{\includegraphics[width=0.24\textwidth]{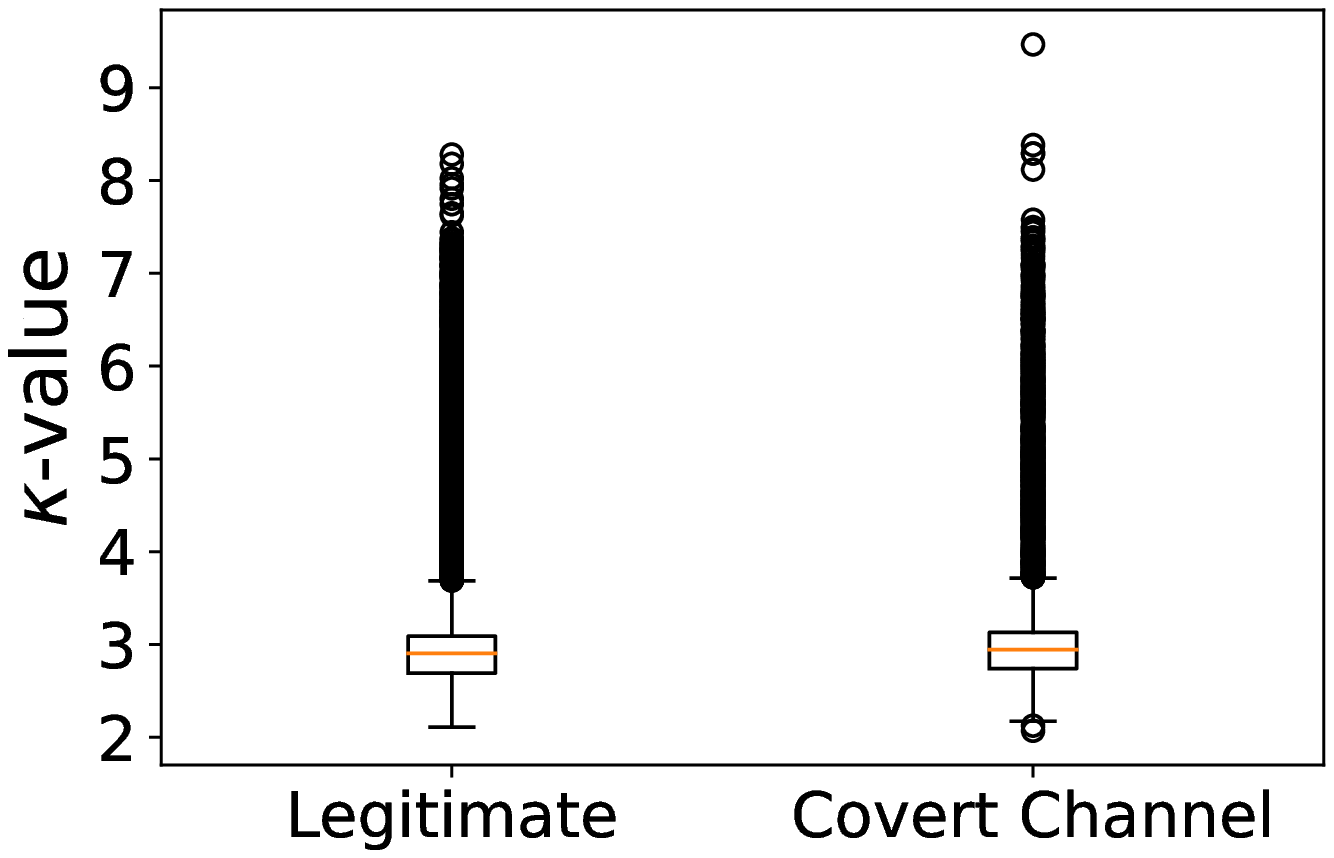}}
    \subfloat[Covert vs.\ Filtered Covert\\(Home Network)\label{fig:comp_cc_vs_filter}]{\includegraphics[width=0.24\textwidth]{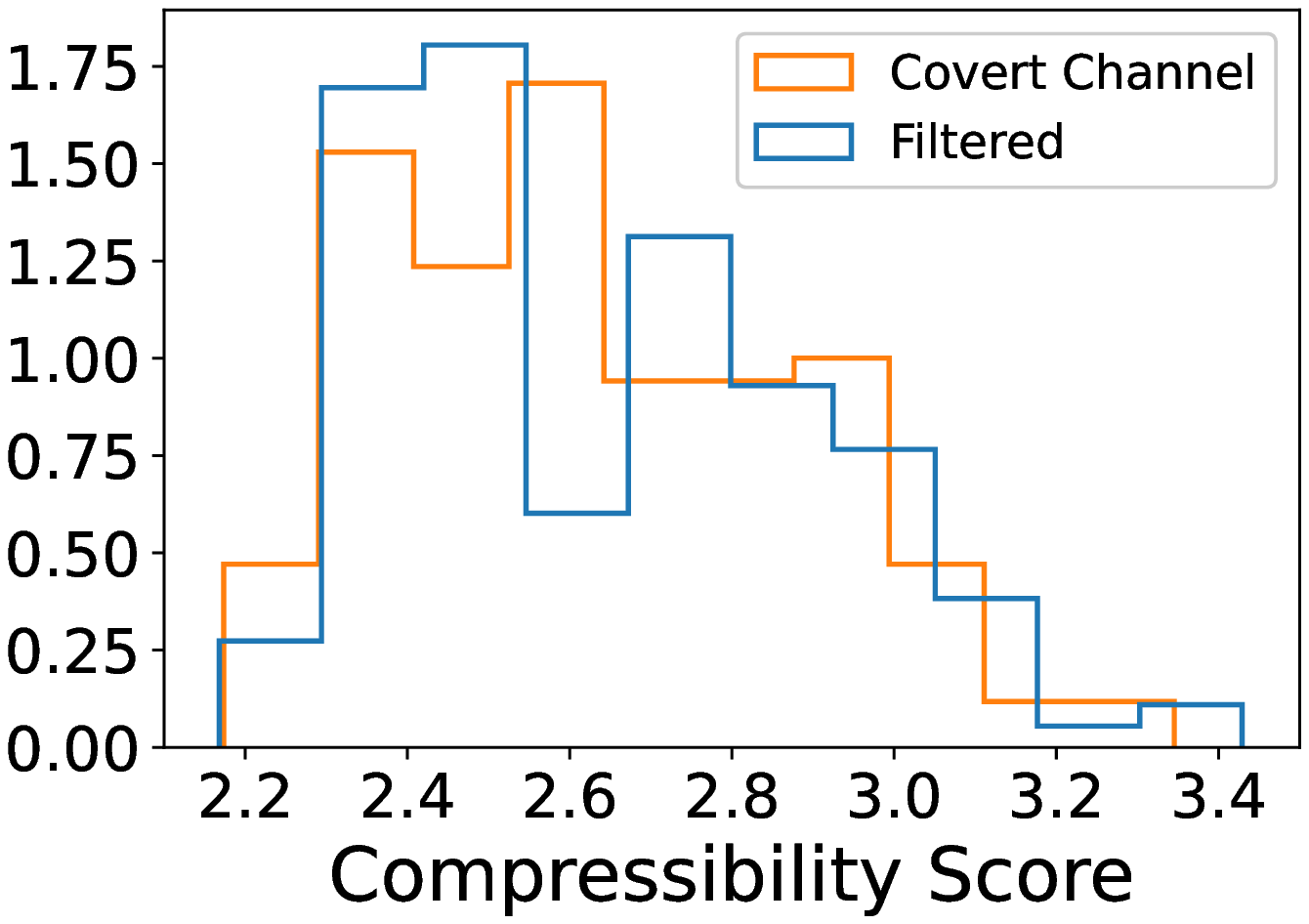}}
    \subfloat[Legitimate vs.\ Legitimate\\(Home Network)\label{fig:comnp_leg_vs_leg}]{\includegraphics[width=0.24\textwidth]{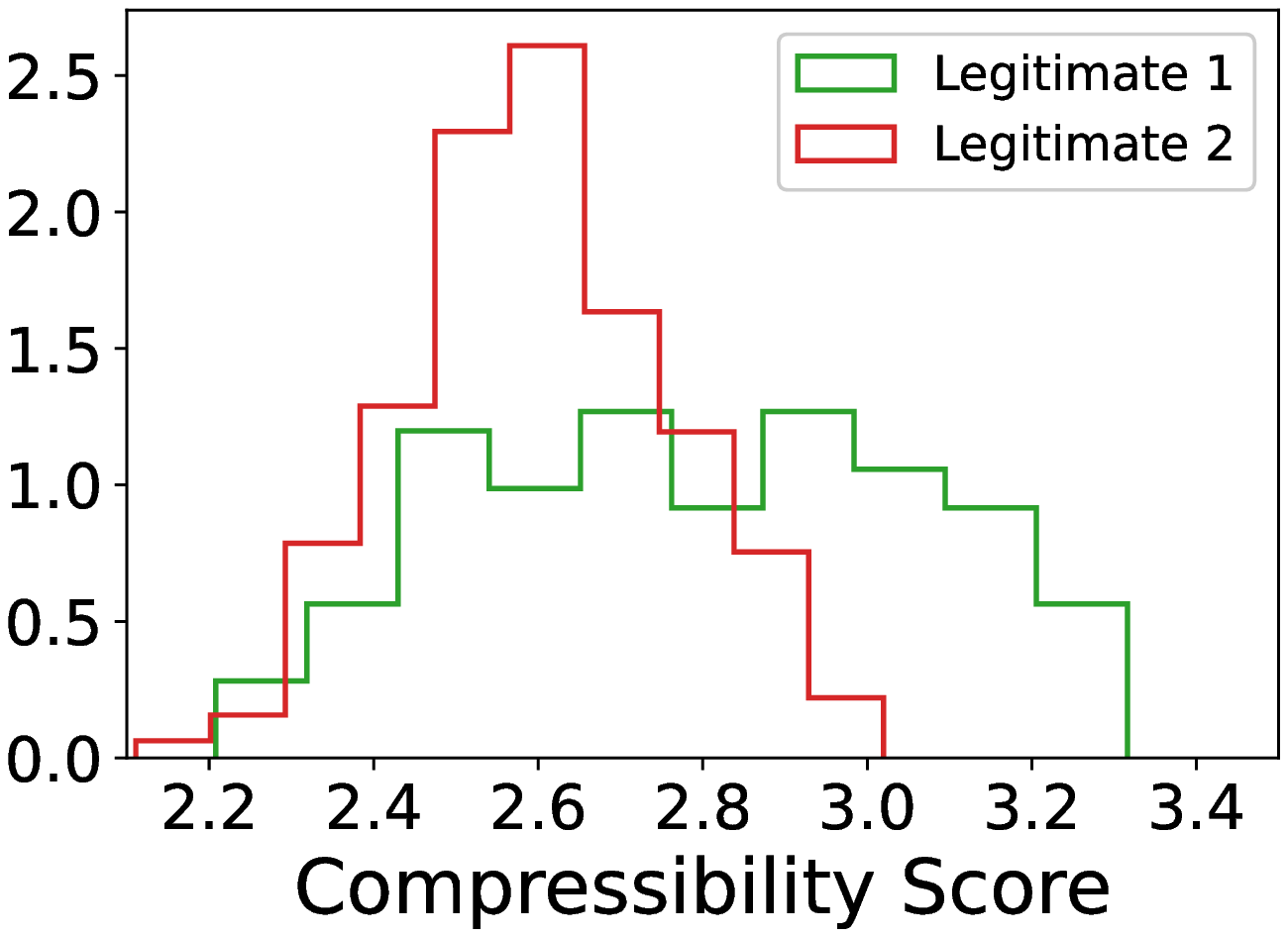}}
    \subfloat[Covert vs.\ Legitimate\\(Home Network)\label{fig:comp_cc_vs_leg}]{\includegraphics[width=0.24\textwidth]{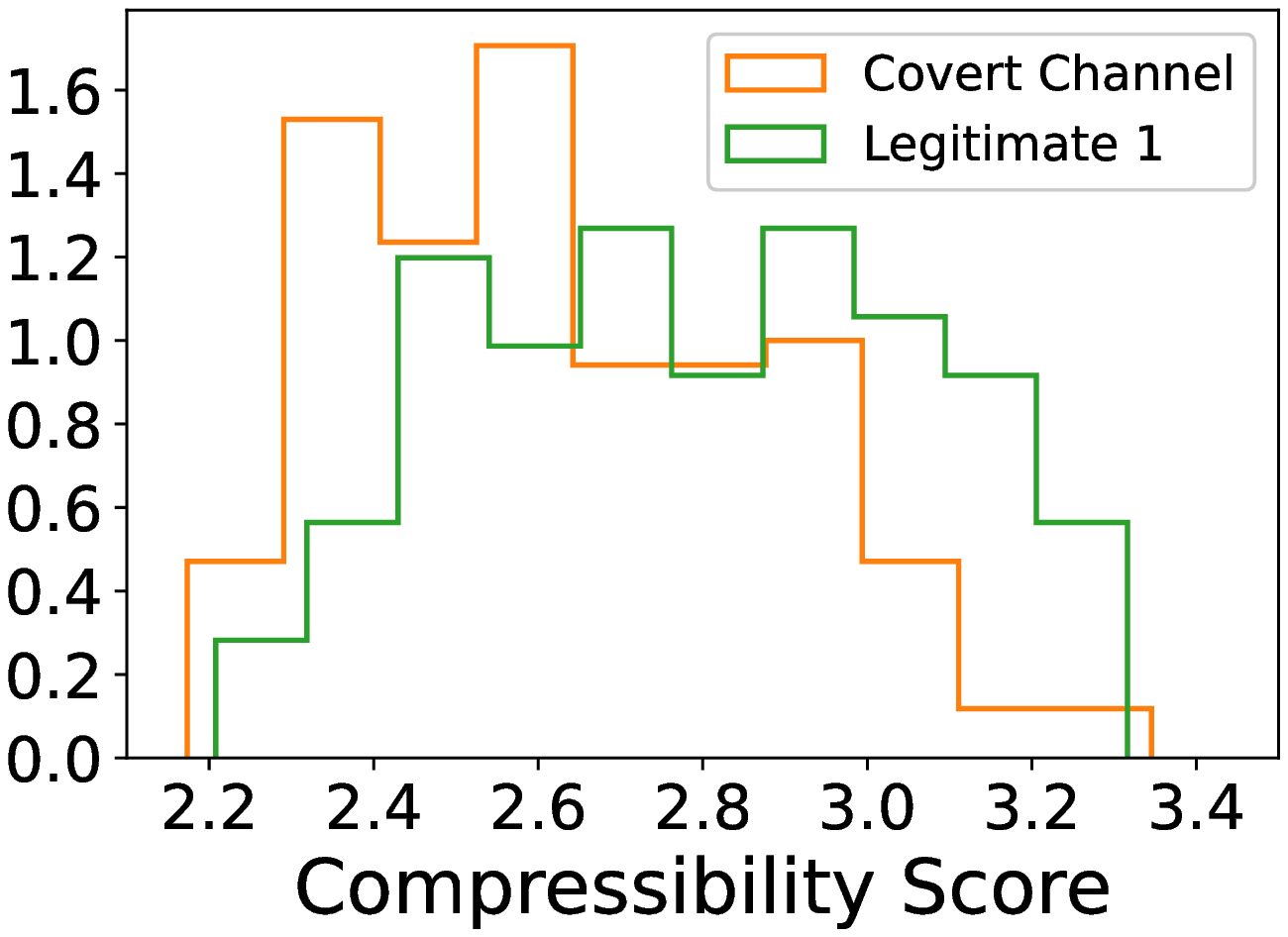}}
    \caption{Overview of legitimate and \CCName{} traffic's compressibility scores, with detailed examples for the home network scenario.}
    \label{fig:comp_score}
\end{figure*}

Fig.~\ref{fig:comp_box} shows the compressibility scores for all recordings. Here we can see a nearly identical values for legitimate and covert channel recordings, which already points towards a challenging detectability of \CCName{}.
Figs.~\ref{fig:comp_cc_vs_filter}, \ref{fig:comnp_leg_vs_leg} and \ref{fig:comp_cc_vs_leg} show exemplary histograms with pairwise plots for the distributions of the compressibility scores of the home network.
As can be seen, we found a large overlap of compressibility scores when comparing covert channel and filtered recordings. This again points towards a strong similarity of the two recordings.
The comparison of two legitimate recordings of the home network shows significant differences in the distributions while we would expect two matching plots, indicating a high dependence on the current status of a network (daytime etc.).
Moreover, the comparison between the covert channel and the legitimate recording shows even smaller differences than the two legitimate recordings.
When determining optimal thresholds for the detection of \CCName{}, our experiments revealed that the compressibility metric (and the related AUC scores for ROC charts) fluctuated based on time of day, network load or other factors, rather than the presence of \CCName{}. This is rooted in the fact that \CCName{} sends only very few messages.
Thus, similar to the KS-test, we observed that there is no clear threshold for the $\kappa$-value to discriminate between legitimate and covert channel recordings. This fact leads to the conclusion that \CCName{} is not detectable with the compressibility score.

\subsubsection{Evaluation of Different Covert Channel Configurations}\label{sect:evalDetectMultipointer}
\begin{figure}
    \centering
    \includegraphics[width=0.8\linewidth]{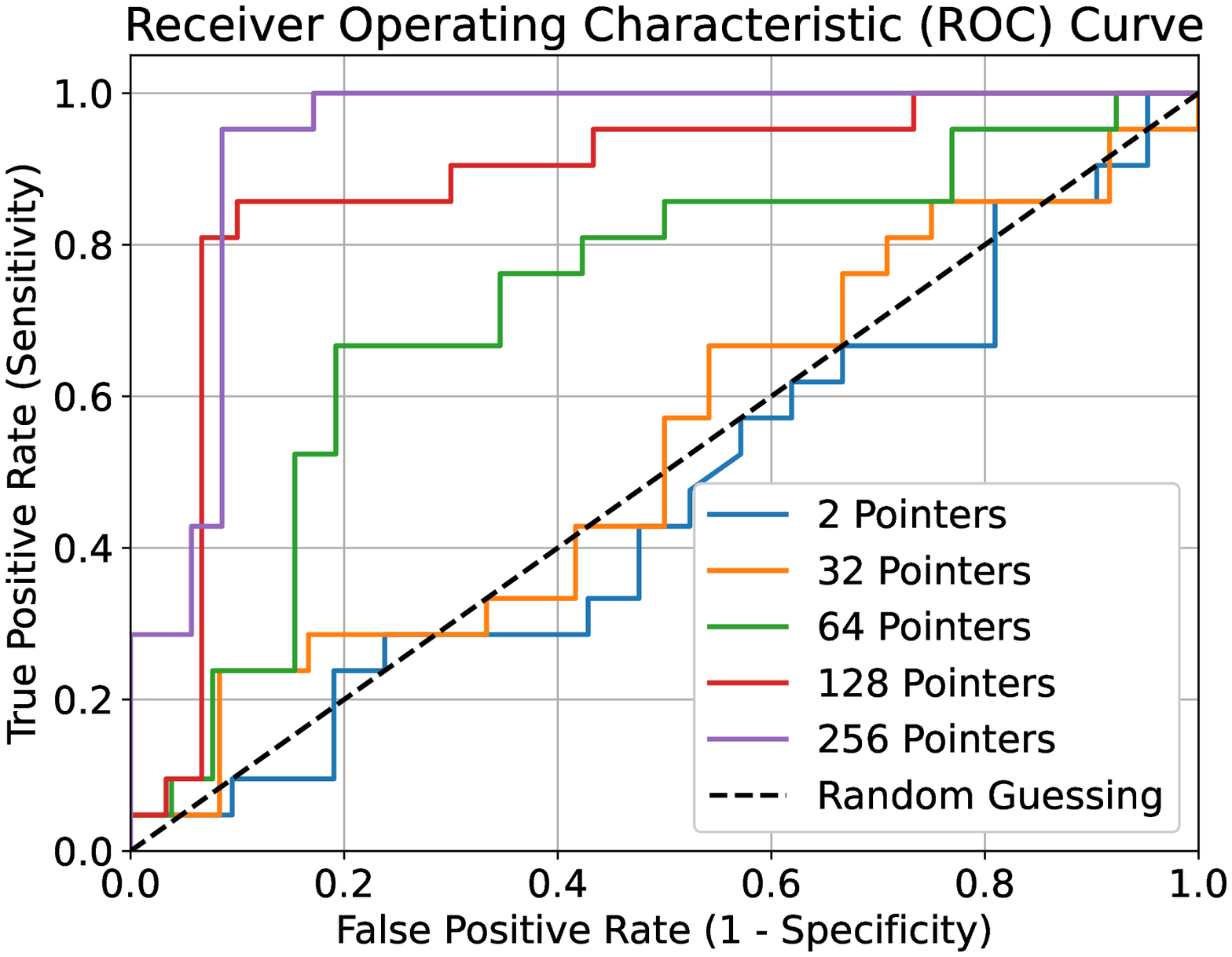}
    \caption{{Detection performance based on pointer count}}
    \label{fig:ROC_multi}
\end{figure}

{We also evaluated the detectability of \CCName{} when only focusing on a single configuration at a time (e.g., 1 byte basic or 2 byte extended modes). We found that the compressibility score performed no different than before, still producing overlapping $\kappa$-values.
Similarly, the KS-test performed comparable (figures omitted due to space reasons).
This was expected, as the detectability of \CCName{} relies primarily on the number of signals sent. 

To further evaluate detectability, we ran our multi-pointer variant of Sect.}~\ref{multihash} {while allowing different pointer counts. Fig.}~\ref{fig:ROC_multi} {shows the detection results using the compressibility score.}

{As can be seen, the approaches with two, and even 32 pointers perform very close to \CCName{}-Basic with only one pointer (AUC close to 0.5). Only when the number of pointers was increased to $\geq 128$, we were able to provide reasonably good detectability that would be useful in practice.}



{We conclude that \CCName{}'s detection depends on the chosen configuration. A configuration with $\leq 32$ pointers (i.e., high CAF, but few matches and low bitrate) was undetectable. A configuration with $\geq 64$ pointers sacrifices is detectable and offers a low CAF but provides a higher bitrate (cf.~Sect.}~\ref{sect:DYSTMultiHash}{).}

\section{Feasibility Study of Remote Variant}\label{sect:eval:remote}

{So far, we analyzed \CCName{}-Basic and \CCName{}-Ext in a local network. We} now study the feasibility of \CCName{}-Remote-Smarthome. In particular, we use a traffic recording from a smart home testbed to evaluate the sending performance of \CCName{} with three different short messages and the number of PoIs that were observed during these tests. 


\paragraph*{Performance Comparison: Local vs.\ Remote}

We utilized traffic recordings from our university's home automation testbed running eleven different smart devices, which are connected to a local router with NAT-based Internet access. The recordings were performed over a duration of several weeks while the network was in \textit{idle} mode, i.e., no user was interacting with these devices.
To simulate the sending process of \CCName{}-Remote-Smarthome, we read the pcap-recordings packet by packet and performed the same calculations as would be done by a real CS. When a match was found, instead of sending out a signal packet, we wrote a corresponding line to a logfile.

To compare the performance of \CCName{}-Remote-Smarthome with \CCName{}-Basic, we transmitted three short messages (13--15 characters) using different configurations.
Tab.~\ref{Tab:remote_trans_times} shows the average time it took to transmit a message (transmitting 1 byte blocks, i.e., $h=8$) for a given scenario, with and without robust mode.
\begin{table}[ht]
    \centering
    \scriptsize
    \caption{Avg.~Remote Transmission Times (13--15 Characters) }
    \label{Tab:remote_trans_times}
    \begin{tabular}{llcc}
        \toprule
        Scenario & Robust & From First Signal & From Start \\
        \midrule
        \multirow[c]{2}{*}{Home Network (local)} & False &  00:26:55 &  00:30:53 \\
         & True & 03:15:08 &  03:38:49 \\
        \midrule
        \multirow[c]{2}{*}{University (local)} & False & 01:11:27 & 01:28:14 \\
         & True &  03:22:14 & 04:06:29 \\
        \midrule
        \multirow[c]{2}{*}{Remote-Smarthome} & False & 00:10:13 & 00:11:40\\
         & True & 08:06:28 & 09:46:30 \\
        \bottomrule
    \end{tabular}
\end{table}
As we can see, \CCName{}-Remote-Smarthome delivers {better} transmission speeds without the robust mode while the transmissions with the robust mode are noticeably slower compared to the other two scenarios.
This is to be expected, as we found more potential PoIs in the remote scenario (12,700 vs.~7,300 and 2,100 PoI per hour), compared to the local mode which results in more potential matches (non-robust mode), but also more potential collisions (robust mode).

\paragraph*{Further Notes}
The robustness measures and the optimization measures for \CCName{}-Remote-Smarthome work the same way as for the LAN scenario; the major difference relies solely on the type of packets that are considered PoI and the type of packets that are signaling packets. For this reason, and because of the simulation-only data, we skip an evaluation of robustness and optimization at this point. The detectability of \CCName{}-Remote-Smarthome can be considered similarly infeasible as with \CCName{}-Basic as we only sent 1,345 signaling packets per day (on average, $\sigma=364$), and even fewer messages while in robust mode.


\section{Throughput Optimization Through Multi-Pointer Method}\label{sect:DYSTMultiHash}
\begin{figure}[hbt]
    \centering
    \subfloat[1 byte/match \label{fig:multi_pointer_1byte}]{\includegraphics[width=0.65\linewidth]{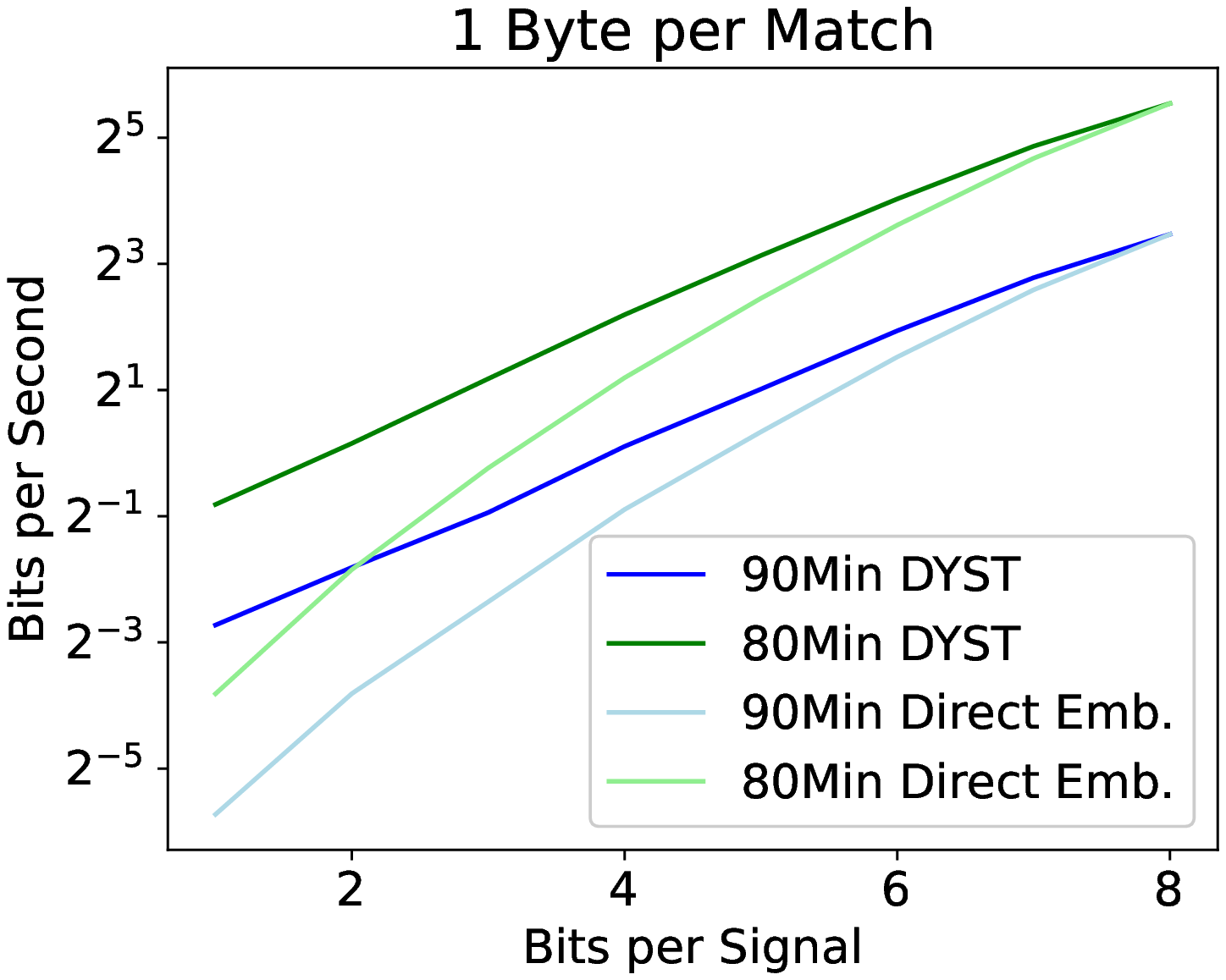}}\\
    \subfloat[2 byte/match \label{fig:multi_pointer_2byte}]{\includegraphics[width=0.65\linewidth]{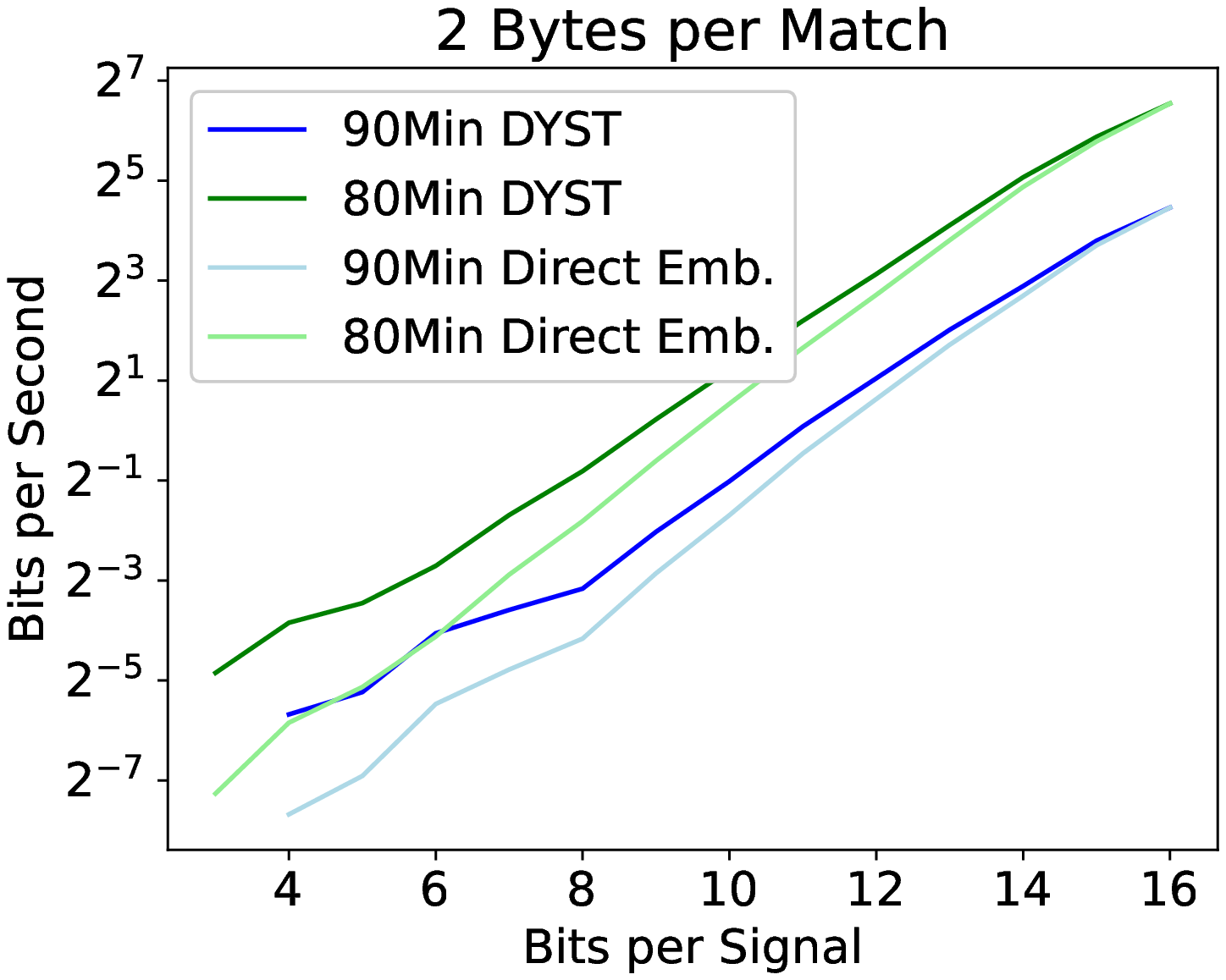}}
    \caption{\label{fig:multihashthroughput}{Bandwidth evaluation results for the multi-pointer approach and two different recordings from the office network (different times of the day).}}
\end{figure}

{The previous experiments focused on a \CCName{} setup with maximized amplification (CAF). We additionally evaluated the throughput-optimized setup that was introduced in Sect.~}\ref{multihash}. 
{To evaluate this approach against the performance of a traditional covert storage channel, we implemented a synthetic version that works on PCAP recordings and analysed the behaviour on two recordings from a busy office network. The results are shown in Figs.}~\ref{fig:multi_pointer_1byte} {and} \ref{fig:multi_pointer_2byte}. 
{The \emph{direct embedding} represents a covert storage channel that instead of embedding the pointer embeds the data itself. Note that we chose to only embed covert data in packets that would have been used as signal packet by DYST. 
As can be seen, we can increase the sending performance by sacrificing the CAF up to a point where we have no amplification but the full bitrate of a normal covert storage channel. In other words, one can configure \CCName{} so that one either optimizes for stealthiness (maximize CAF) or throughput. Note that the CAF becomes one if the pointer size equals the message size. 
We can observe that for both scenarios (i.e., pointing to either 1 or 2 bytes per match), \CCName{} has the advantage over the direct embedding up to the point where the pointer size matches (or exceeds) the size of the message being pointed to. The maximum bitrate \emph{with} amplification was 29.12 (7 bit pointer to 8 bit message) and 58.75 (15 bit pointer to 16 bit message), respectively.}

\section{Discussion}\label{sect:disc}
{Our approach provides room for several follow-up developments in terms of alternative methodology. We have studied how one can point to the \emph{content} of previously seen packets. However, other hiding patterns }\cite{CSURpaper}{ could probably be useful, too. For instance, the \emph{inter-packet times} hiding pattern could be used to point to a series of inter-packet gaps, the \emph{size modulation} pattern could be used to point to a sequence of packet sizes and so forth.} 
{Further, by neglecting the context of network traffic and discarding the use of a hash function, we have recently shown that a history channel can be used to point to textual elements on the Internet (e.g., content from Wikipedia pages) to bypass censorship }\cite{OPPRESSION}{.} 
{A similar scenario could be} a social media service, where legitimate users submit postings, such as \emph{Facebook}, \emph{Twitter}, or any form of blog or platform that generates a massive amount of content {(e.g., \emph{Dropbox}, \emph{Google Drive} or \emph{Github})}. 
Signaling for these scenarios is not bound to the same service, which means that signals can be sent out-of-band, for example via a different platform or a network protocol. 
%
%
Moreover, one could use history covert channels to point to previously seen content in audio/video streams.
Finally, \CCName{} could be applied for stealthy communication between local processes of a secure operating system by monitoring hardware events.


\section{Conclusion}\label{sect:concl}

We introduced \emph{history} covert channels, jointly with the paradigm of covert channel \emph{amplification}. History covert channels send secret messages by \emph{pointing} to unaltered legitimate data.
They can be applied in extremely hostile environments, i.e., where a covert sender is unable to send any other message than previously whitelisted ones.
We introduced different variants and an implementation called \CCName{}. Our results indicate a limited but variable throughput dependent on the number of bits signaled at once.
We analyzed the robustness and also have shown that traditional heuristics are unable to provide satisfying results unless many pointers are employed. 
Future work will focus on improving the sending performance and developing additional countermeasures.
Code and data: \url{https://github.com/NIoSaT/DYST}

\bibliographystyle{IEEEtran}
\bibliography{sample-base}

\section{Acknowledgements}
S.~Wendzel and S.~Zillien have been supported by the ``Innovative University'' programme, a joint initiative of the Federal Government and the German States (project \textit{EMPOWER}, FKZ 03IHS242D). 
This open access article was supported by the HS Worms library funds.

\input{supplement_n_rebuttal}

\clearpage
\section{Biographies}

\begin{IEEEbiography}[{\includegraphics[width=1in,height=1.25in,clip,keepaspectratio]{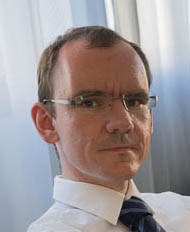}}]{Steffen Wendzel} is a professor at Hochschule Worms, Germany, where he is the scientific director of the Center for Technology and Transfer. He is also a lecturer at the Faculty of Mathematics \& Computer Science at the Fern\-Universität in Hagen, Germany, from which he also received his PhD (2013) and Habilitation (2020). Before joining Hochschule Worms, he led a smart building security research team at Fraunhofer FKIE in Bonn, Germany. He (co-)authored 180+ publications and (co-)organized several conferences 
as well as special issues for major journals. 
Steffen's major research focus is on censorship circumvention, 
scientific taxonomy, and IoT security. Website: \url{https://www.wendzel.de}.
\end{IEEEbiography}

\begin{IEEEbiography}[{\includegraphics[width=1in,height=1.25in,clip,keepaspectratio]{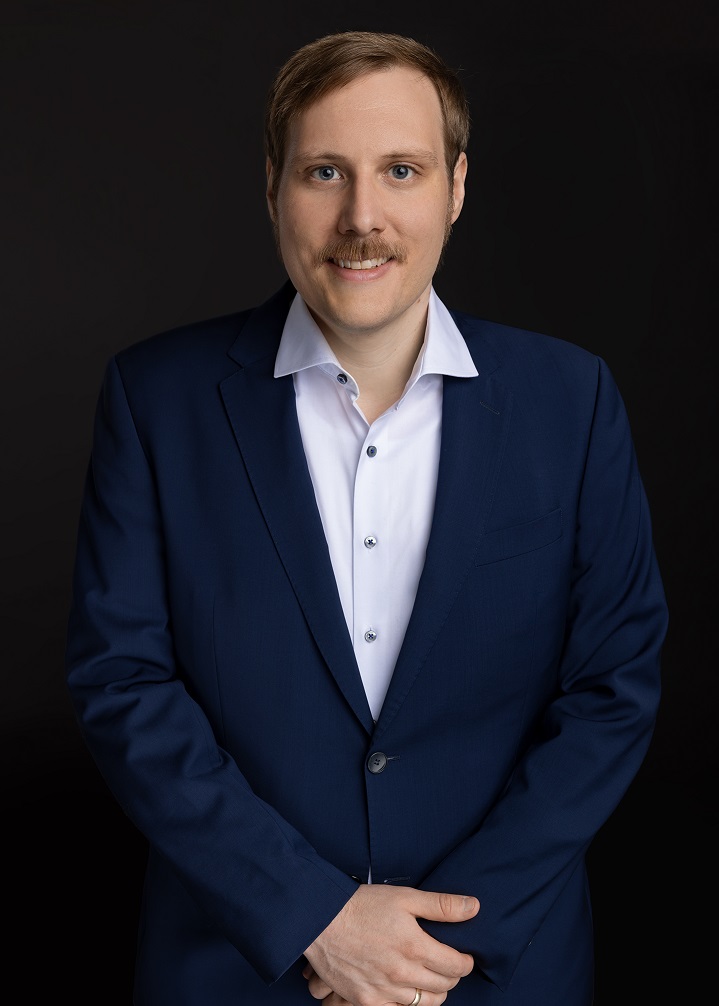}}]{Tobias Schmidbauer} is a professor at the Nuremberg Institute of Technology and received his PhD (2023) and M.Sc.\ (2019) from the Faculty of Mathematics \& Computer Science at FernUniversit\"at in Hagen, Germany, and his diploma degree in computer science for public management from the University of Applied Sciences for Public Administration in Bavaria in cooperation with the University of Applied Science Hof, Germany, in 2016. Alongside his studies, he worked from 2007 until 2017 for the datacenter of the tax administration of the German federal state of Bavaria. Since 2017, he worked full-time for the Bavarian State Office for Information Security until April 2024.
\end{IEEEbiography}

\begin{IEEEbiography}[{\includegraphics[width=1in,height=1.25in,clip,keepaspectratio]{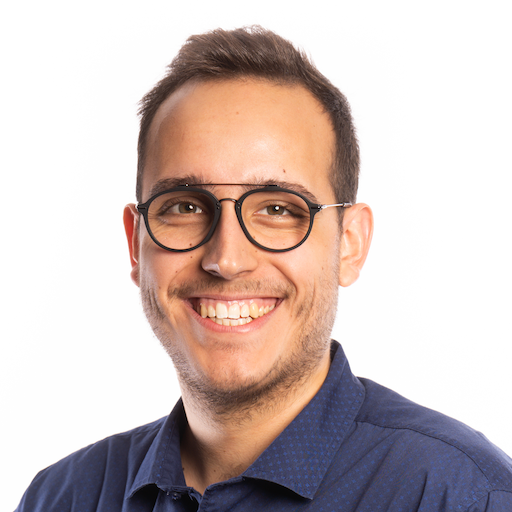}}]{Sebastian Zillien} is a researcher and PhD student at the Hochschule Worms, Germany, where he works for the projects WoDiCoF+ and ATTRIBUT. He received his BSc in Applied Informatics in 2018 and his MSc in Mobile Computing in 2020 from Hochschule Worms. His research focus includes network and protocol-level security, anomaly detection, covert channels and reliability. He has had publications, among others, at AsiaCCS, TDSC, IFIP SEC, NordSec and 
J.UCS.
\end{IEEEbiography}

\begin{IEEEbiography}[{\includegraphics[width=1in,height=1.25in,clip,keepaspectratio]{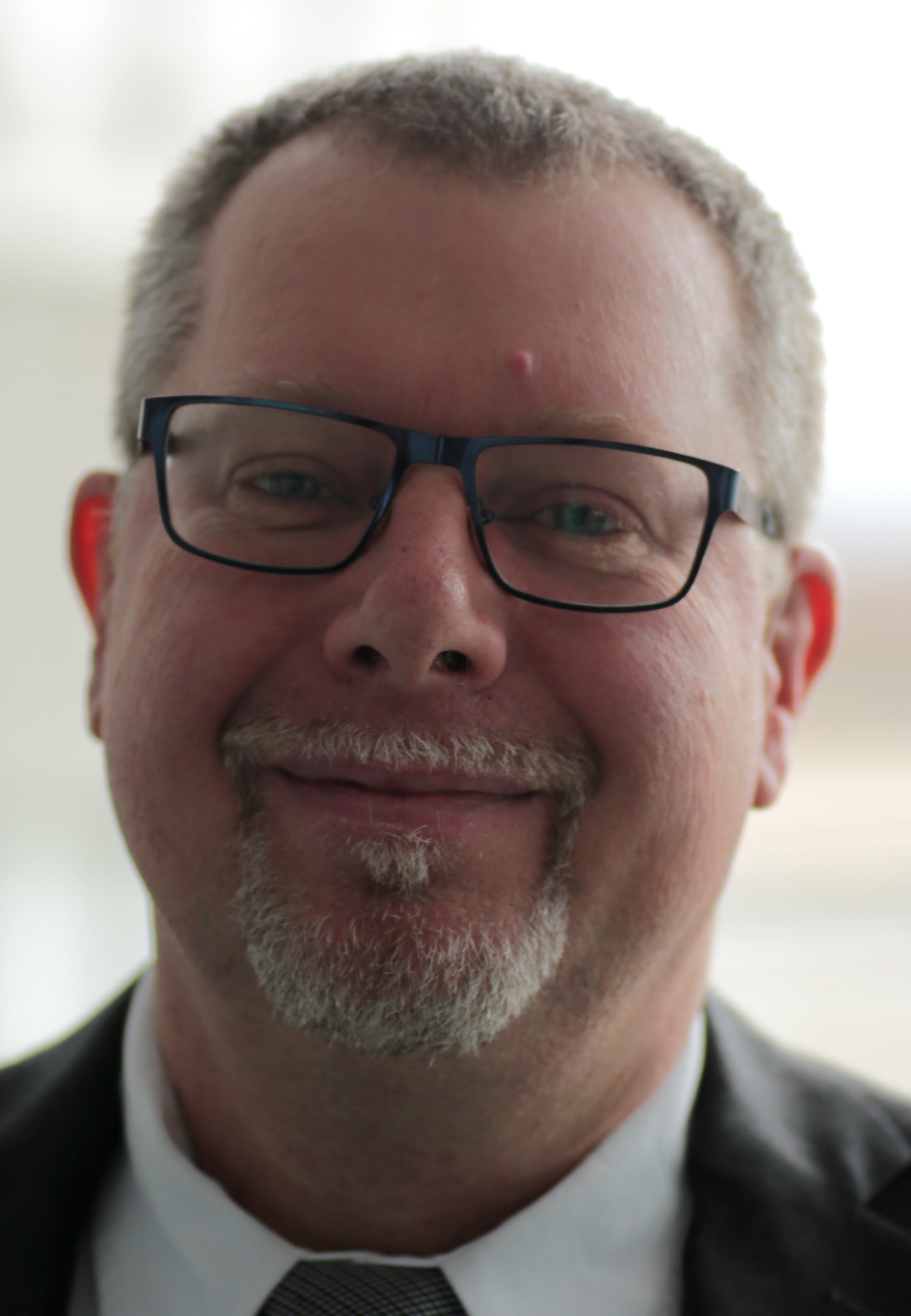}}]{Jörg Keller} is a professor of computer engineering at FernUniversität in Hagen, Germany, where he leads the Parallelism \& VLSI research group since 1996. He received MSc and PhD degrees and habilitation from Saarland University, Saarbrücken, in 1989, 1992, and 1996, respectively. His research interests include network steganography, cryptographic primitives for embedded systems, energy-efficient and fault-tolerant parallel computation, and blended and virtual laboratories. He is author or co-author of 2 books and more than 180 refereed articles in journals and conference proceedings. He has co-organized numerous conferences and workshops, and special issues. 
Website: \url{https://feu.de/pv/en}
\end{IEEEbiography}
\end{document}

%% file: supplement_n_rebuttal.tex
\section{Further Notes on the Optimization of DYST}
\label{sub:opti}
\begin{figure*}[!b]
    \centering
    \subfloat[Necessary legitimate packets \label{fig:necessary_packets}]{\includegraphics[width=0.33\textwidth]{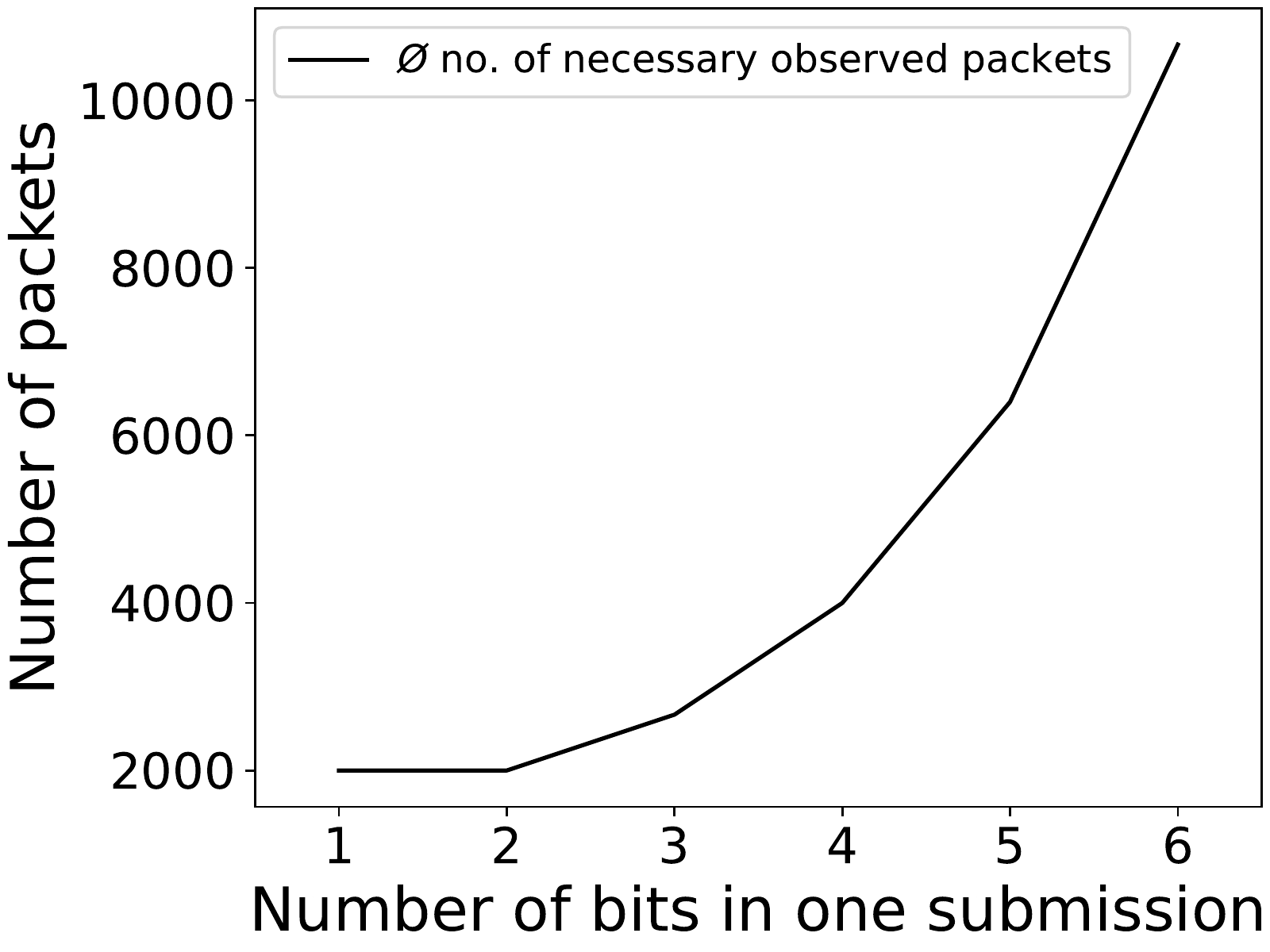}}
    \subfloat[Optimizing throughput\label{fig:opt_thr}]{\includegraphics[width=0.33\textwidth]{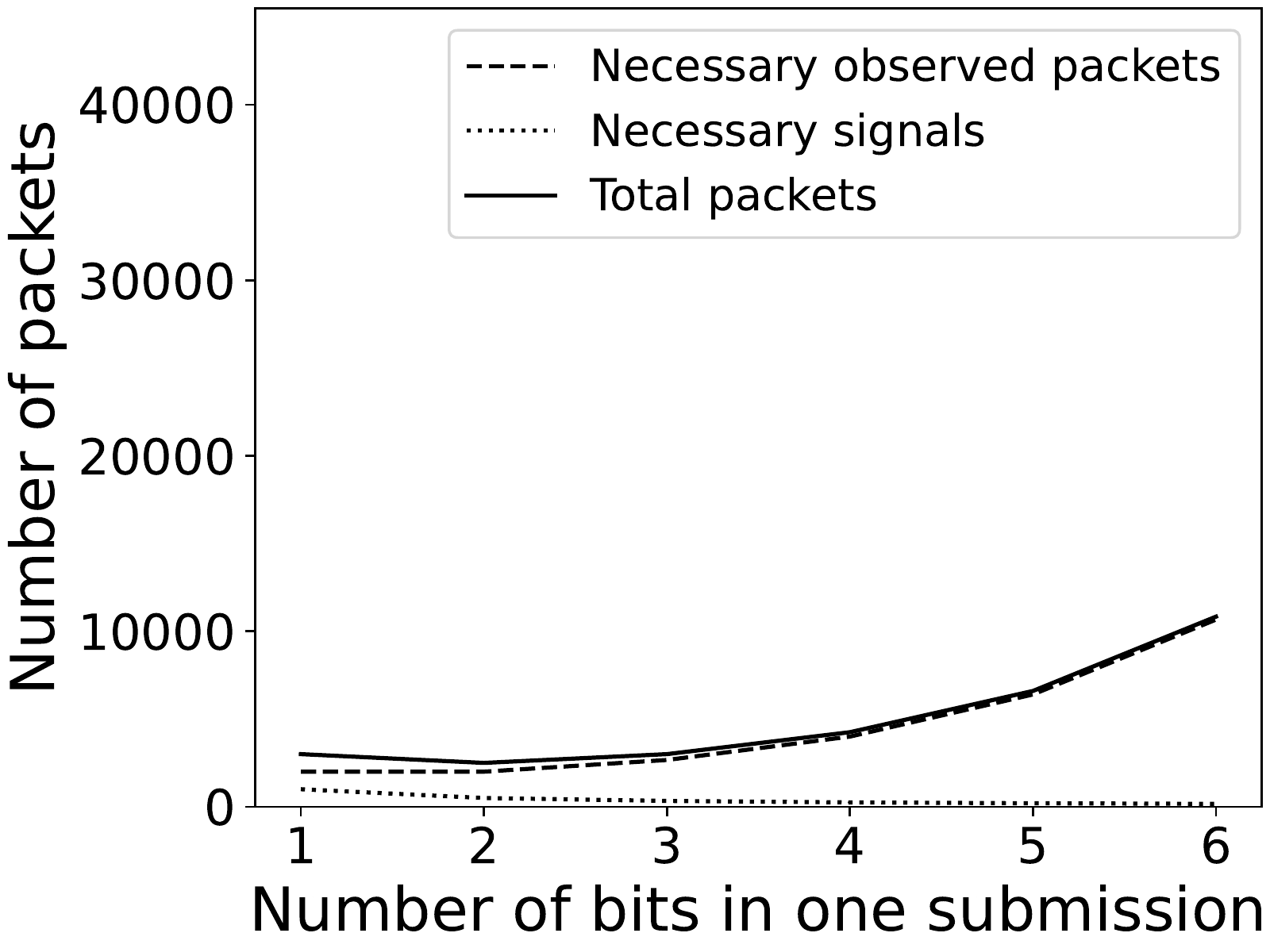}}\\
    \subfloat[Optimizing robustness\label{fig:opt_robust}]{\includegraphics[width=0.33\textwidth]{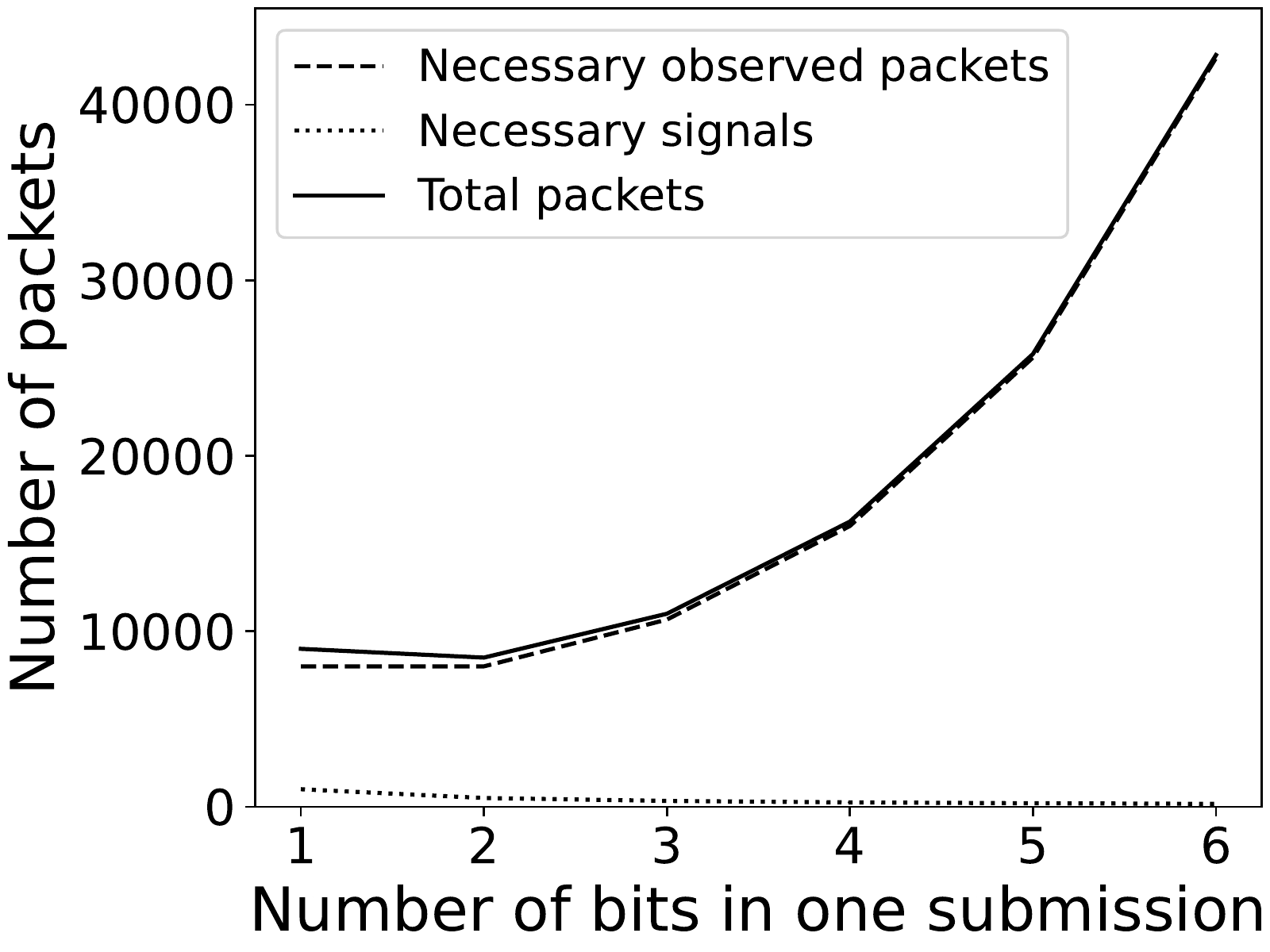}}
    \subfloat[Signals in total traffic\label{fig:opt_share}]{\includegraphics[width=0.33\textwidth]{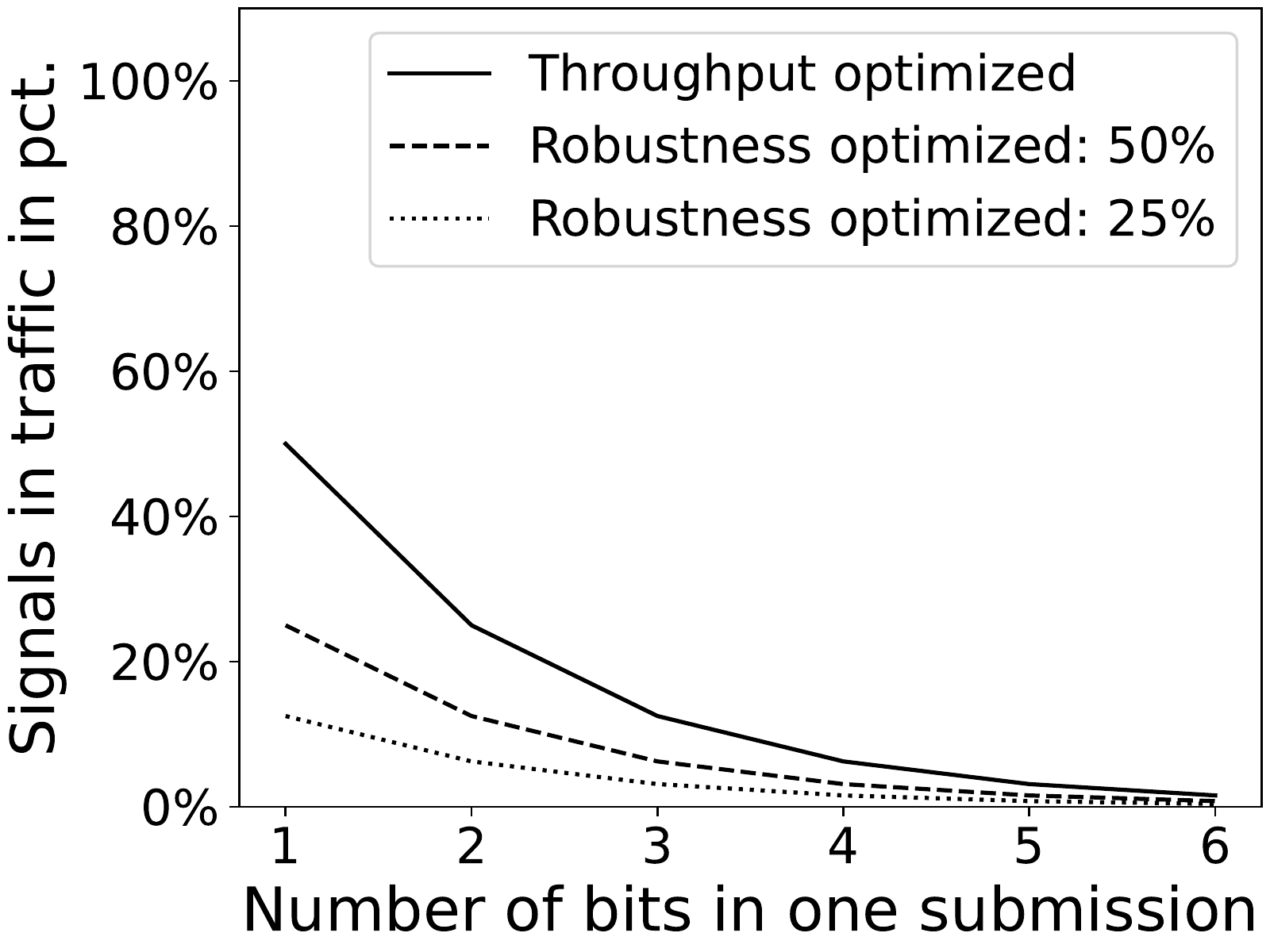}}
    \caption{\label{fig:optimizing}Transmission of 1,000 bits with \CCName{}}
\end{figure*}
Fig.~\ref{fig:optimizing} presents the transmission of 1,000 bits with \CCName{} with 1 to 6 bits transferred at once.
The theoretical number of hashes necessary to transmit a Message $M$ with \CCName{}-Basic 
can be calculated by dividing $\frac{len(M)}{h}$ and  $P_h(X\geq h)$, resulting in $len(M)\cdot 2^h / h$. The number of necessary packets to transmit a Message $M$ with a length of 1,000 bits is visualized in Fig.~\ref{fig:necessary_packets} and shows that the more bits shall be submitted at once, the more packets need to be observed. The number of packets grows exponentially with the number of bits to transmit in one chunk. This represents one packet for each observed match, disregarding robustness and stealthiness considerations. Further, the number of necessary signals decreases with the number of bits transferred in one chunk. This leads to a tradeoff between stealthiness and throughput. The stealthiness also increases with robustness, as for a robust transmission from CS to CR some packets need to be ignored as already described. Henceforth, the more robust the channel, also the stealthier \CCName{} will be, however, the throughput suffers.
This trade-off is also presented in Figs.~\ref{fig:opt_thr}-\ref{fig:opt_share} for the transmission of the message $M$ consisting of 1,000 bits with a maximum number of 6 bits signaled at once.
In the case of a throughput-optimized \CCName{} implementation, for each matching hash, a signal is sent. Fig.~\ref{fig:opt_thr} visualizes this configuration where the number of signals to be sent decreases as the number of bits encoded in one matching hash increases (dotted line), and nears 1 for a theoretical signal, pointing at 1,000 bits at once. The number of necessary packets to observe increases (dashed line) as the probability of a match also decreases because the total number of packets in a recording increases. These are represented by the solid line (showing the sum of necessarily observed packets and necessary signals).
For a robust approach, we assume in our example that only each fourth packet fits our previously described requirements (see Fig.~\ref{fig:opt_robust}). The number of necessary packets to observe increases while the number of signals stays constant, compared to the throughput-optimized scenario. Thus, compared to the total number of packets observed, \textit{fewer} signals are included in an experiment.
The share of signals in percent is shown in Fig.~\ref{fig:opt_share}. The solid line represents a throughput-optimized approach, while the dashed line shows a robust approach where each second packet fits the robustness requirements. The dotted line visualizes the approach where each fourth packet fits these requirements and equals the scenario in Sect.~\ref{ssec:robustness}.
The share of signals in percent decreases with the number of packets utilized for signaling and with the number of bits signaled at once. This leads to the conclusion that the fewer packets are utilized and the more bits are signaled at once, the less likely a detection. If CS and CR fear the presence of a warden, 
they can simply ignore each certain number of packets or increase the number of bits signaled at once to decrease the noticeability of a signal.

\paragraph*{Alternative Improvement of Sending Performance}
While the paper at hand presents an initial methodology, our \emph{OPPRESSION} extension \cite{OPPRESSION} applies the history covert channel method by pointing to publicly accessible (historic) textual web content (instead of network packets). This allowed us to improve the sending performance of popular censorship circumvention tools and underpins the flexibility of the history covert channel method. \hl{OPPRESSION's pointers refer to nodes in a Patricia Trie, which is a form of a tree representing the possible sentences Alice wants to transfer to Bob.}

\paragraph*{Improving Over Current Research}
\hl{However, we assume that one could enhance the sending performance provided by OPPRESSION similarly like Unix-like systems use \emph{inodes} (i.e., filesystem metadata entries) as follows. Instead of pointing to only one data chunk (or sentence), Alice and Bob could pre-compute trees that refer to \emph{multiple} data chunks (multiple sentences), which is handled by \emph{indirect} data pointers in inodes. Thus, one pointer could refer to significantly more data as it refers to several more pointers (that again refer to more pointers) that point to several different tree nodes. However, this would require larger trees and more pre-computation operations for Alice and Bob. There is currently \emph{no} implementation or evaluation of such an approach and it is thus left as future work for the research community.}

%% file: sample-authordraft.bbl
\begin{thebibliography}{10}
\providecommand{\url}[1]{#1}
\csname url@samestyle\endcsname
\providecommand{\newblock}{\relax}
\providecommand{\bibinfo}[2]{#2}
\providecommand{\BIBentrySTDinterwordspacing}{\spaceskip=0pt\relax}
\providecommand{\BIBentryALTinterwordstretchfactor}{4}
\providecommand{\BIBentryALTinterwordspacing}{\spaceskip=\fontdimen2\font plus
\BIBentryALTinterwordstretchfactor\fontdimen3\font minus
  \fontdimen4\font\relax}
\providecommand{\BIBforeignlanguage}[2]{{%
\expandafter\ifx\csname l@#1\endcsname\relax
\typeout{** WARNING: IEEEtran.bst: No hyphenation pattern has been}%
\typeout{** loaded for the language `#1'. Using the pattern for}%
\typeout{** the default language instead.}%
\else
\language=\csname l@#1\endcsname
\fi
#2}}
\providecommand{\BIBdecl}{\relax}
\BIBdecl

\bibitem{Lampson:1973}
B.~W. Lampson, ``A note on the confinement problem,'' \emph{Commun. {ACM}},
  vol.~16, no.~10, pp. 613--615, 1973.

\bibitem{DoD:1985}
{DoD}, ``Trusted computer system evaluation criteria,'' in \emph{The `Orange
  Book' Series}.\hskip 1em plus 0.5em minus 0.4em\relax Palgrave Macmillan UK,
  1985, pp. 1--129.

\bibitem{petitcolas1999information}
F.~A.~P. Petitcolas, R.~J. Anderson, and M.~G. Kuhn, ``Information hiding-a
  survey,'' \emph{Proc. {IEEE}}, vol.~87, no.~7, pp. 1062--1078, 1999.

\bibitem{ZanderAB07}
S.~Zander, G.~J. Armitage, and P.~Branch, ``A survey of covert channels and
  countermeasures in computer network protocols,'' \emph{{IEEE} Commun. Surv.
  Tutorials}, vol.~9, no. 1-4, pp. 44--57, 2007.

\bibitem{OutOfBandSurvey}
B.~Carrara and C.~Adams, ``Out-of-band covert channels - {A} survey,''
  \emph{{ACM} Comput. Surv.}, vol.~49, no.~2, pp. 23:1--23:36, 2016.

\bibitem{LCWM:StegoMalware}
L.~Caviglione and W.~Mazurczyk, ``Never mind the malware, here's the
  stegomalware,'' \emph{{IEEE} Secur. Priv.}, vol.~20, no.~5, pp. 101--106,
  2022.

\bibitem{Luca:StegoMalwareRepo}
\BIBentryALTinterwordspacing
L.~Caviglione, ``steg-in-the-wild,'' 2023. [Online]. Available:
  \url{https://github.com/lucacav/steg-in-the-wild}
\BIBentrySTDinterwordspacing

\bibitem{barradas2020towards}
D.~Barradas and N.~Santos, ``Towards a scalable censorship-resistant overlay
  network based on {WebRTC} covert channels,'' in \emph{Proc. Int. Workshop on
  Distributed Infrastructure for Common Good}.\hskip 1em plus 0.5em minus
  0.4em\relax {ACM}, 2020, pp. 37--42.

\bibitem{mileva2014covert}
A.~Mileva and B.~Panajotov, ``Covert channels in {TCP/IP} protocol stack -
  extended version-,'' \emph{Central Eur. J. Comput. Sci.}, vol.~4, no.~2, pp.
  45--66, 2014.

\bibitem{CSURpaper}
S.~Wendzel, S.~Zander, B.~Fechner, and C.~Herdin, ``Pattern-based survey and
  categorization of network covert channel techniques,'' \emph{{ACM} Comput.
  Surv.}, vol.~47, no.~3, pp. 50:1--50:26, 2015.

\bibitem{IoTStego17}
S.~Wendzel, W.~Mazurczyk, and G.~Haas, ``Don't you touch my nuts: Information
  hiding in cyber physical systems,'' in \emph{2017 {IEEE} Security and Privacy
  Workshops, {SP} Workshops}.\hskip 1em plus 0.5em minus 0.4em\relax {IEEE},
  2017, pp. 29--34.

\bibitem{Krishnamurthy:2018}
P.~Krishnamurthy, F.~Khorrami, R.~Karri, D.~Paul{-}Pena, and H.~Salehghaffari,
  ``Process-aware covert channels using physical instrumentation in
  cyber-physical systems,'' \emph{{IEEE} Trans. Inf. Forensics Secur.},
  vol.~13, no.~11, pp. 2761--2771, 2018.

\bibitem{hildebrandt2020threat}
M.~Hildebrandt, R.~Altschaffel, K.~Lamsh{\"o}ft, M.~Lange, M.~Szemkus,
  T.~Neubert, C.~Vielhauer, Y.~Ding, and J.~Dittmann, ``Threat analysis of
  steganographic and covert communication in nuclear {I\&C} systems,'' in
  \emph{Int. Conf. Nuclear Security: Sust. \& Strength. Efforts}.\hskip 1em
  plus 0.5em minus 0.4em\relax IAEA, 2020.

\bibitem{Lamshoeft22:CPSStego}
K.~Lamsh{\"{o}}ft, T.~Neubert, C.~Kr{\"{a}}tzer, C.~Vielhauer, and J.~Dittmann,
  ``Information hiding in cyber physical systems: Challenges for embedding,
  retrieval and detection using sensor data of the {SWAT} dataset,'' in
  \emph{Proc. IH{\&}MMSec '21}.\hskip 1em plus 0.5em minus 0.4em\relax {ACM},
  2021, pp. 113--124.

\bibitem{Millen:20Years}
J.~K. Millen, ``20 years of covert channel modeling and analysis,'' in
  \emph{1999 {IEEE} Symposium on Security and Privacy}.\hskip 1em plus 0.5em
  minus 0.4em\relax {IEEE}, 1999, pp. 113--114.

\bibitem{XuNWAA19}
Q.~Xu, H.~Naghibijouybari, S.~Wang, N.~B. Abu{-}Ghazaleh, and M.~Annavaram,
  ``{GPUGuard}: mitigating contention based side and covert channel attacks on
  {GPUs},'' in \emph{Proc. Int. Conf. Supercomp.}\hskip 1em plus 0.5em minus
  0.4em\relax {ACM}, 2019.

\bibitem{urbanski2017detecting}
M.~Urbanski, W.~Mazurczyk, J.~Lalande, and L.~Caviglione, ``Detecting local
  covert channels using process activity correlation on {Android}
  smartphones,'' \emph{Comput. Syst. Sci. Eng.}, vol.~32, no.~2, 2017.

\bibitem{BlockNN17}
K.~Block, S.~Narain, and G.~Noubir, ``An autonomic and permissionless android
  covert channel,'' in \emph{Proc. Conf. Security and Privacy in Wireless and
  Mobile Networks, WiSec 2017}.\hskip 1em plus 0.5em minus 0.4em\relax {ACM},
  2017, pp. 184--194.

\bibitem{hanspach2014covert}
M.~Hanspach and M.~Goetz, ``On covert acoustical mesh networks in air,''
  \emph{J. Commun.}, vol.~8, no.~11, pp. 758--767, 2013.

\bibitem{guri2015bitwhisper}
M.~Guri, M.~Monitz, Y.~Mirski, and Y.~Elovici, ``{BitWhisper}: Covert signaling
  channel between air-gapped computers using thermal manipulations,'' in
  \emph{{IEEE} 28th Computer Security Foundations Symposium, {CSF} 2015}.\hskip
  1em plus 0.5em minus 0.4em\relax {IEEE}, 2015, pp. 276--289.

\bibitem{guri2017led}
M.~Guri, B.~Zadov, and Y.~Elovici, ``{LED-it-GO}: Leaking {(A} lot of) data
  from air-gapped computers via the (small) hard drive {LED},'' in \emph{Proc.
  {DIMVA} 2017}, ser. LNCS, vol. 10327.\hskip 1em plus 0.5em minus 0.4em\relax
  Springer, 2017, pp. 161--184.

\bibitem{guri2018mosquito}
M.~Guri, Y.~A. Solewicz, and Y.~Elovici, ``{MOSQUITO:} covert ultrasonic
  transmissions between two air-gapped computers using speaker-to-speaker
  communication,'' in \emph{{IEEE} Conference on Dependable and Secure
  Computing, {DSC} 2018}.\hskip 1em plus 0.5em minus 0.4em\relax {IEEE}, 2018,
  pp. 1--8.

\bibitem{gong2021enhancing}
C.~Gong, X.~Yue, Z.~Zhang, X.~Wang, and X.~Dai, ``Enhancing physical layer
  security with artificial noise in large-scale {NOMA} networks,'' \emph{IEEE
  Trans. on Vehicular Technology}, vol.~70, no.~3, pp. 2349--2361, 2021.

\bibitem{mucchi2022security}
L.~Mucchi, S.~Caputo, P.~Marcocci, G.~Chisci, L.~Ronga, and E.~Panayirci,
  ``Security and reliability performance of noise-loop modulation: Theoretical
  analysis and experimentation,'' \emph{IEEE Transactions on Vehicular
  Technology}, vol.~71, no.~6, pp. 6335--6350, 2022.

\bibitem{Cabuk06}
S.~Cabuk, ``Network covert channels: Design, analysis, detection, and
  elimination,'' Ph.D. dissertation, Purdue University, 2006.

\bibitem{CabukTimingChanDet_2009}
S.~Cabuk, C.~E. Brodley, and C.~Shields, ``{IP} covert channel detection,''
  \emph{{ACM} Trans. Inf. Syst. Secur.}, vol.~12, no.~4, pp. 22:1--22:29, 2009.

\bibitem{Cabuk:2004:ICT:1030083.1030108}
------, ``{IP} covert timing channels: design and detection,'' in \emph{Proc.
  {CCS} 2004}.\hskip 1em plus 0.5em minus 0.4em\relax {ACM}, 2004, pp.
  178--187.

\bibitem{JitterBug}
G.~Shah and A.~Molina, ``Keyboards and covert channels,'' in \emph{Proc.
  {USENIX} Security Symposium}.\hskip 1em plus 0.5em minus 0.4em\relax {USENIX}
  Assoc., 2006.

\bibitem{Walls2011Liquid}
R.~J. Walls, K.~Kothari, and M.~K. Wright, ``Liquid: {A} detection-resistant
  covert timing channel based on {IPD} shaping,'' \emph{Comput. Networks},
  vol.~55, no.~6, pp. 1217--1228, 2011.

\bibitem{ModelbasedCTC}
S.~Gianvecchio, H.~Wang, D.~Wijesekera, and S.~Jajodia, ``Model-based covert
  timing channels: Automated modeling and evasion,'' in \emph{Proc. {RAID}
  2008}, ser. LNCS, vol. 5230.\hskip 1em plus 0.5em minus 0.4em\relax Springer,
  2008, pp. 211--230.

\bibitem{DBLP:conf/prdc/YarochkinDLHK08}
F.~V. Yarochkin, S.~Dai, C.~Lin, Y.~Huang, and S.~Kuo, ``Towards adaptive
  covert communication system,'' in \emph{14th {IEEE} Pacific Rim International
  Symp. on Dependable Comp., {PRDC} 2008}.\hskip 1em plus 0.5em minus
  0.4em\relax {IEEE}, 2008, pp. 153--159.

\bibitem{Fridich2009}
J.~Fridrich, \emph{Steganography in Digital Media: Principles, Algorithms, and
  Applications}.\hskip 1em plus 0.5em minus 0.4em\relax Cambridge University
  Press, 2009.

\bibitem{CoverSelection2}
M.~Kharrazi, H.~T. Sencar, and N.~D. Memon, ``Cover selection for
  steganographic embedding,'' in \emph{Proc. {ICIP}}.\hskip 1em plus 0.5em
  minus 0.4em\relax {IEEE}, 2006, pp. 117--120.

\bibitem{CoverSelectionImageSimilarity}
Z.~Wang, G.~Feng, L.~Shen, and X.~Zhang, ``Cover selection for steganography
  using image similarity,'' \emph{{IEEE} Trans. Dependable Secur. Comput.},
  vol.~20, no.~3, pp. 2328--2340, 2023.

\bibitem{CoverSelection3}
O.~Evsutin, A.~Kokurina, and R.~V. Meshcheryakov, ``Approach to the selection
  of the best cover image for information embedding in {JPEG} images based on
  the principles of the optimality,'' \emph{J. Decis. Syst.}, vol.~27, no.
  Supp, pp. 256--264, 2018.

\bibitem{CoverSelection4}
H.~Sajedi and M.~Jamzad, ``Cover selection steganography method based on
  similarity of image blocks,'' in \emph{Proc. Int. Conf. Computer and
  Information Technology Workshops}.\hskip 1em plus 0.5em minus 0.4em\relax
  IEEE, 2008, pp. 379--384.

\bibitem{CoverSelection5}
------, ``Using contourlet transform and cover selection for secure
  steganography,'' \emph{Int. J. Inf. Sec.}, vol.~9, no.~5, pp. 337--352, 2010.

\bibitem{CoverSelection6}
S.~Wu, Y.~Liu, S.~Zhong, and Y.~Liu, ``What makes the stego image
  undetectable?'' in \emph{Proc. Int. Conf. Internet Multimedia Computing and
  Service, {ICIMCS} 2015}.\hskip 1em plus 0.5em minus 0.4em\relax {ACM}, 2015,
  pp. 47:1--47:6.

\bibitem{CoverlessImageStego}
Z.~Zhou, Y.~Mu, and Q.~M.~J. Wu, ``Coverless image steganography using
  partial-duplicate image retrieval,'' \emph{Soft Comput.}, vol.~23, no.~13,
  pp. 4927--4938, 2019.

\bibitem{CoverlessImageStego2}
H.~Liu, C.~Zhang, Z.~Wang, C.~Guo, P.~Gou, L.~Shan, and Z.~Lu, ``To deliver
  more information in coverless information hiding,'' \emph{Multim. Tools
  Appl.}, vol.~83, no.~3, pp. 7215--7229, 2024.

\bibitem{DBLP:conf/wcnis/JiFM10}
L.~Ji, Y.~Fan, and C.~Ma, ``Covert channel for local area network,'' in
  \emph{Proc. {IEEE} Int. Conf. Wireless Communications, Networking and
  Information Security, {WCNIS} 2010}.\hskip 1em plus 0.5em minus 0.4em\relax
  {IEEE}, 2010, pp. 316--319.

\bibitem{Schmidbauer:DeadDrops:ARP}
T.~Schmidbauer, S.~Wendzel, A.~Mileva, and W.~Mazurczyk, ``Introducing dead
  drops to network steganography using {ARP}-caches and {SNMP}-walks,'' in
  \emph{Proc. ARES 2019}.\hskip 1em plus 0.5em minus 0.4em\relax ACM, 2019.

\bibitem{DBLP:conf/otm/MazurczykS08}
W.~Mazurczyk and K.~Szczypiorski, ``Steganography of {VoIP} streams,'' in
  \emph{Proc. {OTM} 2008}, ser. LNCS, R.~Meersman and Z.~Tari, Eds., vol.
  5332.\hskip 1em plus 0.5em minus 0.4em\relax Springer, 2008, pp. 1001--1018.

\bibitem{DBLP:conf/iih-msp/BaiHHX08}
L.~Y. Bai, Y.~Huang, G.~Hou, and B.~Xiao, ``Covert channels based on jitter
  field of the {RTCP} header,'' in \emph{Proc. {IIH-MSP} 2008}.\hskip 1em plus
  0.5em minus 0.4em\relax {IEEE}, 2008, pp. 1388--1391.

\bibitem{lizhiRTP}
L.~Ying, Y.~Huang, J.~Yuan, and L.~Y. Bai, ``A novel covert timing channel
  based on rtp/rtcp,'' \emph{Chinese Journal of Electronics}, vol.~21, no.~4,
  pp. 711--714, 2012.

\bibitem{wendzel2014hidden}
S.~Wendzel and J.~Keller, ``Hidden and under control - {A} survey and outlook
  on covert channel-internal control protocols,'' \emph{Ann. des
  T{\'{e}}l{\'{e}}communications}, vol.~69, no. 7-8, pp. 417--430, 2014.

\bibitem{BotnetBookCh2019}
L.~Caviglione, W.~Mazurczyk, and S.~Wendzel, ``Advanced information hiding
  techniques for modern botnets,'' in \emph{Botnets: Architectures,
  Countermeasures, and Challenges}.\hskip 1em plus 0.5em minus 0.4em\relax CRC
  Press, 2019, pp. 165--188.

\bibitem{HICCUPS}
J.~Lubacz, W.~Mazurczyk, and K.~Szczypiorski, ``Hiding data in {VoIP},'' in
  \emph{Proc. 26th Army Science Conference}, 2008.

\bibitem{BerkEtAl}
V.~Berk, A.~Giani, and G.~Cybenko, ``Detection of covert channel encoding in
  network packet delays,'' Dartmouth College, Department of Computer Science,
  Technical Report TR2005-536, 2005.

\bibitem{zhang2018covert}
X.~Zhang, Y.~Tan, C.~Liang, Y.~Li, and J.~Li, ``A covert channel over {VoLTE}
  via adjusting silence periods,'' \emph{{IEEE} Access}, vol.~6, 2018.

\bibitem{SchmidbauerWendzel:IndirectCCSurvey}
T.~Schmidbauer and S.~Wendzel, ``{SoK}: {A} survey of indirect network-level
  covert channels,'' in \emph{{ASIA} {CCS} '22: {ACM} Asia Conference on
  Computer and Communications Security}.\hskip 1em plus 0.5em minus 0.4em\relax
  {ACM}, 2022, pp. 546--560.

\bibitem{RFC3550}
H.~Schulzrinne, S.~L. Casner, R.~Frederick, and V.~Jacobson, ``{RTP: A
  Transport Protocol for Real-Time Applications},'' RFC 3550, Jul. 2003.

\bibitem{interprotocolstego}
F.~Lehner, W.~Mazurczyk, J.~Keller, and S.~Wendzel, ``Inter-protocol
  steganography for real-time services and its detection using traffic coloring
  approach,'' in \emph{Proc. {LCN} 2017}.\hskip 1em plus 0.5em minus
  0.4em\relax {IEEE}, 2017, pp. 78--85.

\bibitem{OvadiaOMGO19}
A.~Ovadya, R.~Ogen, Y.~Mallah, N.~Gilboa, and Y.~Oren, ``Cross-router covert
  channels,'' in \emph{Proc. {WOOT} '19}.\hskip 1em plus 0.5em minus
  0.4em\relax {USENIX} Assoc., 2019.

\bibitem{ClarkECC1981}
G.~C. {Clarke, Jr.} and J.~B. Cain, \emph{Error-Correction Coding for Digital
  Communication}.\hskip 1em plus 0.5em minus 0.4em\relax Springer
  Science+Business Media, 1981.

\bibitem{Vermani1996}
L.~R. Vermani, \emph{Elements of Algebraic Coding Theory}.\hskip 1em plus 0.5em
  minus 0.4em\relax Routledge, 1996.

\bibitem{SemiPassiveSemiActive}
K.~lamshöft and J.~Dittmann, ``Assessment of hidden channel attacks:
  Targetting {M}odbus/{TCP},'' \emph{IFAC-PapersOnLine}, vol.~53, no.~2, 2020.

\bibitem{KS_Test}
F.~J. Massey, ``The {K}olmogorov-{S}mirnov test for goodness of fit,''
  \emph{Journal of the Amer. Stat. Assoc.}, vol.~46, no. 253, pp. 68--78, 1951.

\bibitem{NIHbook}
W.~Mazurczyk, S.~Wendzel, S.~Zander, A.~Houmansadr, and K.~Szczypiorski,
  \emph{Information Hiding in Communication Networks: Fundamentals, Mechanisms,
  and Applications}.\hskip 1em plus 0.5em minus 0.4em\relax Wiley, 2016.

\bibitem{OPPRESSION}
S.~Zillien, T.~Schmidbauer, M.~Kubek, J.~Keller, and S.~Wendzel, ``Look what's
  there! {Utilizing} the {Internet}'s existing data for censorship
  circumvention with {OPPRESSION},'' in \emph{Proc. 19th ACM ASIA Conf. on
  Comp. \& Comm. Secur. (AsiaCCS'24)}.\hskip 1em plus 0.5em minus 0.4em\relax
  ACM, 2024, in press.

\end{thebibliography}
